%% file: main.tex
\newif\iftr
\newif\ifconf
\newif\ifconfformat

\newif\ifsq
\newif\ifsqCAP
\newif\ifsqVS
\newif\ifsqEN
\newif\ifsqTIT

\trtrue

\iftr
\conffalse
\confformatfalse
\else
\conftrue
\confformattrue
\fi

\ifconfformat
\documentclass[lettersize,journal]{IEEEtran}
\else
\documentclass[lettersize,journal]{IEEEtran_custom}

\usepackage[sort&compress, numbers]{natbib}
\usepackage[hidelinks]{hyperref}
\fi

\usepackage{amsmath,amssymb,amsfonts}
\usepackage[linesnumbered,ruled]{algorithm2e}
\usepackage{graphicx}
\usepackage{textcomp}
\usepackage{xcolor}
\definecolor{rpegrevision}{RGB}{176,36,74}
\newcommand{\revisionnote}[1]{%
  \par\smallskip\noindent{\footnotesize\color{rpegrevision}%
  \textbf{[Pierre]} #1}\par\smallskip}

\usepackage{fontawesome}
\usepackage{enumitem}
\usepackage{moresize}

\renewcommand{\revisionnote}[1]{}

\usepackage{booktabs}
\usepackage{makecell}
\usepackage{xurl}
\usepackage{colortbl}

\newcommand{\faY}[0]{\faBatteryFull}
\newcommand{\faH}[0]{\faBatteryHalf}
\newcommand{\faN}[0]{\faTimes}

\newcommand{\nameS}{RLM Pipeline Execution Graph\ }
\newcommand{\nameA}{RPEG}
\newcommand{\nameAS}{RPEG\ }

\ifconfformat
\fi

\usepackage{amsmath,amssymb,mathrsfs}
\usepackage{graphicx,booktabs,tabularx,array,multirow,longtable,makecell}
\usepackage{enumitem,float,placeins,rotating,listings,seqsplit}
\usepackage{tikz}
\usetikzlibrary{positioning,arrows.meta,fit,backgrounds,calc,shapes.geometric}
\usepackage[nameinlink,capitalise]{cleveref}

\newcommand{\newdevelopment}[1]{{\textcolor{violet}\ifmmode\boldsymbol{#1}\else\bfseries\boldmath #1\fi}}

\newcolumntype{Y}{>{\raggedright\arraybackslash}X}
\makeatletter
\@ifundefined{iftr}{\newif\iftr\trfalse}{}
\makeatother

\newtheorem{rpegdefinition}{Definition}[section]
\newtheorem{rpeglemma}{Lemma}[section]
\newtheorem{rpegexample}{Example}[section]
\newenvironment{rpegproof}[1][Proof]{\par\noindent\textit{#1.}\ }{\hfill\(\square\)\par}
\graphicspath{{PDFs/}{Figures/}{figures/generated/}{../papers/RPEG/PDFs/}}

\makeatletter
\AtBeginDocument{\@ifundefined{crefname}{}{%
  \crefalias{paragraph}{section}%
  \Crefname{paragraph}{Paragraph}{Paragraphs}%
  \crefname{rpegdefinition}{Definition}{Definitions}%
  \Crefname{rpegdefinition}{Definition}{Definitions}%
  \crefname{rpeglemma}{Lemma}{Lemmas}%
  \Crefname{rpeglemma}{Lemma}{Lemmas}%
}}
\makeatother

\usepackage{moresize}
\usepackage{inconsolata}

\input{mac-compression}

\begin{document}

\title{\nameA: Training Reasoning Language Models \\ Scalably and Cost-Effectively}

\ifconfformat
\author{
Maciej Besta, Leonard Schmidt, Pierre Pang, Patrik Okanovic, Robert Gerstenberger, Torsten Hoefler
\IEEEmembership{Fellow, IEEE}
\thanks{M. Besta is with the Department of Computer Science, ETH Zurich, Zurich, Switzerland}
\thanks{L. Schmidt is with the Department of Computer Science, ETH Zurich, Zurich, Switzerland}
\thanks{P. Pang is with the Department of Computer Science, ETH Zurich, Zurich, Switzerland}
\thanks{P. Okanovic is with the Department of Computer Science, ETH Zurich, Zurich, Switzerland}
\thanks{R. Gerstenberger is with the Department of Computer Science, ETH Zurich, Zurich, Switzerland}
\thanks{T. is with the Department of Computer Science, ETH Zurich, Zurich, Switzerland}
}
\else
\author{Maciej Besta$^{*\dagger}$\thanks{$^*$Corresponding author}\thanks{$^\dagger$Core contributions}, 
Leonard Schmidt$^{\dagger}$, 
Pierre Pang$^{\dagger}$, 
Patrik Okanovic, 
Robert Gerstenberger, 
Torsten Hoefler\\ 
{\small Department of Computer Science, ETH Zurich}
}
\fi

\maketitle

\begin{abstract}
Reinforcement learning (RL) is now central to training Reasoning Language Models (RLMs), but RLM pipelines are difficult to scale and costly to run because each RL step couples multiple large models, heterogeneous stages, cross-step state, and complex asynchronous dependencies. Existing systems typically optimize only parts of this design space. We address this complexity by jointly optimizing the pipeline structure as well as its execution using a novel data-centric graph intermediate representation called \nameS (\nameA). \nameAS exposes structural choices such as bounded asynchrony and actor--critic sharing together with execution optimizations such as stage fusion, fine-grained scheduling, and multi-dimensional model parallelism. It is coupled with a comprehensive analytical performance model for RLM training, jointly capturing compute, communication, memory, and hierarchical rack-, node-, and GPU-level effects. Across systems with up to 256 GH200 GPUs, our distributed runtime engine implementing \nameAS improves performance by up to 2$\times$ over ReaL and reduces step latency by up to 2.48$\times$ over verl, while increasing hardware utilization by up to 30\%. Overall, \nameAS provides a graph-driven foundation for next-generation training pipelines for frontier reasoning LLMs.
\end{abstract}

\ifconfformat
\fi

\noindent \textbf{Code:} \url{https://github.com/spcl/RPEG}

\input{intro}
\input{background}
\input{overview}
\input{formulation}
\input{model}

\input{optimization}
\input{runtime}
\input{eval}
\input{rw}
\input{conc}

\iftr
\input{acks}
\fi

\ifconfformat
\else
\phantomsection
\addcontentsline{toc}{section}{References}
\bibliographystyle{IEEEtranS}
\bibliography{references.complete}
\fi

\begin{appendices}

\input{appendix-pipeline}
\input{appendix-eval}
\input{appendix-model}
\input{appendix-derivations}

\end{appendices}

\ifconfformat
\input{acks}
\fi

\end{document}

%% file: mac-compression.tex
\newif\ifsq
\newif\ifsqCAP
\newif\ifsqVS
\newif\ifsqEN
\newif\ifsqTIT
\newif\ifsqALG

\sqALGtrue

\ifconf

\sqtrue
\sqCAPtrue
\sqENtrue
\sqVStrue
\sqTITtrue

\else

\sqfalse
\sqCAPfalse
\sqENfalse
\sqVSfalse
\sqTITfalse

\fi

\ifsqCAP
\usepackage[font={normalfont, footnotesize}]{caption}
\usepackage[font={normalfont, footnotesize}]{subcaption}

\SetAlCapFnt{\footnotesize}
\SetAlCapNameFnt{\footnotesize}
\else
\usepackage[font={normalfont, small}]{caption}
\fi

\ifsqALG
\SetAlFnt{\footnotesize}
\SetAlCapFnt{\footnotesize}
\SetAlCapNameFnt{\footnotesize}
\fi

\newcommand{\vspaceSQ}[1]{\ifsqVS\vspace{#1}\fi}
\newcommand{\enlargeSQ}[1]{\ifsqEN\enlargethispage{\baselineskip}\fi}

\ifsqTIT
\usepackage[compact]{titlesec}
\titlespacing*{\section}{0pt}{4pt}{2pt}
\titlespacing*{\subsection}{0pt}{4pt}{2pt}
\titlespacing*{\subsubsection}{0pt}{2pt}{1pt}
\fi

%% file: intro.tex
\section{Introduction}

\IEEEPARstart{R}{einforcement} learning (RL) has become central to post-training large language models (LLMs) for human preferences, safety constraints, and task-specific objectives~\cite{stiennon2020learning, ouyang2022training, bai2022helpful}. More recently, RL-based post-training has become the foundation of \emph{reasoning language models} (RLMs, also called \textit{reasoning models}): models that perform extended reasoning, produce chain-of-thought (CoT) trajectories~\cite{wei2022chain}, exploit richer structures~\cite{yao2023tree, besta2025reasoning, besta2023graph, yao2023beyond, besta2025demystifying}, self-correct~\cite{besta2024checkembed}, and allocate more computation to harder problems~\cite{guo2025deepseekr1, openai2025o3systemcard, du2025kimi}. Examples include DeepSeek-R1~\cite{guo2025deepseekr1}, GPT-5 Thinking~\cite{singh2025openai}, Kimi K2.5 Thinking~\cite{kimiteam2026kimik25visualagentic}, and Grok 4.7~\cite{xai2026grok47}. Here, RL becomes a mechanism by which models learn to reason. The computational footprint is enormous and continues to grow: frontier-scale models consume millions of GPU-hours, while scaling both train-time RL and inference-time reasoning continues to improve model capabilities~\cite{liu2025deepseekv3, openai2024o3, cottier2024frontiercost}.
These costs grow further in agentic systems~\cite{yao2023react,wang2024agentsurvey,
mialon2024gaia,wu2024autogen,li2023camel,yang2024sweagent, besta2025affordable, besta2024multi, besta2025psychologically, besta2026graphseek}: long-horizon tool use and
multi-agent execution increase system overhead~\cite{kim2026cost}, with
reported token usage reaching \(4\times\)--\(15\times\) that of ordinary
chat~\cite{anthropic2025multiagent} and roughly three orders of magnitude
higher for agentic coding workloads~\cite{bai2026agents}.

Yet, RLM post-training is {substantially more complex than a single-model workload}. In the canonical Proximal Policy Optimization (PPO) pipeline, each step comprises a \emph{generation} stage that samples responses with a trainable \textbf{actor} (also called \textbf{policy}), an \emph{assessment} stage that evaluates them using a \textbf{reward model}, \textbf{critic}, and frozen \textbf{reference model}, and a \emph{training} stage that updates the trainable components~\cite{mei2024real, zheng2023secrets}. Other widely used online RLM-training paradigms, such as Group Relative Policy Optimization (GRPO)~\cite{shao2024deepseekmath}, simplify parts of this pipeline (for example by removing the critic) but retain essentially the same generation--assessment--training template and its heterogeneous execution characteristics. In PPO, generation is autoregressive and latency-sensitive, assessment is largely teacher-forced and throughput-oriented, while training adds backward passes, optimizer state, and parameter updates. RLM training is therefore simultaneously {multi-model}, {multi-stage}, {heterogeneous}, and {iterative} -- and fundamentally a parallel and distributed computing problem~\cite{weingram2025hpc}.

\begin{figure*}[t]
    \centering
    \includegraphics[width=1.0\textwidth]{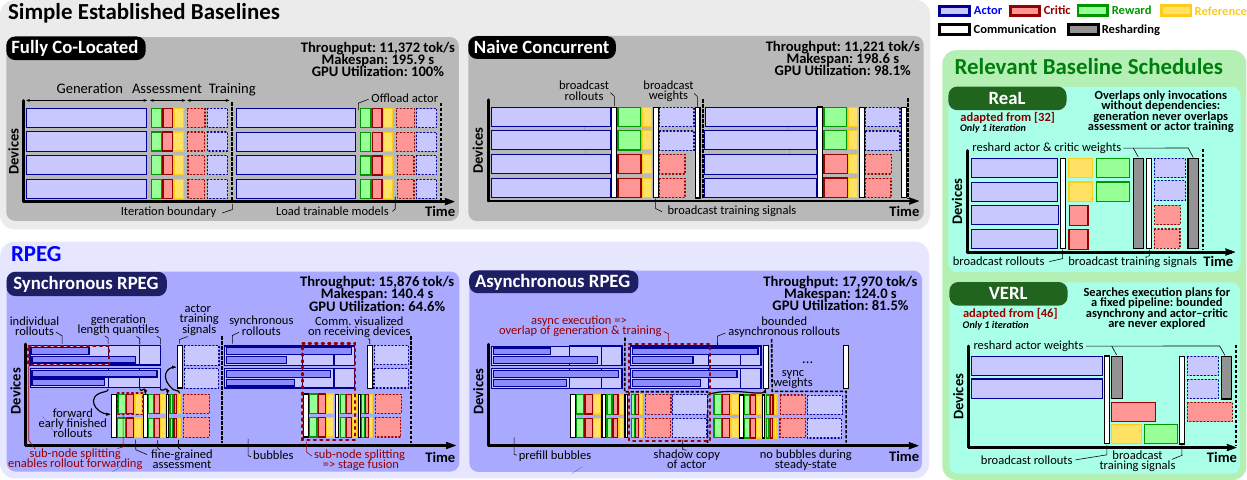}
    \caption{\textbf{Representative PPO schedules selected by different planners.} \nameAS exposes stage-level and fine-grained overlap, enabling disaggregated placements and bounded-asynchrony that are not available under fully colocated or coarse-stage baselines. \textit{No AI was used to conceive or draw the picture.}}
    \label{fig:schedules}
    \vspace{-1em}
\end{figure*}

Scaling individual models does not by itself optimize an RLM training
pipeline. Data (DP), tensor (TP), pipeline (PP), context (CP), sequence (SP), and expert parallelism (EP)
must be coordinated with cross-stage placement, state exchange, and
execution order~\cite{besta2026performance, ben2019demystifying}. Existing systems pursue three roughly complementary directions:
\textbf{(1)} {effective execution planning}, including placement and
parallel-strategy selection in ReaL~\cite{mei2024real} and
HybridFlow/\textsc{verl}~\cite{sheng2025hybridflow}, and workflow compilation
in RLinf~\cite{yu2025rlinf};
\textbf{(2)} {fine-grained and asynchronous execution}, including
sample- and microbatch-level fusion in RLHFuse~\cite{zhong2025rlhfuse}
and decoupled rollout--training pipelines in AReaL~\cite{fu2025areal},
Laminar~\cite{sheng2025laminar}, and StaleFlow~\cite{li2026unleashing};
and \textbf{(3)} {adaptive resource orchestration}, which adjusts
parallelism or reallocates resources as workloads
evolve~\cite{tan2026orchestrl,chen2026libra,tan2026bidirl}.
These mechanisms increasingly coexist within the same system.

The key design question is therefore how broadly these decisions can be
optimized together. Aspects such as overlap, placement, parameter freshness, model sharing, parallelism, sharding vs.~replication, and memory cannot be optimized independently: improving one may increase communication, resharding, or memory pressure elsewhere. This interaction becomes even more important on hierarchical supercomputers, where communication costs differ substantially within a node, across nodes, and across racks. Finally, maximizing throughput alone may still leave expensive hardware poorly utilized. Taken together, these couplings make it difficult to achieve both high performance and high hardware utilization while respecting communication and memory constraints.

We address this challenge by optimizing not only \emph{how} an RLM pipeline executes, but also selected aspects of \emph{the pipeline structure itself} (\textbf{contribution~1}). We represent both jointly as a single optimization object \(C=(G^\circ,P)\), exposing structural and execution decisions within one searchable configuration space. Specifically, the compact steady-state graph \(G^\circ\) captures the recurring pipeline structure and cross-step state semantics, including structural variants such as bounded asynchrony and actor--critic (AC) sharing~\cite{besta2026performance}; the execution plan \(P\) specifies per-invocation hardware mapping, multi-dimensional model parallelism, and microbatching. Thus, the pipeline itself becomes an optimization variable rather than merely a fixed graph to be mapped onto hardware.

We realize this abstraction in the \textbf{\nameS (\nameA)}, a data-centric
graph intermediate representation and optimization framework
(\textbf{contribution~2}). Inspired by property graphs from the graph database
domain~\cite{besta2022neural, besta2023gdi, besta2023demystifying}, RPEG is property-rich rather than a bare dependency
graph, compactly attaching rich data (e.g., the model, role, staleness,
execution semantics) needed for subsequent optimization. A staged graph-transformation pipeline materializes
execution refinements such as stage fusion and fine-grained sub-invocation
execution, and converts each candidate configuration \(C\) into a schedulable
hardware-aware graph. This allows the optimizer to evaluate structural and
execution choices jointly without directly searching a fully materialized
multi-step graph. 
We showcase RPEG execution plans in Figure~\ref{fig:schedules}.

We further develop a distributed-memory runtime
(\textbf{contribution~3}) that executes the selected RPEG plans faithfully on
supercomputers, connecting the graph-level optimization decisions to their
physical execution.

RPEG is coupled with a calibrated comprehensive analytical performance model for end-to-end RLM training (\textbf{contribution~4}). The model jointly captures compute, communication, and memory costs of heterogeneous RLM operators, including resharding, pipeline bubbles, optimizer traffic, co-resident model state, and hierarchical GPU-, node-, rack-, and inter-rack communication. Automated hardware calibration replaces reliance on nominal vendor peaks, enabling rapid evaluation of large numbers of candidate plans.

Finally, we comprehensively evaluate RPEG from analytical modeling to large-scale distributed execution (\textbf{contribution~5}). We validate the performance and memory models, study optimizer stability and ablations, and execute RPEG plans on up to 256 GH200 GPUs. Optimized synchronous plans are 1.250--1.471$\times$ faster than their executable fully colocated inputs; against ReaL, RPEG improves end-to-end performance by up to 2$\times$; and in the controlled synchronous comparison with separate actor and critic models, RPEG achieves up to 2.477$\times$ lower time per step than \textsc{verl} (median 1.488$\times$). RPEG also improves hardware utilization by up to 30\%, while experiments with AC sharing, stage fusion, and bounded asynchrony expose further performance gains and tradeoffs enabled by the representation, all while preserving training semantics and accuracy (within 0.47\% of the \textsc{verl} baseline), and with negligible optimizer overheads even at 2k GPUs.

%% file: background.tex
\section{Basics of RLM Pipelines}
\label{rpegx:sec-basics}

We call one generation--assessment--training cycle an \emph{RL step}.
An \emph{invocation} is one model's execution in a particular role, such as
actor generation or critic training; it may itself contain many
token-generation or forward/backward operations. A \emph{rollout} is a
sampled response to a prompt, and its successive token-generation decisions
form a trajectory.

In PPO-style RLHF, the trainable \emph{actor} generates responses;
the frozen \emph{reward model} scores them; the trainable \emph{critic}
estimates token-level future returns; and the frozen \emph{reference model}
supplies log-probabilities for policy regularization. One RL step therefore
contains six invocations: actor generation, reward/reference/critic
inference, and actor/critic training~\cite{mei2024real,zheng2023secrets}.
PPO compares current-policy probabilities with those of the policy that
generated the responses and uses reward/value estimates to construct
advantages, which guide the update~\cite{schulman2017proximal,ouyang2022training}.
GRPO removes the critic and estimates advantages from rewards within groups
of responses, but retains the generation--assessment--training
template~\cite{shao2024deepseekmath}.

These invocations have different computational profiles. Generation first
processes the prompt (\emph{prefill}) and then decodes tokens sequentially,
making it latency-sensitive and increasingly memory-bound as the KV cache
grows. Assessment instead processes complete sequences with highly parallel
forward passes and is typically throughput-oriented. Training adds
backpropagation, optimizer state, and parameter updates, substantially
increasing both computation and memory requirements. \textbf{Intra-model parallelism} distributes one invocation using standard
techniques such as data (DP), tensor (TP), pipeline (PP), sequence (SP),
context (CP), and expert parallelism (EP)~\cite{ben2019demystifying}. A \emph{microbatch} is a smaller batch
partition processed within an invocation. A \emph{microtask}, such as one
pipeline stage's forward or backward computation for one microbatch, is a
finer execution unit. Sub-invocation execution exposes units smaller than a whole model invocation,
such as microbatches, pipeline microtasks, or groups of responses that finish
at similar lengths, to the scheduler.

\textbf{Inter-model strategies} organize different model
invocations~\cite{besta2026performance}. \emph{Co-location} places them on
overlapping GPU sets; \emph{disaggregation} uses disjoint sets. Different
invocations of one model may use different parallel layouts, requiring
\emph{resharding}, i.e., redistribution of parameter tensors.
\emph{Stage fusion} overlaps work within an RL step: assessment of completed
responses can overlap continuing generation, or actor/critic training
microtasks can interleave while retaining separate parameters.
\emph{Bounded asynchrony} instead overlaps different RL steps by allowing
serving invocations (generation or assessment) to use parameter versions
delayed by at most a configured number of RL steps.
\emph{Actor--critic (AC) sharing} uses one backbone with policy \& value
heads instead of two separate models. Unlike fusion, AC sharing changes
the trainable model; asynchrony changes parameter freshness. Both are
thus explicit configuration choices rather than universally
interchangeable implementations of synchronous, separate-model PPO.

%% file: overview.tex
\begin{figure*}[t]
    \centering
    \iftr
    \includegraphics[width=1.0\textwidth]{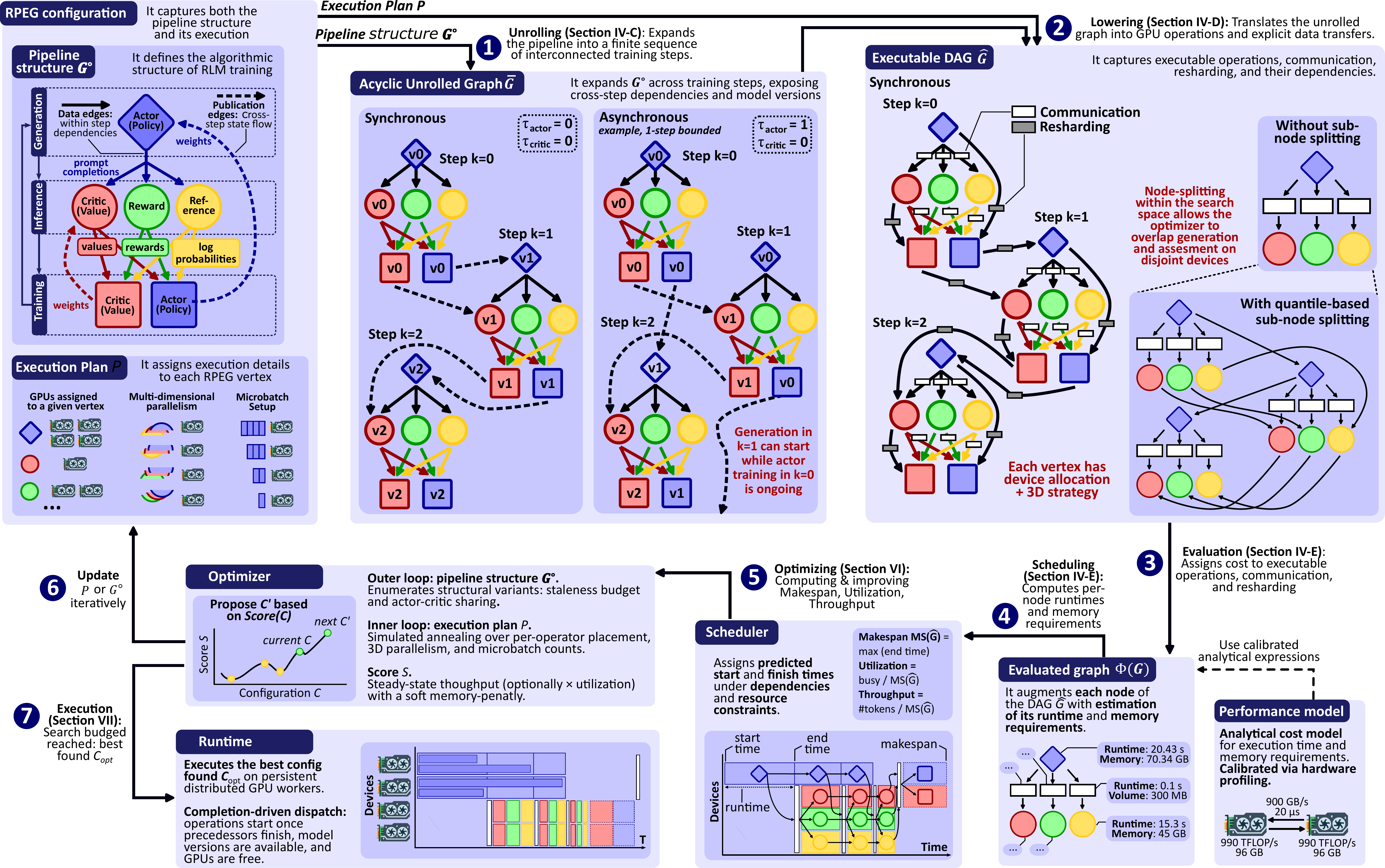}
    \else
    \includegraphics[width=1.0\textwidth]{rpeg_overview_v12.pdf}
    \fi
    \caption{\textbf{Overview of the \nameA.} A steady-state graph template and execution plan are unrolled, lowered to an executable DAG with explicit communication and resharding, evaluated with local analytical costs, and scheduled to obtain a steady-state score. \textit{No AI was used to conceive or draw the picture.}}
    \label{fig:example}
\end{figure*}

\section{Overview of \nameA}
\label{sec:overview}

\nameA's key idea is to optimize not only \emph{how} an RLM training
pipeline executes, but also selected aspects of \emph{the pipeline structure
itself}. We represent the complete optimization object as an \textbf{\nameA\ configuration}
\(C=(G^\circ,P)\); \(G^\circ\) captures the
pipeline structure while \(P\) specifies its execution plan. 
Symbols used in the paper are listed in Table~\ref{tab:notation}.

\input{table-notation}

\textbf{The graph \(G^\circ\)} captures the
steady-state algorithmic structure of one RL step and cross-step semantics. Its vertices are
\emph{operators}, i.e., individual model invocations in specific roles such
as generation, inference, or training. Its edges encode how data and
updated model state flow between them within and across steps.
The representation is richer than a bare dependency graph.
Operators carry their role, associated model, and declared input/output sets used for consistency checks,
while edges distinguish within-step dataflow from cross-step model-state
publication and carry typed data containers.

\textbf{The execution plan $P$} specifies how each operator is mapped to hardware through
device allocation, multi-dimensional model parallelism, and microbatching.

During \textbf{staged graph transformation}, $C$ is progressively refined. For example, a model invocation may be
decomposed into finer-grained executable phases, while communication and
redistribution implied by the selected hardware mapping become explicit graph nodes.
Thus, the same graph-centric abstraction evolves from a compact logical
description into a detailed executable representation that
specifies the operations, data movement, and resource requirements
on a given cluster (Figure~\ref{fig:example}).
This transformation consists of the following steps:
\small
\[
C \xrightarrow{\textsc{Unroll}} \bar G
\xrightarrow{\textsc{Lower}} \hat G
\xrightarrow{\textsc{LocalEval}} \Phi(\hat G)
\xrightarrow{\textsc{Schedule+Score}} \text{Score}(C).
\]
\normalsize

\textit{{Unrolling}}~\includegraphics[scale=0.2,trim=0 16 0 0]{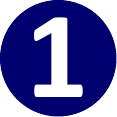} expands the cyclic steady-state graph IR over a
finite horizon \(K_{\mathrm{eval}}\), yielding an acyclic graph \(\bar G\)
in which cross-step publication dependencies become explicit. By
representing bounded-staleness publication already at the compact IR level,
this stage exposes asynchronous inter-step overlap while keeping the
search space compact. \textit{{Lowering}}~\includegraphics[scale=0.2,trim=0 16 0 0]{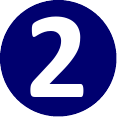} then combines \(\bar G\) with the
execution plan \(P\) and produces a hardware-aware executable DAG
\(\hat G\), inserting explicit \textsc{Comm} and \textsc{Reshard} nodes
whenever the selected mapping requires communication or redistribution.
Lowering also materializes finer-grained executable units, such as
response-length phases, when they enable scheduling below a whole model
invocation.

\textit{{Local evaluation}}~\includegraphics[scale=0.2,trim=0 16 0 0]{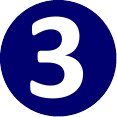} assigns analytical cost semantics to \(\hat G\):
it produces a node-level time map \(\mathcal{T}\) and estimates memory
requirements under the selected hardware mapping, without yet resolving
global resource contention. We denote the resulting graph, annotated with time/memory
estimates, by \(\Phi(\hat G)\). \textit{{Scheduling}}~\includegraphics[scale=0.2,trim=0 16 0 0]{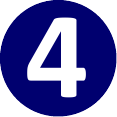} then constructs a
resource-aware execution order under precedence and device-reservation
constraints, yielding the predicted makespan \(\mathrm{MS}(\hat G)\) and scheduled
resource usage.
Finally,
the \textbf{optimizer} \textit{{scores}}~\includegraphics[scale=0.2,trim=0 16 0 0]{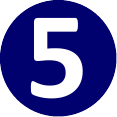} the configuration using steady-state throughput or,
optionally, a throughput--utilization objective, together with a soft
penalty for memory-infeasible candidates.
The optimizer treats this score as a search signal and iteratively \textit{{updates}}~\includegraphics[scale=0.2,trim=0 16 0 0]{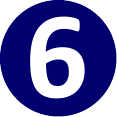}
both \(G^\circ\) and \(P\). In this way, the framework jointly optimizes
pipeline structure and execution mapping rather than treating the RLHF
graph as fixed.

After optimization, the \textbf{runtime} reconstructs the selected configuration and
its physical mapping and \textit{{executes}}~\includegraphics[scale=0.2,trim=0 16 0 0]{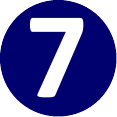} it on persistent distributed workers.
Operations start only when their actual predecessors have completed, the
required model versions are available, and the necessary resources are free.
Runtime traces verify the executed dependencies, model versions, transfers,
and resource ownership.

%% file: table-notation.tex
\iftr
\begin{table*}[t]
    \centering
    \small
    \setlength{\tabcolsep}{4pt}
\else
\begin{table}[t]
    \centering
    \sf
    \ssmall
    \setlength{\tabcolsep}{1pt}
    \renewcommand{\arraystretch}{0.8}
\fi
    \caption{\textbf{Main notation used throughout the paper.}}
    \label{tab:notation}
    \ifconf
    \vspace{-0.9em}
    \fi
    \begin{tabular}{@{} l l @{}}
        \toprule

        \multicolumn{2}{@{}l}{\textbf{Models \& their architecture and workload}} \\
        \midrule
        $\mathcal{M};\; \mathcal{M}_\mathrm{train}$ & Set of all models; a trainable subset of models ($\mathcal{M}_\mathrm{train} \subset \mathcal{M}$). \\
        $B;\; R$ & Batch size (input prompts per step); generated rollouts per prompt. \\ 
        $L_p;\; L_r$ & Average prompt length; average rollout length. \\ 
        $s_b$ & Number of bytes per scalar in working precision ($s_b = 2$ for BF16). \\
        \midrule
        \multicolumn{2}{@{}l}{\textbf{Hardware \& its architecture}} \\
        \midrule
        $D$ & Device space; $D = [N_r] \times [N_n] \times [N_g]$. \\
        $N_r;\ N_n;\ N_g$ & Number of racks; nodes/rack; GPUs/node. \\
        $\alpha;\; \beta$ & Latency; inverse bandwidth ($\alpha$--$\beta$ model). \\
        $F_\mathrm{peak}$ & Workload-effective GPU throughput (FLOP/s per GPU). \\
        $\mathrm{BW}_\mathrm{mem}$ & HBM bandwidth (bytes/s per GPU). \\
        $\gamma_M,\; \gamma_K$ & GEMM SM-utilization constants (batch dim.; hidden dim.). \\
        $M_\mathrm{GPU}, M_\mathrm{eff}$ & Per-GPU HBM capacity, effective capacity after runtime overhead. \\
        \midrule
        \multicolumn{2}{@{}l}{\textbf{RPEG}} \\
        \midrule
        $C$ & An RPEG configuration: $C = (G^\circ, P)$. \\
        $G^\circ$ & \makecell[l]{An RPEG: $G^\circ = (V^\circ, E^\circ_\mathrm{data}, E^\circ_\mathrm{pub}, \mathcal{A}, \mathcal{M}, \tau)$.\\ $V^\circ$: RPEG vertices, $E^\circ_\mathrm{data}$: data edges, $E^\circ_\mathrm{pub}$: publication edges.} \\
        $\mathcal{A}, a$ & A set of containers and a single container $a =(|a|,s(a))\in \mathcal{A}$.    \\
        $\bar G$ & The unrolling of $G^\circ$: $\bar G = (\bar V,\bar E_{\mathrm{data}},\bar E_{\mathrm{pub}})$. \\
        $\hat G$ & The lowering of $\bar G$: $\hat G = (\hat V,\hat E)$. \\
        $\Phi(\hat G)$ & Evaluation of $\hat G$ (estimating runtimes of all vertices). \\
        $K_{eval}$ & Unrolling horizon (\#steps considered when unrolling $G^\circ$ into $\bar G$). \\
        $\tau$ & Staleness budget per trainable model; $\tau : \mathcal{M}_\mathrm{train} \to \mathbb{N}$. \\
        \midrule
        \multicolumn{2}{@{}l}{\textbf{RPEG's Execution Plan}} \\
        \midrule
        $P$ & \makecell[l]{RPEG's execution plan: $P=(\sigma,\Pi,mb)$.} \\
        $\sigma(v)$ & Device allocation of operator $v \in V^\circ$ (the set of GPUs reserved by $v$).\\ 
        $|\sigma(v)|$ & The number of GPUs in this allocation. \\
        $\Pi(v), mb(v)$ & Data/pipeline/tensor parallelism $\Pi(v) = (P_d, P_p, P_t)$; microbatch count. \\
        \midrule
        
        \multicolumn{2}{@{}l}{\textbf{Evaluation Pipeline}} \\
        \midrule
        $\mathcal T$ & Node-level time map (runtime estimations for operators), $\mathcal T : \hat{V} \to \mathbb{R}_{\ge 0}$. \\
        $W_\mathrm{shard}(v)$ & Per-GPU parameter count (weight shard) for operator $v$. \\
        $\mathrm{peak}(d)$ & Per-device peak memory (persistent + worst-case transient). \\
        $\mathrm{Score}(C)$ & Throughput estimate for configuration $C$. \\
\iftr 
        $F_{\mathrm{eff}}(v)$ & Achieved per-GPU GEMM throughput of operator $v$. \\
        $I(v), \eta(v)$ & Arithmetic intensity and calibrated GEMM utilization of operator $v$. \\
\fi
        \bottomrule
    \end{tabular}
\iftr
\end{table*}
\else
\end{table}
\fi

%% file: formulation.tex
\section{RPEG Transformation Pipeline}\label{rpegx:sec-formulation}

We now detail the RPEG transformation pipeline that transforms a compact high-level graph-driven algorithmic description of the RL post-training into a low-level representation schedulable on a give cluster; see also Figure~\ref{fig:example}.
Extensive mathematical details of all the subsections are in Appendix~\ref{sec:app:pipeline}.

\subsection{Graph-Centric RLM Training Representation}

\(G^\circ\) defines the \textbf{algorithmic structure} of RLM training:
\[
G^\circ=(V^\circ,E^\circ_{\mathrm{data}},E^\circ_{\mathrm{pub}},\mathcal{A},\mathcal{M},\tau).
\]

Here, \(V^\circ\) is the finite set of logical operators in one steady-state training step; \(E^\circ_{\mathrm{data}}\) is the set of within-step data dependencies; \(E^\circ_{\mathrm{pub}}\) is the set of cross-step \emph{model-state publication} dependencies, which determine how updated model parameters become available to later consumers; \(\mathcal{A}\) is the set of typed dataflow containers, such as completions, rewards, log-probabilities, or model weights; $\mathcal{M}$ is a set of all the used models (here, \(\mathcal{M}_{\mathrm{train}}\) is the set of trainable models); and \(\tau:\mathcal{M}_{\mathrm{train}}\rightarrow\mathbb{N}_0\) is a per-trainable-model staleness budget. Intuitively, \(G^\circ\) specifies \emph{what} recurring RLM computation is performed and which cross-step state dependencies are semantically admissible, while deliberately leaving \emph{how} it is mapped to hardware to the execution plan.

Each \textbf{vertex} \(v\in V^\circ\) denotes one model invocation. It carries
a \textit{role}
\(r(v)\in\{\text{generation},\text{inference},\text{training}\}\), its
associated \textit{model} \(\mu(v)\in \mathcal{M}\), and declared
\textit{input/output sets} \(I(v),O(v)\subseteq \mathcal{A}\) specifying which typed
containers the operator may consume and produce.
For instance, a generation vertex consumes prompts and actor
weights and produces completions, while a reward-inference vertex
consumes prompt--completion pairs and produces rewards.

Every \textbf{edge} refers to one typed container \(a\in \mathcal{A}\), whose
size metadata \((|a|,s(a))\) specifies its element count and bytes per
element. Its byte volume is \(|a|s(a)\). Containers represent both transient products such as completions, log-probabilities, rewards, and values, and persistent state such as model weights. Data edges \(E^\circ_{\mathrm{data}}\) encode within-step dependencies, whereas model-state publication edges \(E^\circ_{\mathrm{pub}}\) encode cross-step state flow from training to future--step consumers. This distinction makes the cross-step reuse of updated model state explicit and exposes bounded asynchrony through the staleness map \(\tau\).

The graph must satisfy four \textbf{well-formedness conditions}. First, \textit{input/output consistency} requires every container carried by an edge to belong to the source output set and the destination input set. Second, \textit{model-state publication edges} must originate at training vertices and carry model parameters. Third, \textit{boundary consistency} requires every container without a producer or consumer to be represented as an explicit graph input or output. Fourth, the within-step graph \((V^\circ,E^\circ_{\mathrm{data}})\) must be \textit{acyclic}; cycles may arise only through publication edges representing cross-step model updates. These conditions ensure acyclicity after valid unrolling; physical feasibility additionally depends on the execution plan, model versions, memory, and resource constraints.

We use a cyclic representation instead of searching directly on a flat multi-step DAG. First, it keeps the search space compact. Suppose invocation \(v\) has
\(q_v\) possible execution-plan assignments. In the compact RPEG, this
choice is made once for \(v\) and reused by all of its copies after
unrolling, giving \(\prod_{v\in V^\circ} q_v\) combinations. If instead
each of the \(K_{\mathrm{eval}}\) unrolled copies could choose its plan
independently, invocation \(v\) alone would have \(q_v^{K_{\mathrm{eval}}}\)
combinations, and the full plan space would grow to
\(\prod_{v\in V^\circ} q_v^{K_{\mathrm{eval}}}\). Second, staleness becomes a first-class structural variable: synchronous and bounded-asynchronous variants are obtained by modifying \(\tau\) and re-unrolling rather than by manually rewiring a materialized multi-step graph. Third, the cyclic form expresses steady-state semantics directly, independently of the finite horizon used for evaluation.

The \textbf{execution plan} \(P=(\sigma,\Pi,mb)\) assigns hardware-dependent attributes to each RPEG vertex. Here, \(\sigma(v)\) is a device allocation, \(\Pi(v)\) is a multi-dimensional model-parallel configuration, and \(mb(v)\) is the number of microbatches. In the evaluated implementation, \(\Pi(v)=(P_d,P_p,P_t)\) instantiates nested data, pipeline, and tensor parallelism. We model the cluster as a hierarchical device space \(D=[N_r]\times[N_n]\times[N_g]\), comprising racks, nodes per rack, and GPUs per node. Placement therefore determines not only co-residency and disaggregation but also whether communication remains intra-node, crosses nodes within a rack, or crosses racks.

\subsection{Joint Structural and Execution Configuration}
\label{rpegx:sec-hierarchy}

Crucially, \(G^\circ\) is itself part of the optimization space rather than a fixed workload graph. The current optimizer explicitly searches structural variants defined by AC sharing and the per-model staleness map \(\tau\). AC sharing changes the model structure and associated dependencies, while changing \(\tau\) changes the cross-step state dependencies and hence the admissible degree of asynchronous overlap. For every such structural candidate \(G^\circ\), the optimizer searches its execution plan variants \(P\). \(G^\circ\) and \(P\) are thus evaluated jointly rather than optimizing only a hardware mapping for one fixed pipeline.

Additional execution refinements are materialized during lowering. Stage fusion can realize multiple operations through a composite fine-grained execution structure, while response-length-aware splitting replaces generation and downstream inference invocations by matching executable phases when their selected placements permit overlap. These transformations enlarge the executable design space without treating arbitrary graph topology as an outer-loop search variable.

\subsection{Unrolling: Making Asynchrony Explicit}

The cyclic RPEG cannot be scheduled directly because model-state publication edges connect training in one step to consumers in later steps. Unrolling expands \(G^\circ\) over a finite horizon \(K_{\mathrm{eval}}\) into an acyclic graph
\(\bar G=(\bar V,\bar E_{\mathrm{data}},\bar E_{\mathrm{pub}})\), where each vertex \(v\in V^\circ\) is replicated once for every step index \(k\in[0,K_{\mathrm{eval}}-1]\). Each within-step edge is copied inside the same replicated step, whereas each publication edge is instantiated across steps according to the staleness budget. Version zero is the initial checkpoint; training at step \(j\)
produces version \(j+1\), and successive updates of the same optimizer
state remain ordered.
A serving consumer (generation or inference) at step \(k\) requires
version \(\max\{0,k-\tau(m)\}\): training step \(k-\tau(m)-1\) supplies it
when \(k>\tau(m)\), and otherwise the initial checkpoint supplies it.
The synchronous case \(\tau(m)=0\) consumes the immediately preceding
update for \(k>0\), and initialization for \(k=0\).

This construction makes bounded asynchrony explicit in the graph. If \(k\le\tau(m)\), no publication edge is created and the corresponding early steps consume the initial parameter version; because they have no training predecessor, their serving operations may overlap subject to all remaining constraints (e.g., resource or update order). Generally, publication edges always point from smaller to larger step indices, while within-step edges preserve the step index. Hence, if the within-step dependency graph is acyclic, the unrolled graph is also acyclic. Unrolling thus exposes inter-step overlap without sacrificing the compact steady-state search space at the RPEG level.

\subsection{Lowering: Delivering Executable Representation}

Unrolling makes semantic dependencies explicit, but not the physical
consequences of the selected execution plan. Lowering combines \(\bar G\)
with the candidate execution plan \(P\) and produces an executable DAG
\(\hat G=(\hat V,\hat E)\) with three kinds of vertices:
\(\mathrm{Op}\) nodes for unrolled operators,
\(\mathrm{Comm}\) nodes for explicit data transfers, and
\(\mathrm{Reshard}\) nodes for parameter redistribution between incompatible
layouts. Every \(\mathrm{Op}\) node carries its placement, model-parallel
configuration, and microbatch count; every \(\mathrm{Comm}\) or
\(\mathrm{Reshard}\) node carries an explicit byte payload
and the source and destination GPU allocations.

The component \(\sigma(v)\) specifies the set of GPUs
assigned to invocation \(v \in V^\circ\), with
\(|\sigma(v)|=P_d(v)P_p(v)P_t(v)\); its unrolled \(\mathrm{Op}\) nodes
inherit this allocation. For analytical scheduling, \(\sigma(x)\) denotes
the GPUs reserved by an executable node \(x\): the assigned GPUs for
\(\mathrm{Op}\) nodes and the destination (and optionally source) GPUs for
\(\mathrm{Comm}\)/\(\mathrm{Reshard}\) nodes.

Lowering also realizes the principal mechanisms behind
\textbf{fine-grained overlap} and \textbf{asynchrony}. First, when
generation and downstream inference are placed on disjoint devices,
generation can be split into quantile-based sub-nodes and inference into
matching sub-nodes using a profiled response-length distribution.
Inference for early-finishing responses can then overlap with later
generation phases. Second, lowering detects \emph{shadow pairs}: if a
trainable model is consumed asynchronously (\(\tau(m)>0\)) and its
producer and consumer use disjoint allocations, the two sides retain
distinct parameter copies and intra-step resharding can be avoided,
enabling cross-step overlap. For non-shadow pairs, lowering inserts
\(\mathrm{Reshard}\) nodes when layouts differ on the same devices,
or precedence edges and explicit model-state transfers when mappings
use different device sets. Transfers also account for partially
overlapping source and destination allocations.

\subsection{Scoring and Scheduling}
\label{sec:scoring-scheduling}

A schedule assigns a start time to {every} node of the
executable DAG. Formally, it is a function
\(s:\hat V\to\mathbb{R}_{\ge0}\); the finish-time function is
\(f(v)=s(v)+\mathcal{T}(v)\), where the analytical model of
Section~\ref{sec:model} supplies \(\mathcal{T}\).
Precedence requires \(f(u)\le s(v)\) for every \((u,v)\in\hat E\).
Resource exclusivity requires that nodes with overlapping
reserved GPU sets, \(\sigma(u)\cap\sigma(v)\neq\varnothing\),
do not execute concurrently.
The predicted makespan is
\[\mathrm{MS}(\hat G)=\max_{v\in\hat V}f(v).\]

We construct this schedule with a greedy scheduler,
whose input is the DAG, a map $\mathcal{T}$, and
GPU reservations \(\sigma(v)\) defined during lowering,
and whose output is the start/finish times of each node.
At each step, it considers nodes whose predecessors have
already been assigned finish times, computes their earliest possible
start times subject to precedence and GPU availability, and schedules
the node that can start earliest.

Ties first favor \(\mathrm{Comm}\)/\(\mathrm{Reshard}\) nodes,
reducing unnecessary delays in transfers needed by downstream
computation. Remaining ties favor nodes with the longest remaining
dependency path, measured by the sum of predicted execution times
along a path to a terminal node (one without successors), since
delaying such work is more likely to increase the makespan.
Generation phases inherit the priority of their first phase,
discouraging unrelated work from interrupting consecutive phases
that reuse the same KV cache.

We score a configuration \(C=(G^\circ,P)\) by
\[
\mathrm{Score}(C)=
\frac{K_{\mathrm{eval}}BR(L_p+L_r)}{\mathrm{MS}(\hat G)},
\]
where \(B\) is the number of prompts per RL step,
\(R\) is the number of rollouts per prompt, and
\(L_p,L_r\) are the average prompt and rollout lengths.
The numerator counts logical prompt--response tokens, not repeated
processing by every model role. The score amortizes execution over the
unrolled horizon.

Memory infeasibility is handled by a soft penalty,
{\footnotesize
\[
\mathrm{Score}_{\mathrm{pen}}(C)=\mathrm{Score}(C)
\exp\!\left[-\beta_{\mathrm{oom}}
\max\!\left(0,\max_{d\in D}
\frac{\mathrm{peak}(d)}{M_{\mathrm{eff}}}-1\right)\right].
\]}
Here \(\beta_{\mathrm{oom}}>0\) controls the penalty strength and
\(M_{\mathrm{eff}}=(1-\varepsilon_{\mathrm{rt}})M_{\mathrm{GPU}}\)
reserves part of physical HBM for runtime overhead and fragmentation
(\(\varepsilon_{\mathrm{rt}}=0.10\) in the analytical baseline).
A candidate slightly above this usable capacity may be explored, but the
returned configuration must satisfy
\(\mathrm{peak}(d)\le M_{\mathrm{eff}}\) for every GPU.
``Nearly feasible'' candidates are not accepted as executable plans.

For a fixed GPU budget, throughput is the default objective.
The optional efficiency objective also uses
\[\mathrm{Util}(C)=
\sum_{d\in D}\frac{t_{\mathrm{busy}}(d)}{|D|\,\mathrm{MS}(\hat G)},\]
where \(t_{\mathrm{busy}}(d)\) is the time GPU \(d\) is occupied by
scheduled nodes. For a user-specified throughput tolerance \(\delta\),
it is
{\footnotesize
\[
\begin{gathered}
\mathrm{Score}_{\mathrm{eff}}(C)
=\mathrm{Score}(C)\,\times
\begin{cases}
\mathrm{Util}(C),
& \mathrm{Score}(C)\ge(1-\delta)\mathrm{Score}^{\star},\\
\lambda_{\mathrm{floor}}, & \text{otherwise},
\end{cases}
\end{gathered}
\]}
where \(\mathrm{Score}^{\star}\) is the highest raw throughput
found so far and \(\lambda_{\mathrm{floor}}=10^{-12}\) strongly penalizes
plans below the throughput floor. The same memory penalty is then applied
to \(\mathrm{Score}_{\mathrm{eff}}\). Thus, the floor is a search
preference, whereas the final memory bound is mandatory. GPU-hours are
reported separately as reserved GPUs multiplied by elapsed hours.

%% file: model.tex
\section{Analytical Performance Model for RLMs}
\label{sec:model}

We now describe the performance model for RLM training. Given the executable DAG \(\hat G\) produced by unrolling and lowering, we assign each node an analytical execution-time estimate and check whether the resulting configuration is memory-feasible. The performance model therefore consists of \textbf{(1)} a node-level time map \(\mathcal{T}:\hat V\rightarrow\mathbb{R}_{\ge 0}\), and \textbf{(2)} a memory-feasibility predicate \(\mathrm{MemFeasible}(C)\) with $C = (G^\circ, P)$ being a given configuration. Together, these model the runtime consequences of different placement strategies, multi-dimensional parallelism, microbatching, communication effects such as resharding, bounded asynchrony, and co-residency on a hierarchical supercomputing machine.

Due to space constraints, we focus on key aspects specific to
RL post-training; detailed derivations appear in
Appendix~\ref{app:ad-derivations}, while calibration,
modeling assumptions, and implementation details appear in
Appendix~\ref{rpegx:appendix-model}.

\subsection{Model Configuration}

We first discuss several setup aspects of the used model.

\subsubsection{Modeling Communication}

Communication arises both \emph{between} executable DAG nodes, through explicit \textsc{Comm} and \textsc{Reshard} nodes, and \emph{within} operator \textsc{Op} nodes (through multi-dimensional parallelism). We use a standard \(\alpha\)--\(\beta\) model~\cite{hockney1994alphabeta} $T(M)=\alpha+\beta M$, for a message of \(M\) bytes. The appropriate \((\alpha,\beta)\) pair is selected from the tier spanned by the communicating allocations: {intra-node}, {inter-node within a rack}, or {inter-rack}. If an operation spans multiple tiers, we conservatively use the slowest one.

For collectives, we use {ring-based reduction/gather estimates and a
pairwise-exchange estimate for all-to-all}, summarized in
Table~\ref{tab:cost-model-collectives}.
{The message-size convention depends on the primitive: \(M\) is the
buffer size for AllReduce/ReduceScatter, the final gathered buffer size for
AllGather, and the aggregate exchanged bytes per GPU for AlltoAll.
P2P uses the transferred bytes. Collective costs are zero for \(p=1\).}

\begin{table}[h]
\centering\setlength{\tabcolsep}{3pt}
\renewcommand{\arraystretch}{0.8}
\scriptsize
\caption{{Communication estimates for \(p\) participating GPUs; \(M\) is the message size;
\((\alpha,\beta)\) are calibrated for the primitive and communication tier.}}
\label{tab:cost-model-collectives}
\begin{tabular}{l l l}
\toprule
\textbf{Primitive} & \textbf{Time} & \textbf{Application} \\
\midrule
AllReduce
& $2(p{-}1)\alpha + 2\tfrac{p-1}{p}M\,\beta$
& TP partial-result aggregation \\[2pt]
AllGather
& $(p{-}1)\alpha + \tfrac{p-1}{p}M\,\beta$
& ZeRO-2 weight gather \\[2pt]
ReduceScatter
& $(p{-}1)\alpha + \tfrac{p-1}{p}M\,\beta$
& ZeRO-2 gradient reduce \\[2pt]
AlltoAll
& $(p{-}1)\alpha + M\,\beta$
& \textsc{Reshard} nodes \\[2pt]
P2P
& $\alpha + M\,\beta$
& PP send/recv, \textsc{Comm} nodes \\
\bottomrule
\end{tabular}
\end{table}

\subsubsection{Model Input}
\label{sec:model-input}

The model is parameterized by three input groups. The \textbf{workload} specifies batch size \(B\),
rollouts per prompt \(R\), average prompt length \(L_p\),
and average rollout length \(L_r\). Then, for each model \(m\in \mathcal{M}\), the \textbf{model architecture} specification provides layer count, hidden dimension, feed-forward dimension, vocabulary size, head dimension, and working precision \(s_b\) (i.e., the number of bytes per scalar in the runtime tensor format, e.g., $s_b=2$ for BF16). The \textbf{hardware} description consists of a hierarchical device space \(D=[N_{\mathrm{r}}]\times[N_{\mathrm{n}}]\times[N_{\mathrm{g}}]\), with \(N_{\mathrm{r}}\) racks, \(N_{\mathrm{n}}\) nodes per rack, and \(N_{\mathrm{g}}\) GPUs per node, plus calibrated performance parameters.

\subsubsection{Hardware Calibration}
\label{sec:calibration}

Rather than relying on vendor peak numbers, we calibrate \emph{effective} hardware parameters on the target system. These include workload-effective compute throughput \(F_{\mathrm{peak}}\), HBM bandwidth \(\mathrm{BW}_{\mathrm{mem}}\), separate \((\alpha,\beta)\) parameters for collective and point-to-point communication at all three tiers, GEMM efficiency constants \(\gamma_M,\gamma_K\), and GPU memory capacity \(M_{\mathrm{GPU}}\). This distinction is important because both compute and communication are strongly topology-dependent on modern hierarchical systems, and because NCCL collectives and P2P transfers can differ substantially in achieved bandwidth.

\subsection{Intra-Model Runtime}
\label{sec:model-intra-model}

For each operator node \(v\in\hat V\), we model runtime under the
{evaluated 3D (DP/PP/TP) strategy \((P_d,P_p,P_t)\) and microbatch
count \(mb\)} by the decomposition
\footnotesize
\[
T_{\mathrm{op}}(v)=T_{\mathrm{gemm}}(v)+T_{\mathrm{bubble}}(v)+T_{\mathrm{comm}}(v)+T_{\mathrm{non\mbox{-}gemm}}(v)+T_{\mathrm{opt}}(v),
\]
\normalsize
where the five terms capture GEMM execution, pipeline bubbles, intra-operator communication, non-GEMM overheads, and optimizer traffic. The dominant GEMM compute term is obtained from the operator work $W(v)$, i.e., the total FLOP count induced by the operator role and model type. For an operator with execution strategy $(P_d,P_p,P_t)$, we define $P_dP_pP_t=|\sigma(v)|$ as the number of GPUs allocated to that operator. Assuming balanced sharding across this 3D mesh, the GEMM work is divided across these devices, giving
\[
T_{\mathrm{gemm}}(v)=\frac{W(v)}{P_dP_pP_t\,F_{\mathrm{eff}}(v)},
\]
The denominator states that the devices allocated to $v$ execute its data, pipeline, and tensor shards concurrently (other operators may use disjoint allocations at the same time if the schedule permits). $F_{\mathrm{eff}}(v)$ is the achieved per-GPU GEMM throughput for $v$ after accounting for arithmetic intensity and calibrated GEMM shape efficiency. Thus, \(W(v)\) specifies how much computation the operator performs, while \(F_{\mathrm{eff}}(v)\) specifies how fast that computation can be sustained in practice.

\iftr
We model \(F_{\mathrm{eff}}(v)\) with a roofline-style throughput corrected
for both bandwidth limits and calibrated GEMM shape under-utilization.
{This captures inefficient GEMM execution at small local batch sizes
or narrow tensor-parallel dimensions. The separate terms
\(T_{\mathrm{non\mbox{-}gemm}}\) and \(T_{\mathrm{opt}}\) account for layer
operations whose latency does not decrease proportionally with tensor
parallelism (e.g., normalization, residual additions, positional
embeddings, attention softmax, activation functions) and
memory-bandwidth-bound optimizer traffic.}
Concretely, $F_{\mathrm{eff}}(v) = \min\!\left(F_{\mathrm{peak}}\eta(v),\, I(v)BW_{\mathrm{mem}}\right)$; $I(v)$ is the operator's arithmetic intensity and $\eta(v)$ is a
calibrated GEMM shape efficiency factor. In our implementation, $\eta(v)$
decreases for small local batch/sequence dimensions and for narrow
tensor-parallel hidden dimensions, parameterized by $\gamma_M$ and
$\gamma_K$.
\fi

The work term \(W(v)\) is itself role- and model-specific: generation, inference, and training have different FLOP counts, and actor/reference, critic/value, and reward operators differ further through their output heads and {whether backward and recomputation phases are present;
optimizer traffic is accounted for separately}. Finally, generation receives an additional RLM-specific refinement by splitting its work into a compute-oriented prefill phase and a memory-bandwidth-bound decode phase sensitive to weight and KV-cache traffic, which is essential to capture long-response reasoning workloads. We use work terms for different RLM model roles as presented in a recent extensive work--depth analysis~\cite{besta2026performance}.

\subsection{Inter-Model Communication and Resharding}

Explicit communication and redistribution nodes are computed separately from operator nodes. If a \textsc{Comm} node transfers container \(a\) between operators \(u\) and \(v\), its runtime is $\alpha+\frac{|a| s(a)}{\min(|\sigma(u)|,|\sigma(v)|)}\beta$; $\sigma(u)$ is the device allocation of operator $u$, $|\sigma(u)|$ is its size in \#GPUs (for $\alpha-\beta$, the tier-specific point-to-point values are used). The division by the smaller allocation reflects the limiting side in concurrent send/recv. If a \textsc{Reshard} node transforms weights between two incompatible layouts on the same device set, we model it as an all-to-all over that allocation $T_{\mathrm{A2A}}\!\left(\frac{W_m s_b}{|\sigma(u)|},\, |\sigma(u)|\right)$, where $W_m s_b$ is the full parameter size of a given model in bytes and $T_{\mathrm{A2A}}(M,p)$ is the collective all-to-all cost for per-GPU message size $M$ across $p$ GPUs. Together with the operator-node formulas from Section~\ref{sec:model-intra-model}, these define the complete time map \(\mathcal{T}\) over \(\hat G\).

\subsection{Memory Model}

Many configurations that look attractive under the time model are infeasible
in practice as they exceed HBM capacity. We thus use an analytical memory
model both to exclude clearly infeasible candidates during search and to
{softly penalize memory-violating candidates} instead of discovering
OOMs only at runtime. We assume mixed-precision
training~\cite{micikevicius2018mixedprecision} with BF16 parameters and FP32
Adam~\cite{kingma2015adam} optimizer state, and activation
checkpointing~\cite{chen2016sublinear}. Optimizer state and gradients are
sharded across data-parallel ranks via ZeRO-2~\cite{rajbhandari2020zero}.

We target high-throughput RLM training on GPU-rich HPC systems, where
3D parallelism and ZeRO-2 sharding enable GPU-resident execution
~\cite{narayanan2021megatron,rajbhandari2020zero}.
CPU offloading primarily addresses insufficient GPU memory
~\cite{ren2021zerooffload,rajbhandari2021zeroinfinity}, but introduces
host--device transfers and potentially CPU-side optimizer work that can
reduce throughput and disrupt the pipeline overlap targeted by RPEG
~\cite{sheng2023flexgen}.

\subsubsection{Parameter and Persistent-State Accounting}

We use the dominant parameter counts derived
in a recent work~\cite{besta2026performance} and retain the placement-specific
byte accounting required here.
Let \(W_{\mathrm{shard}}(v)\) denote the maximum per-GPU parameter
shard size of invocation \(v\), accounting for its pipeline and tensor
partitioning and any replicated embeddings or heads. For device-level
accounting, \(W_{\mathrm{shard}}(m,d)\) denotes the corresponding parameter
shard of model \(m\) on GPU \(d\) under the relevant execution layout.

The persistent footprint depends on how each GPU uses the model.
A training GPU retains FP32 master weights and two Adam moment arrays,
sharded across \(P_d\) data-parallel ranks under ZeRO-2 (12 bytes per
parameter before sharding). Its BF16 weights are instead materialized
on demand. A GPU serving a trainable model with \(\tau(m)>0\) on an
allocation disjoint from training retains a persistent BF16
\emph{shadow copy}; frozen models such as reference and reward likewise
retain BF16 weights. If training and serving allocations overlap,
no additional shadow copy is needed. Synchronous serving-only
allocations retain no persistent weights between invocations.

Thus, the per-model persistent footprint \(S(m,d)\) is
\vspaceSQ{-0.5em}
\small
\[
S(m,d)=
\begin{cases}
12\,W_{\mathrm{shard}}(m,d)/P_d, & \text{training},\\
W_{\mathrm{shard}}(m,d)\,s_b, & \text{shadow/frozen},\\
0, & \text{no persistent copy}.
\end{cases}
\]
\normalsize
%
Here, \(P_d\) is the training layout's data-parallel degree,
and \(s_b\) is the BF16 scalar size in bytes.

\subsubsection{Transient-State Accounting}

In addition to persistent state, each invocation \(v\) requires
a peak transient allocation \(A(v)\), including activations, generation
KV cache, BF16 weights materialized on demand, training gradients, and
local working buffers. Under ZeRO-2, the sharded BF16 gradient term is
\(W_{\mathrm{shard}}(v)s_b/P_d(v)\). State already included in \(S(m,d)\)
is excluded from \(A(v)\) to avoid double counting. Since invocations
sharing a GPU cannot execute simultaneously, only the largest transient
allocation on that device needs to be considered.

\subsubsection{Co-Residency-Aware Device Peaks}
\label{sec:coresidency}

The key inter-model issue is that RL pipelines place multiple models on
the same physical devices. A GPU may participate, for example, in actor
generation, reference inference, and actor training within one overlapping
steady-state schedule. Thus, memory cannot be evaluated operator by
operator in isolation.

We thus define \textbf{device-level peak memory} as persistent co-resident
state plus the largest transient term among operators mapped to that device:
\small
\begin{equation}
\mathrm{peak}(d)
=
\underbrace{\sum_{\substack{m \text{ placed}\\ \text{on } d}} S(m,d)}_{\text{persistent}}
+
\underbrace{\max_{\substack{v \text{ placed}\\ \text{on } d}} A(v)}_{\text{worst transient}}.
\label{eq:mem:peak_device}
\end{equation}
\normalsize

To account for runtime memory not explicitly captured by the
analytical model (e.g., allocator reservations, communication
workspaces, fragmentation) we reserve a calibrated safety fraction
\(\varepsilon_{\mathrm{rt}}\) of GPU memory. The effective capacity is
$M_{\mathrm{eff}} = (1-\varepsilon_{\mathrm{rt}})\,M_{\mathrm{GPU}}$.
Finally, a given configuration is \textbf{memory-feasible} iff every device
respects the effective capacity bound:
\small
\begin{equation}
\mathrm{MemFeasible}(C)=\mathrm{true}
\iff
\forall d\in D:\ \mathrm{peak}(d)\le M_\mathrm{eff}.
\label{eq:mem:feasible}
\end{equation}
\normalsize

\begin{figure}[h]
\vspaceSQ{-1em}
    \centering
    \includegraphics[width=\linewidth]{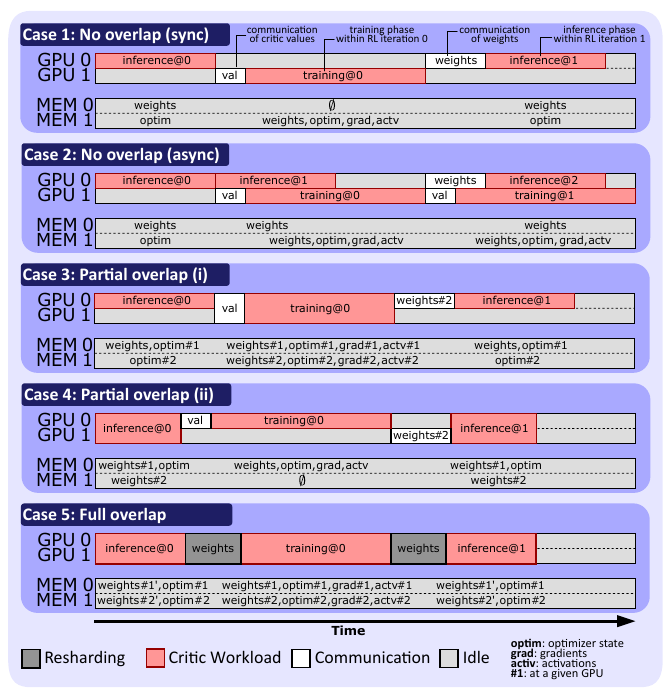}
    \vspaceSQ{-1.5em}
    \caption{\textbf{Per-device memory classification for 5 allocation and staleness configurations} of one trainable model (critic) on two GPUs. In Case~1 (sync, disjoint), GPU~0 receives weights transiently via a \textsc{Comm} node. In Case~2 (async, disjoint), both GPUs hold persistent weight copies (shadow pair). Cases~3--4 show partial device overlap. Case~5 shows full colocation where a \textsc{Reshard} node transforms the weight layout between inference and training. \textit{No AI was used to conceive or draw the picture.}}
    \label{fig:memory-model}
    \vspaceSQ{0em}
\end{figure}

\iftr
Figure~\ref{fig:memory-model} illustrates this persistent/transient classification for five representative two-GPU cases: (1) synchronous disjoint execution, where inference receives weights transiently via communication and holds no persistent copy; (2) asynchronous disjoint execution, where a shadow pair is active and both sides keep persistent weight copies; (3) partial overlap where one device is shared between training and inference while the other remains inference-only; (4) partial overlap with the complementary overlap pattern; and (5) full colocation with resharding, where the same devices execute both roles and a layout transformation bridges inference and training. These cases show how bounded asynchrony changes memory behavior: enabling a shadow pair increases static memory by keeping persistent copies on both sides, but it avoids repeated redistribution and allows cross-step overlap.
\else
Figure~\ref{fig:memory-model} illustrates this accounting for five
representative placements: synchronous and asynchronous disjoint
allocations, two cases of partial overlap, and full colocation with
resharding. In particular, shadow copies increase persistent memory
consumption but enable cross-step overlap and avoid repeated
redistribution.
\fi

%% file: optimization.tex
\section{Joint Structural and Execution Optimization}
\label{rpegx:sec-optimization}

The preceding sections define how to evaluate a configuration
\(C=(G^\circ,P)\). For each such candidate, the transformation pipeline
(Section~\ref{rpegx:sec-formulation}) delivers an executable DAG \(\hat G\). The analytical
performance model (Section~\ref{sec:model}) estimates node execution
times and per-GPU memory demands, while scheduling
(Section~\ref{sec:scoring-scheduling}) determines the predicted makespan
and score, including memory penalties. During optimization, this
evaluation is repeated for each newly proposed configuration; the
resulting scores guide the search, while the best memory-feasible
configuration found is retained.

The optimization problem is defined by four fixed inputs: \textbf{(1)} the RL
training algorithm (e.g., PPO, GRPO, or DPO), \textbf{(2)} workload parameters
such as batch and sequence sizes, \textbf{(3)} model specifications, and
\textbf{(4)} the available hardware. The RL algorithm determines the initial
RPEG structure; for example, PPO contains actor, critic, reward, and reference
invocations, whereas GRPO has no critic. Together, these inputs
establish the initial compact RPEG and its optimization constraints.

This search has two levels. First~\includegraphics[scale=0.2,trim=0 16 0 0]{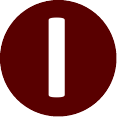}, we enumerate a small set of
\emph{pipeline-structure candidates} \(G^\circ\). Second, for each such graph,
we search its much larger space of \emph{execution plans} \(P\)~\includegraphics[scale=0.2,trim=0 16 0 0]{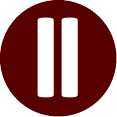}.
Algorithm~\ref{rpegx:alg-plan-search} summarizes the complete procedure.
In the following, we detail the key aspects of the algorithm; further details are in Appendix~\ref{ls:ch:search}.

\subsection{Pipeline Structure Candidates}
\label{sec:pipeline-str-cand}

For a fixed RL algorithm, the explicitly searched structural choices are
actor--critic sharing and bounded staleness. We represent them by
\(y=(ac,\tau)\), where \(ac\in\{0,1\}\) determines whether actor and critic
parameters are shared, and \(\tau(m)\) is the allowed staleness of each
trainable model \(m\). AC sharing is considered only when both an actor and
critic exist. Each \(\tau(m)\) is chosen from a small user-defined finite set.
We denote the resulting set of legal tuples by \(Y_{\mathrm{cand}}\).
Each \(y\in Y_{\mathrm{cand}}\) determines one RPEG \(G^\circ(y)\).

\subsection{Execution Plan Candidates}
\label{sec:exec-plan-cand}

For every model invocation \(v\) in a fixed \(G^\circ\), we enumerate its
legal execution choices as
$X_{\mathrm{cand}}(v) =
\left\{(\sigma,\Pi,mb)\;:\;\text{execution constraints hold}\right\}$,
where \(\sigma\) is the GPU allocation, \(\Pi=(P_d,P_p,P_t)\) is the
DP/PP/TP configuration, and \(mb\) is the microbatch count. ``Legal'' means
that the choice can be mapped to the given workload, model, and machine:
for example, pipeline stages must receive valid layer partitions, tensor
parallelism is kept within a node, and the allocation must contain the GPUs
required by \(P_dP_pP_t\). We enumerate aligned contiguous GPU blocks
satisfying these constraints and discard obviously infeasible choices early.

An execution plan \(P\) assigns one legal choice to each invocation \(v\):
$P(v) = \bigl(\sigma(v),\Pi(v),mb(v)\bigr) \in X_{\mathrm{cand}}(v)$.
The search starts from an initial plan and repeatedly replaces the assignment
of one or two invocations, as described next.

\subsection{Stochastic Plan Search}
\label{sec:stochastic-plan-search}

For each structural candidate \(G^\circ(y)\), we search its execution plans
using \textbf{simulated annealing}. Starting from the current plan \(P\), one
search round proposes up to \(S_n=6\) nearby plans. Each proposal defines a
complete configuration \(C=(G^\circ,P)\), which is evaluated using the
pipeline from Section~\ref{sec:overview}. Thus, unrolling, lowering,
analytical modeling, scheduling, and scoring are performed for every newly
proposed configuration.

At the beginning of each search round, the optimizer chooses one of
two proposal mechanisms. With probability
\(p_{\mathrm{pair}}=0.28\), it uses the two-invocation mechanism described
below; otherwise, with probability \(1-p_{\mathrm{pair}}=0.72\), it uses the
single-invocation mechanism.

{For a \textbf{single-invocation} move, the optimizer first chooses the
invocation \(v\) whose execution-plan assignment will be modified. With
probability \(p_{\mathrm{cp}}=0.6\), \(v\) is sampled from the current
schedule's critical (bottleneck) path with probability proportional to its predicted
execution time; with probability \(1-p_{\mathrm{cp}}=0.4\), it is sampled
uniformly from all invocations. The optimizer then samples alternative
assignments for $v$ from \(X_{\mathrm{cand}}(v)\), producing up to \(S_n\)
neighboring plans. The subsequent evaluation determines
whether any of them is better than the current plan.}

For a \textbf{two-invocation} move, the optimizer instead changes two
allocations jointly. It samples a \textit{donor invocation} \(v_d\) with probability
proportional to its current GPU count \(|\sigma(v_d)|\), so invocations
occupying more GPUs are more likely to donate resources. It then chooses a
\textit{receiver invocation} \(v_r\) among invocations whose current allocations overlap with
the donor's allocation. The donor is assigned a legal smaller allocation,
while the receiver is resampled from \(X_{\mathrm{cand}}(v_r)\) so that it
may use some of the released GPUs. A receiver can already be feasible and
still benefit from additional GPUs: feasibility only means that its assignment
is legal, whereas a larger allocation may reduce its execution time and hence
the overall makespan. The coordinated move allows such a resource
redistribution to be tested directly instead of first passing through a plan
that leaves the donor's released GPUs unused.

\subsection{Proposal Evaluation and Acceptance}
\label{sec:proposal-eval-and-accept}

Regardless of which proposal mechanism is used, one search round
produces up to \(S_n\) complete candidate configurations
\(C'_1,\ldots,C'_{S_n}\). They are evaluated independently, with uncached
evaluations executed in parallel when possible. The optimizer then selects
\[
C'=\arg\max_{C'_i}\mathrm{Score}_{\mathrm{pen}}(C'_i)
\]
and only this highest-scoring proposal is considered for updating the
current plan.

Let
\(s=\mathrm{Score}_{\mathrm{pen}}(C)\) and
\(s'=\mathrm{Score}_{\mathrm{pen}}(C')\).
If \(s'>s\), \(C'\) is accepted. Otherwise it is accepted with probability
\[
p_{\mathrm{acc}}
=
\min\left\{
1,\,
\exp\left(\frac{s'-s}{|s|t}\right)
\right\}.
\]
The \textbf{temperature} \(t>0\) controls how readily the search accepts a
temporarily worse configuration: larger \(t\) increases exploration, whereas
smaller \(t\) makes the search more greedy. The normalization by \(|s|\)
makes the temperature largely independent of the absolute score scale.

A sequence of proposal-and-acceptance rounds forms a \textbf{search chain}.
The temperature remains fixed within each chain. The first chain uses
temperature \(t_{\mathrm{init}}\). Each subsequent, warm-restarted chain uses
\(t_{\mathrm{warm}}>t_{\mathrm{init}}\), allowing broader exploration around
a previously discovered high-scoring plan. If the selected proposal is
rejected in ten consecutive rounds, the chain is considered stuck and a new
one is started.

The \textbf{search budget} is specified either as wall-clock time or as a
maximum number of full configuration evaluations. The budget is allocated equally across structural candidates:
in wall-clock mode, the remaining time is divided evenly among the remaining
\(y\in Y_{\mathrm{cand}}\); in evaluation-budget mode, each candidate receives
the same fixed budget of uncached configuration evaluations.

\begin{algorithm}[t]
\DontPrintSemicolon
\caption{\textbf{RPEG optimization}}
\label{rpegx:alg-plan-search}

\KwIn{{Fixed RL algorithm, workload, model, and hardware inputs;
legal structural choices \(Y_{\mathrm{cand}}=\{(ac,\tau)\}\);
wall-clock or uncached-evaluation search budget;
temperatures \(t_{\mathrm{init}}, t_{\mathrm{warm}}\)}}

\KwOut{{\(C^\star\): highest-scoring memory-feasible configuration
found within the search budget}}

{{Initialize cache and \(C^\star\gets\bot\).
Define \(\mathrm{Eval}(C)\) to return (penalized score $\mathrm{Score}_{\mathrm{pen}}(C)$,
memory feasibility, schedule) via cached unrolling/lowering
(Sec.~\ref{rpegx:sec-formulation}), performance modeling
(Sec.~\ref{sec:model}), and scheduling/scoring
(Sec.~\ref{sec:scoring-scheduling})}}\;

\ForEach{{\(y\in Y_{\mathrm{cand}}\)}
\includegraphics[scale=0.2,trim=0 16 0 0]{I.pdf}}{

    Construct \(G^\circ(y)\) (Sec.~\ref{sec:pipeline-str-cand});
    For each \(v\in V^\circ(y)\),
    enumerate/reuse \(X_{\mathrm{cand}}(v)\) by filtering
    \((\sigma,\Pi,mb)\) combinations for legal parallelism,
    device packing, and HBM feasibility
    (Sec.~\ref{sec:exec-plan-cand},
    App.~\ref{sec:app:search:plan-space})\;

    {\(\mathrm{first}\gets\mathrm{true}\)}\;

    \While{{budget remains for plan-search
    on \(G^\circ(y)\)}
    \includegraphics[scale=0.2,trim=0 16 0 0]{II.pdf}}{

        \eIf{{\(\mathrm{first}\)}}{
            {{Initialize \(P\) using the concurrent baseline:
            group independent invocations, partition GPUs, and
            choose \(P(v)\in X_{\mathrm{cand}}(v)\) to minimize
            estimated step time
            (Sec.~\ref{sec:exec-plan-cand},
            App.~\ref{ls:sec:search:restarts})}}\;
            {\(t\gets t_{\mathrm{init}};\
            \mathrm{first}\gets\mathrm{false}\)}\;
        }{
            {{Initialize \(P\) from a cached high-scoring
            feasible plan with a compatible vertex set
            (Sec.~\ref{sec:cache},
            App.~\ref{ls:sec:search:restarts})}}\;
            {\(t\gets t_{\mathrm{warm}}\)}\;
        }

        {{\((s,\mathrm{ok},Q)\gets
        \mathrm{Eval}((G^\circ(y),P))\)}}\;
        {If \(\mathrm{ok}\) and the score improves \(C^\star\),
        set \(C^\star\gets(G^\circ(y),P)\)}\;
        {\(\mathrm{rejected}\gets0\)}\;

        \While{{budget remains and \(\mathrm{rejected}<10\)}}{

            \eIf{{pair kernel selected with
            \(p_{\mathrm{pair}}=0.28\)}}{
                {{Sample donor \(u\) proportional to its GPU
                count and receiver \(v\) from overlapping allocations;
                for \(i\le S_n=6\), copy \(P'_i\gets P\),
                shrink \(P'_i(u)\) using \(X_{\mathrm{cand}}(u)\),
                and resample \(P'_i(v)\) from \(X_{\mathrm{cand}}(v)\)
                (Sec.~\ref{sec:stochastic-plan-search},
                App.~\ref{ls:sec:search:moves})}}\;
            }{
                {{Select \(v\) using critical-path weights
                from schedule \(Q\) with probability
                \(p_{\mathrm{cp}}=0.6\), otherwise uniformly;
                for \(i\le S_n=6\), copy \(P'_i\gets P\) and
                sample \(P'_i(v)\sim
                \mathrm{Unif}(X_{\mathrm{cand}}(v))\)
                (Sec.~\ref{sec:stochastic-plan-search},
                App.~\ref{ls:sec:search:moves})}}\;
            }

            {{For each \(P'_i\), evaluate
            \((s'_i,\mathrm{ok}'_i,Q'_i)\gets
            \mathrm{Eval}((G^\circ(y),P'_i))\)}}\;

            {{\(i^\star\gets\arg\max_i s'_i\);
            \(s'\gets s'_{i^\star}\)}}\;

            {{Update \(C^\star\) from all evaluated
            memory-feasible proposals with higher scores}}\;

            \eIf{{{accept the best proposal with probability
            \(p_{\mathrm{acc}}(s,s',t)\)
            (Sec.~\ref{sec:proposal-eval-and-accept},
            App.~\ref{ls:sec:search:formulation})}}}{
                {{\(P\gets P'_{i^\star}\);
                \(s\gets s'\);
                \(Q\gets Q'_{i^\star}\);
                \(\mathrm{ok}\gets\mathrm{ok}'_{i^\star}\);
                \(\mathrm{rejected}\gets0\)}}\;
            }{
                {\(\mathrm{rejected}\gets\mathrm{rejected}+1\)}\;
            }

            {{If \(\mathrm{ok}\), retain the current \(P\)
            as a possible restart seed}}\;
        }
    }
}

{{Polish \(C^\star\) by repeatedly testing every
\(x\in X_{\mathrm{cand}}(v)\) for each invocation \(v\),
replacing only \(P(v)\), and greedily accepting feasible
score improvements until none remain
(Sec.~\ref{sec:stop-refine},
App.~\ref{ls:sec:search:polish})}}
\includegraphics[scale=0.2,trim=0 16 0 0]{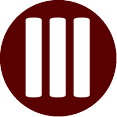}\;

\Return{{\(C^\star\)}}\;
\end{algorithm}

\subsection{Caching}
\label{sec:cache}

Evaluating a configuration is considerably more expensive than generating a
new proposal, so evaluated configurations are \textbf{cached} using their complete
graph and execution-plan description. Repeated proposals therefore reuse their
previous score. Structural candidates with the same invocation set also reuse
the corresponding \(X_{\mathrm{cand}}(v)\) sets.
The implementation additionally retains high-scoring feasible plans
with compatible invocation sets as seeds for subsequent search chains.

\subsection{Stopping and Final Refinement}
\label{sec:stop-refine}

The search stops when its wall-clock or evaluation budget is exhausted and
always retains the the highest-scoring memory-feasible configuration
found so far.
Finally~\includegraphics[scale=0.2,trim=0 16 0 0]{III.pdf}, we deterministically \textbf{refine that plan}. We visit every invocation
in a fixed order and try its legal execution choices while keeping all other
assignments unchanged; if any choice improves the score, we retain the best one.
These passes repeat until an entire pass produces no improvement. Thus, the
final ``greedy coordinate descent'' simply removes any remaining improvement
that can be obtained by changing one invocation at a time.

%% file: runtime.tex
\iftr
\begin{table*}[t]
\centering
\footnotesize
\renewcommand{\arraystretch}{1.0}
\setlength{\tabcolsep}{2pt}
\begin{tabularx}{\textwidth}{ll}
\toprule
\textbf{Optimized RPEG component} & \textbf{Runtime action} \\
\midrule
$G^\circ$ and its typed edges & Instantiate step-indexed model operations, data dependencies, and explicit transfer nodes. \\
Placement $\sigma(v)$ & Bind the invocation's worker ranks to its selected physical GPUs. \\
Parallelism $\Pi(v)$ & Form DP replica groups, PP layer stages, and TP tensor shards/collectives in Megatron. \\
Microbatch count $mb(v)$ & Split each DP replica's local batch into the selected number of microbatches and issue the corresponding model operations. \\
Publication edges and $\tau(m)$ & Resolve each consumer's required parameter version and install its corresponding shards. \\
Cross-layout dependencies & Route tensor fragments between producer and consumer layouts through \textsc{Comm}/\textsc{Reshard} operations. \\
Resource set $\mathcal R(x)$ & Reserve the GPUs and worker groups needed by $x$ until its completion; transfers reserve their participating endpoints. \\
\bottomrule
\end{tabularx}
\caption{Example concrete realization of RPEG components within its runtime.}
\label{pp:tab:runtime-provenance}
\end{table*}
\fi

\section{Completion-Driven Distributed Runtime}
\label{rpegx:sec-runtime}

The RPEG runtime realizes a configuration \(C^\star\) 
selected by the optimized on physical GPUs,
preserving its execution dependencies, placement, parallelism, and
model-version semantics. Crucially, while predicted execution times
guide \emph{plan selection}, actual node completion events govern
\emph{execution}, allowing the runtime to adapt to varying computation
and communication durations without changing the selected plan.

Execution proceeds in three stages, see Alg.~\ref{rpegx:alg-runtime-dispatch}. 
First~\includegraphics[scale=0.2,trim=0 16 0 0]{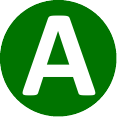},
\textbf{graph realization} 
unrolls and lowers \(C^\star\) into an executable DAG \(\hat G=(\hat V,\hat E)\)
(as in Sec.~\ref{rpegx:sec-formulation}), and assigns concrete GPU/worker
bindings and resource requirements to every executable node \(x\in\hat V\).
Then~\includegraphics[scale=0.2,trim=0 16 0 0]{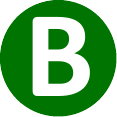},
\textbf{persistent worker groups} instantiate the selected DP/PP/TP
configurations. Each logical RPEG invocation \(v\in V^\circ\) reuses
its assigned workers across unrolled training steps, avoiding repeated
model initialization and distributed-group setup.
Third~\includegraphics[scale=0.2,trim=0 16 0 0]{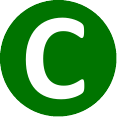},
\textbf{completion-driven dispatch} executes nodes as their
predecessors finish and required physical resources become available.

The controller operates directly on the executable DAG, including
\textsc{Op}, \textsc{Comm}, and \textsc{Reshard} nodes.
A node \(x\) becomes ready when all predecessors in \(\hat G\) have
completed, and may execute when its exclusive resource set
\(\mathcal R(x)\) is available. These resources include assigned
workers and GPUs and, for direct transfers, the participating endpoints.
Among ready nodes, priority is given to earlier RL steps, then
communication/resharding, and finally a fixed topological order.
The graph also determines the exact parameter version consumed by
each invocation and when intermediate data or obsolete parameter
copies can be released, thereby preserving bounded-staleness
semantics while enabling safe overlap.

\begin{algorithm}[t]
\DontPrintSemicolon
\caption{\textbf{Completion-driven RPEG execution}}
\label{rpegx:alg-runtime-dispatch}

\KwIn{Optimized configuration \(C^\star=(G^\circ,P)\),
execution horizon \(K_{\mathrm{run}}\)}
\KwOut{Completed execution and runtime trace}

{\((\hat G,b,\mathcal R)\gets
\mathrm{RealizeAndValidate}(C^\star,K_{\mathrm{run}})\)
\includegraphics[scale=0.2,trim=0 16 0 0]{A.pdf}}\;

{\(\mathcal W\gets\mathrm{LaunchPersistentWorkers}(b)\)
\includegraphics[scale=0.2,trim=0 16 0 0]{B.pdf}}\;

{\(\mathrm{remaining}(x)\gets|\mathrm{pred}_{\hat G}(x)|\)
for every \(x\in\hat V\)}\;
{\(\mathrm{Busy},\mathrm{Launched},\mathrm{Done}
\gets\varnothing\)}\;

\While{\(\mathrm{Done}\neq\hat V\)
\includegraphics[scale=0.2,trim=0 16 0 0]{C.pdf}}{

    {\(\mathrm{Ready}\gets
    \{x\in\hat V\setminus\mathrm{Launched}:
    \mathrm{remaining}(x)=0\}\)}\;

    {Order \(\mathrm{Ready}\) by RL step,
    \textsc{Comm}/\textsc{Reshard} before \textsc{Op},
    then topological order}\;

    \ForEach{\(x\in\mathrm{Ready}\) in this order}{
        \If{\(\mathcal R(x)\cap\mathrm{Busy}=\varnothing\)}{
            {Reserve \(\mathcal R(x)\) in \(\mathrm{Busy}\)}\;
            {Bind inputs and required parameter versions}\;
            {Record dispatch; submit \(x\) asynchronously
            to \(\mathcal W\)}\;
            {\(\mathrm{Launched}\gets
            \mathrm{Launched}\cup\{x\}\)}\;
        }
    }

    {Wait for newly completed nodes}\;

    \ForEach{newly completed node \(x\)}{
        {Record completion; publish outputs and model versions}\;
        {Release \(\mathcal R(x)\) from \(\mathrm{Busy}\)}\;
        {\(\mathrm{Done}\gets\mathrm{Done}\cup\{x\}\)}\;
        \ForEach{successor \(z\) of \(x\)}{
            {\(\mathrm{remaining}(z)\gets
            \mathrm{remaining}(z)-1\)}\;
        }
        {Release artifacts and parameter copies with
        no remaining consumers}\;
    }
}

{Stop workers; return execution trace}\;
\end{algorithm}

Model invocations use \textbf{PyTorch/Megatron} with the selected
DP/PP/TP groups. \textbf{NCCL} supports tensor-native parameter
transfers and redistribution, while the principal experimental
campaign uses \textbf{Gloo} for semantic data transfers.
On Clariden, worker processes share a pooled \textbf{Slurm}
launch with NCCL communication over OFI/libfabric/CXI.
These backend choices realize the graph's operations and transfers
without changing its logical dependencies.

\iftr
\textbf{Physical realization.}
Table~\ref{pp:tab:runtime-provenance} summarizes how RPEG's
optimization decisions are translated into physical execution,
connecting the selected graph and execution plan to concrete
Megatron workers, communication operations, and parameter versions.

\textbf{Execution verification.}
Since RPEG jointly optimizes graph structure, placement, parallelism,
and scheduling, its performance gains are meaningful only if the
runtime executes the configuration actually selected by the optimizer.
Otherwise, unintended changes in execution could invalidate the
predicted-to-measured performance correspondence or alter training
semantics. RPEG therefore identifies each optimized plan (and fused
microtask schedule, if applicable) by a hash and verifies its physical
GPU bindings, parallelism, dependencies, parameter versions, and
execution order against the runtime trace.
\fi

%% file: eval.tex
\section{Evaluation}

We show the advantages of \nameA\ {along four complementary dimensions}:
accuracy of the analytical time/memory model, schedule quality, improvements over
prior systems, and optimizer efficiency \& stability.

Experiments use two HPC platforms: an A100 system
with eight 40-GB GPUs, intra-node NVSwitch, and four 200-Gb/s
InfiniBand interfaces per node; and a GH200 system with four GPUs
and four Slingshot-11 interfaces per node.
Due to space constraints, further setup details are in Appendix~\ref{app:eval}.

\subsection{Model Validation}

We first validate both execution-time and peak-memory predictions against
measurements on NVIDIA GH200 {GPUs}. The study covers 123
configurations spanning three LLaMA model sizes (3B, 7B, 13B), batch sizes
$B \in \{32, \ldots, 2048\}$, GPU counts from 1 to 128, and 25 distinct
$(P_d,P_p,P_t)$ strategies, including DP-only, TP-only, PP-only, and
3D-hybrid configurations. We fix $L_p=512$, $L_r=128$, and $R=1$.
All runs use BF16, FlashAttention-2, activation checkpointing for training,
and ZeRO-2 gradient sharding for $P_d>1$. Hardware parameters come from the
automated calibration suite described in Section~\ref{sec:calibration}.

\begin{figure}[t]
    \centering
    \includegraphics[width=\columnwidth]{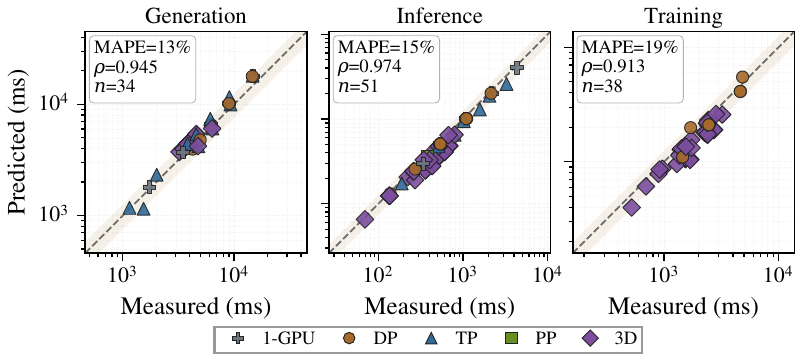}
    \caption{\textbf{Predicted vs.\ measured operator latency.} Each point is one (workload, strategy) configuration. Marker shape indicates the parallelism strategy. The solid line denotes exact agreement and the shaded band marks $\pm 25\%$ error. Overall MAPE is 15.5\% with Spearman $\rho=0.990$ ($n=123$), correctly ranking 95.4\% of all configuration pairs.}
    \label{fig:cost:time-validation}
\end{figure}

\textbf{Execution Time. }
Figure~\ref{fig:cost:time-validation} compares predicted and measured
operator latency for generation, inference, and training. The model attains
15.5\% mean absolute percentage error (MAPE) and Spearman rank correlation
$\rho=0.990$, correctly ranking 95.4\% of all configuration pairs.
NCCL microbenchmarks directly calibrate the communication
parameters $(\alpha,\beta)$; the operator measurements then test their
combined effect with compute and other costs including TP AllReduce,
PP point-to-point transfers, and ZeRO-2 gradient synchronization.

\begin{figure}[t]
    \centering
    \includegraphics[width=\columnwidth]{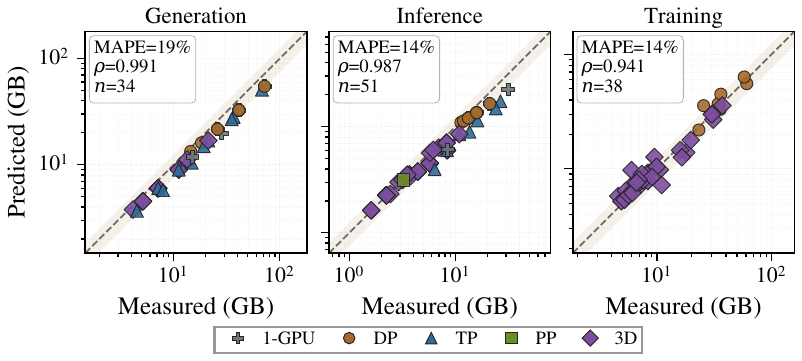}
    \caption{\textbf{Predicted vs.\ measured peak GPU memory.}
    The model achieves
    15.1\% MAPE with Spearman $\rho=0.972$,
    {correctly ordering memory use for 93.7\% of configuration pairs}.
    The remaining underprediction in inference and generation reflects
    unmodeled CUDA context overhead (${ \sim }2$--$3$\,GB per GPU).}
    \label{fig:cost:mem-validation}
\end{figure}

\textbf{Peak Memory. }
Figure~\ref{fig:cost:mem-validation} compares predicted and measured
per-device peak memory.
The memory model achieves 15.1\% MAPE with
Spearman $\rho=0.972$, {correctly identifying which configuration
uses more memory in 93.7\% of pairs}. The systematic underprediction for
inference and generation {reflects} unmodeled CUDA context overhead
of roughly 2--3\,GB per GPU, {for which the
$\varepsilon_{\mathrm{rt}}$ margin in $M_{\mathrm{eff}}$ reserves capacity}.

\subsection{Predicted Schedule Advantages}
\label{sec:eval-predicted}

\begin{figure}[t]
    \centering
    \includegraphics[width=\columnwidth]{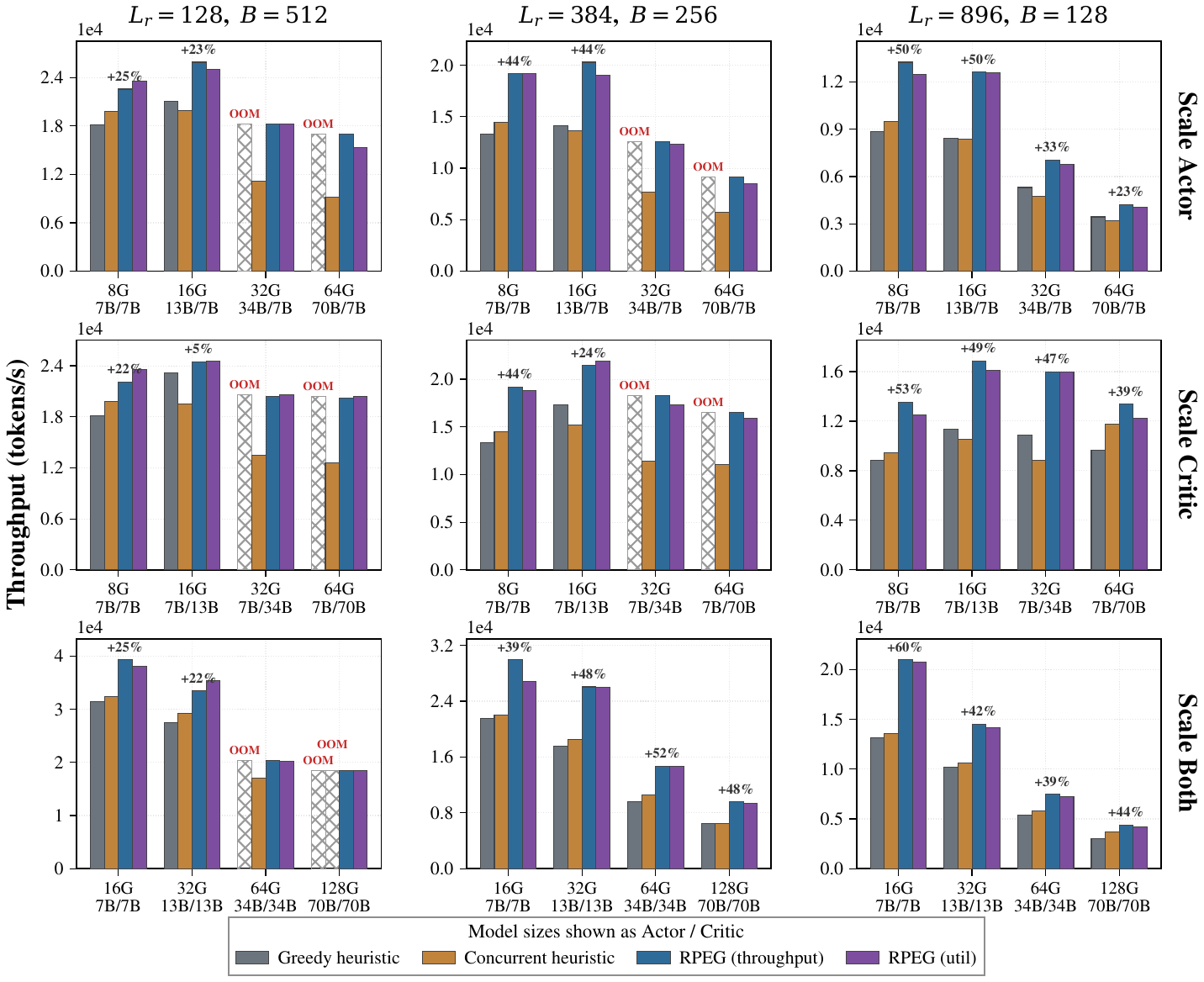}
    \caption{\textbf{Weak scaling: predicted throughput (tokens/s)} across three scaling experiments (rows) and three workload regimes (columns). OOM marks infeasible baselines. Percentage labels indicate the optimizer's gain over the concurrent heuristic. Models: LLaMA-2.}
    \label{fig:scaling-throughput}
\end{figure}

We also compare the execution plans selected by RPEG against heuristic
baselines under the analytical model. This isolates the effect of joint
optimization of placement, parallelism, overlap, and bounded asynchrony
from runtime-system overheads. We compare to two baselines: the \textbf{greedy baseline} chooses a
feasible strategy for each invocation on the full GPU set and executes
invocations sequentially; the \textbf{concurrent baseline} assigns independent
invocations disjoint allocations. 

\subsubsection{Weak Scaling}

\begin{figure}[t]
    \centering
    \includegraphics[width=\columnwidth]{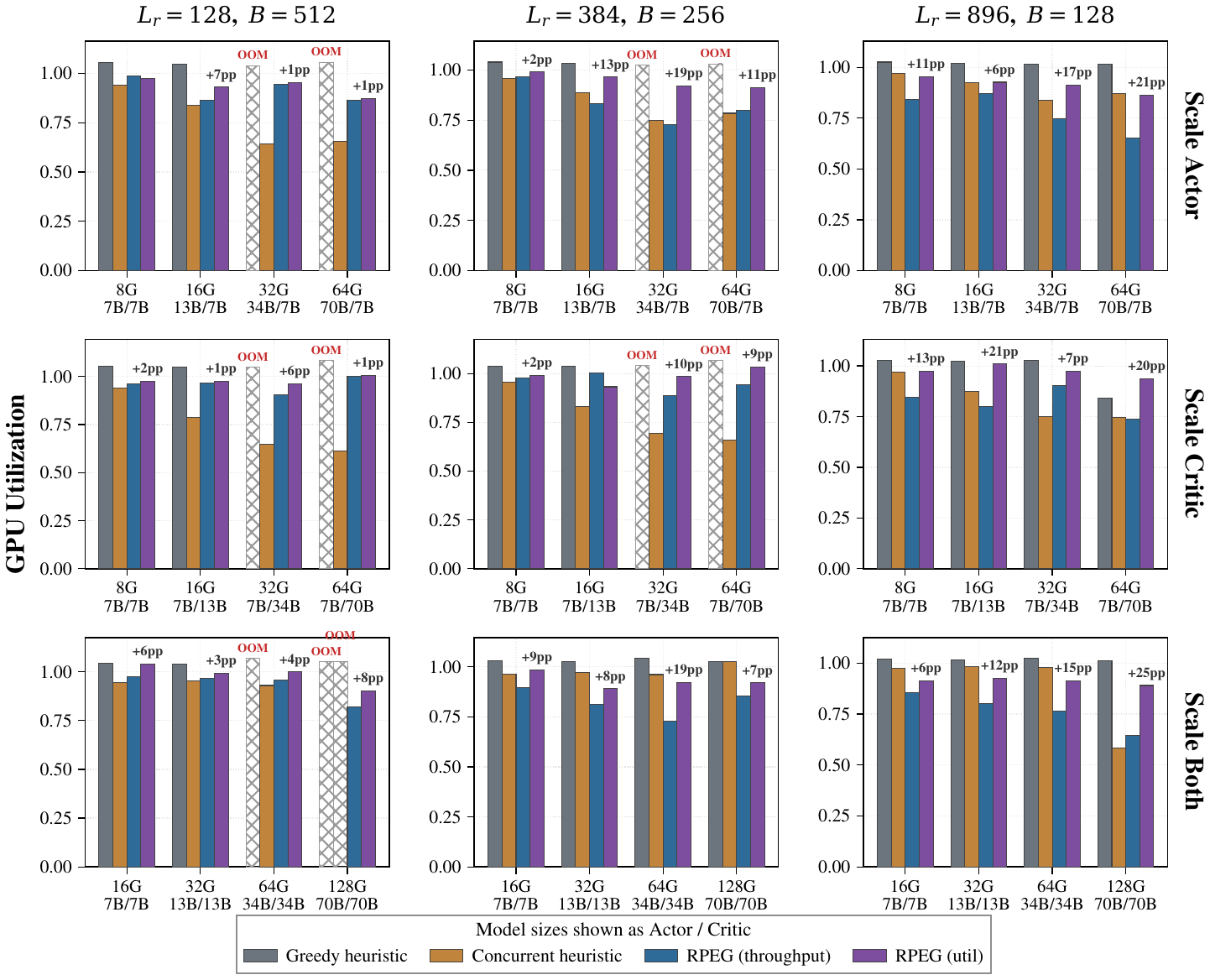}
    \caption{\textbf{Weak scaling: predicted GPU utilization} for the same configurations as Figure~\ref{fig:scaling-throughput}. The throughput$\times$utilization objective (red) consistently improves utilization over the throughput-only objective (green) by up to +25\,pp (absolute difference between utilization percentages). Models: LLaMA-2.}
    \label{fig:scaling-util}
\end{figure}

The weak scaling results are in Figures~\ref{fig:scaling-throughput}
and~\ref{fig:scaling-util};
we increase actor size, critic/reward size, or both as GPU count
grows from 8 to 128. The $(B,L_r)$ settings vary workload characteristics.

RPEG improves
\textbf{throughput} by 5--60\% over the concurrent heuristic where that baseline
is feasible, and finds feasible plans in all tested cases, including
those with OOM baselines. Gains are largest for long responses, which
expose more generation--assessment overlap.
Next, RPEG does not optimize for raw device occupancy alone. Instead, it
{assigns} generation, assessment, and training (and their
subunits) to different GPU subsets with different 3D strategies.
The resulting GPU {utilization-aware} plans attain 78--100\% scheduled
\textbf{GPU utilization}. Relative to throughput-only optimization, they trade
1--3\% throughput for up to 25 percentage points higher utilization.
This is scheduled resource occupancy, not a measurement of GPU arithmetic
utilization.

\subsubsection{Strong Scaling}

\begin{figure}[h]
    \centering
    \includegraphics[width=\linewidth]{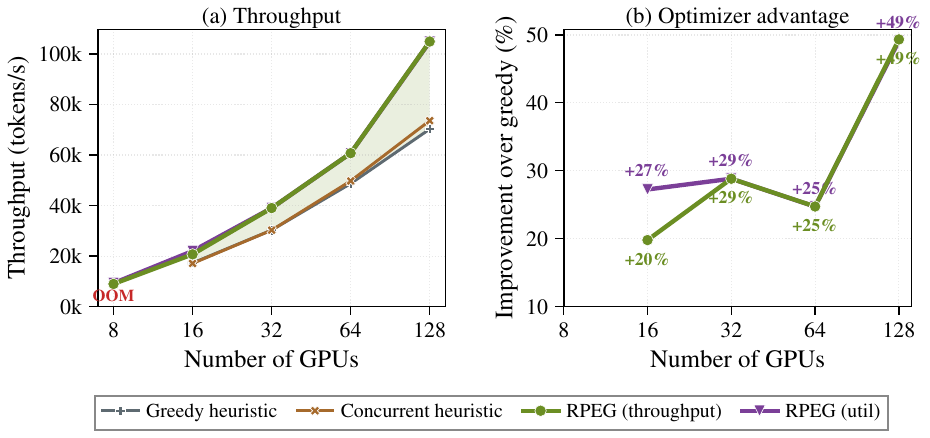}
    \caption{\textbf{Strong scaling.} (a) Absolute throughput (tokens/s). At 8 GPUs, both baselines are OOM. (b) Percentage improvement over the greedy heuristic. The optimizer's advantage grows with GPU count. Model: LLaMA-2 13B.}
    \label{fig:strong-scaling}
\end{figure}

In the {strong-scaling regime} (Figure~\ref{fig:strong-scaling}) the model is fixed and GPU count increases.
With 13B models, $B=1024$, $L_p=128$, and $L_r=384$, RPEG finds a
feasible 8-GPU plan at 8,966 tokens/s while both baselines are OOM.
From 16 to 128 GPUs, predicted throughput is 20--49\% higher than the
greedy baseline, reaching 104,930 tokens/s at 128 GPUs.

\subsubsection{Ablations}

Table~\ref{tab:ablation} varies one feature at a time (out of nine) across four
scenarios: memory-constrained execution (16 GPUs, 50-GB HBM limit), a wide
search space (128 GPUs), heterogeneous model sizes (34B actor and 7B
critic/reward), and deep pipelines (70B critic/reward across two racks).
Each setting uses five seeds, for 200 runs in total. The largest average losses in predicted performance
come from disabling bounded asynchrony (A2, 16.1\%),
generation--inference overlap (A1, 9.2\%), and microbatch
tuning (A3, 4.3\%). A3 fixes $mb=1$ for $P_p=1$;
otherwise, it fixes $mb=P_p$ for generation/inference and
$mb=2P_p$ for training. All A1--A9 modifications are identified in the
table.

\begin{table}[h]
\centering
\ssmall
\sf
\setlength{\tabcolsep}{1pt}
\renewcommand{\arraystretch}{0.8}
\caption{\textbf{Ablation study}. Search space features (\textbf{A1--A3}) dominate. The last row (\textbf{A4--A8}) summarises pair moves, warm restarts, critical-path vertex selection, and soft OOM penalty, which show $< 1\%$ impact across 9 configurations spanning memory-constrained and wide-search-space regimes. We describe the details of each ablation configuration in Appendix~\ref{app:eval}.}
\label{tab:ablation}
\begin{tabular}{@{} l r r r r @{}}
\toprule
\textbf{Ablation} & \textbf{16 GPU, 13B} & \textbf{32 GPU, 34B} & \textbf{64 GPU, 70B} & \textbf{Avg.} \\
 & (PFLOP/s) & (PFLOP/s) & (PFLOP/s) & \\
\midrule
Full system ($\tau{=}1$) & 0.233 & 0.656 & 1.339 & -- \\
\midrule
A1: gen-inf overlap    & 0.203 (-12.8\%) & 0.569 (-13.2\%) & 1.318 (-1.5\%)  & \textbf{-9.2\%} \\
A2: async execution    & 0.189 (-18.8\%) & 0.497 (-24.3\%) & 1.270 (-5.2\%)  & \textbf{-16.1\%} \\
A3: mb tuning          & 0.238 (+2.1\%)  & 0.672 (+2.4\%)  & 1.105 (-17.5\%) & \textbf{-4.3\%} \\
\midrule
A7: polish phase       & 0.233 (+0.0\%)  & 0.647 (-1.5\%)  & 1.269 (-5.2\%)  & \textbf{-2.2\%} \\
A9: geometric cooling  & 0.233 (+0.0\%)  & 0.607 (-7.5\%)  & 1.327 (-0.9\%)  & \textbf{-2.8\%} \\
\midrule
A4/A5/A6/A8            & \multicolumn{3}{c}{$< 1\%$ average impact across all tested configurations} & $< 1\%$ \\
\bottomrule
\end{tabular}
\end{table}

\textbf{The effects are workload-dependent}, e.g., A3 improves
performance by 2.1--2.4\% in the two smaller configurations
but reduces it by 17.5\% in the 70B case. Search-algorithm
changes have smaller average effects: disabling final polishing
(A7) and using geometric cooling (A9) reduce performance by
2.2\% and 2.8\%, respectively, while A4/A5/A6/A8 have less
than 1\% average impact. Thus, most of the gain is due to
the available execution choices, not to any single search heuristic.

\begin{figure}[h]
    \centering
    \includegraphics[width=\linewidth]{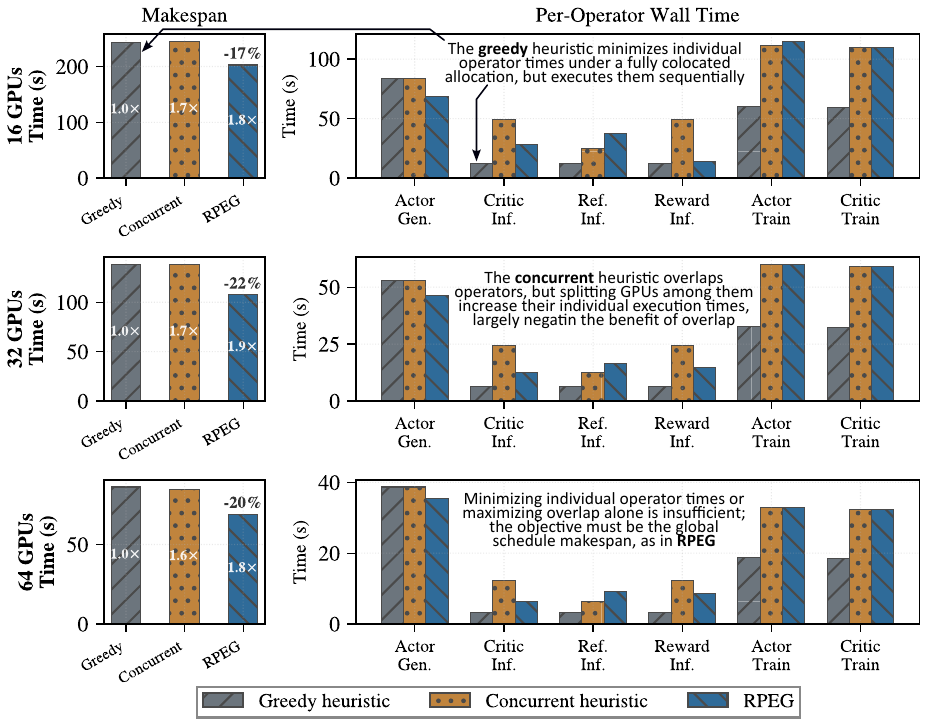}
    \caption{\textbf{Predicted per-operator wall-time breakdown} for LLaMA-2 13B PPO on 16, 32, and 64 GPUs; $B=1024$, $L_p=128$, and $L_r=384$. The greedy heuristic (left) executes all operators sequentially. The concurrent heuristic (center) partitions GPUs but does not reduce makespan significantly. The optimizer (right) reduces makespan by 17--22\% via GPU partitioning, gen-inf overlap, and asynchronous execution.}
    \label{fig:operator-breakdown}
\end{figure}

\textbf{A representative case study} illustrates the trade-off
between minimizing individual operator times and the overall
schedule makespan (Figure~\ref{fig:operator-breakdown}).
We quantify effective operator parallelism as
$P_{\mathrm{eff}} =
\sum_{v\in\hat V_{\mathrm{Op}}}T_{\mathrm{op}}(v)/
\mathrm{MS}(\hat G)$,
i.e., the sum of individual operator durations divided by
the total schedule makespan.
The greedy heuristic minimizes individual operator times
under a fully colocated allocation, but executes them
sequentially ($P_{\mathrm{eff}}=1$).
The concurrent heuristic overlaps operators
($P_{\mathrm{eff}}=1.6$--$1.7$), but splitting GPUs
among them increases their individual execution times,
largely negating the benefit of overlap.
RPEG also makes locally suboptimal choices: most operators
execute more slowly than under greedy, yet their overlap
($P_{\mathrm{eff}}=1.8$--$1.9$) reduces overall makespan
by 17--22\%.
Notably, actor generation is an exception: RPEG accelerates it
despite allocating fewer GPUs, because the larger local batch
improves utilization during memory-bound decoding.
Thus, minimizing individual operator times or maximizing
overlap alone is insufficient; the objective must be
the \textit{global} schedule makespan.

\subsection{Measured Performance Improvements}
\label{sec:eval-runtime}

We next assess end-to-end performance.

\subsubsection{Comparison with Existing Systems}

\begin{figure}[t]
    \centering
    \includegraphics[width=1.0\linewidth]{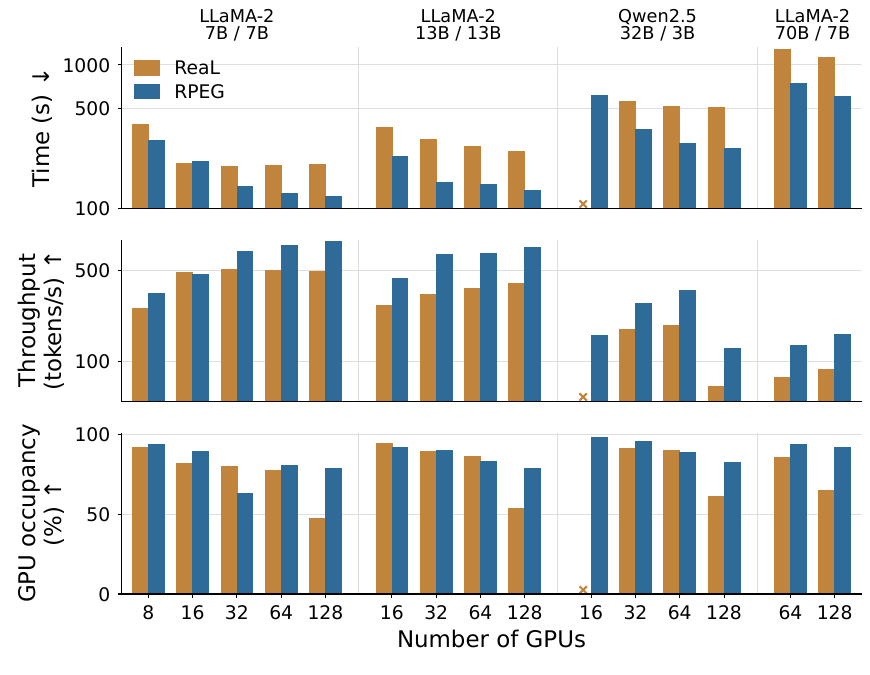}
    \caption{\textbf{RPEG versus ReaL}. System: A100s.}
    \label{fig:runtimes}
\end{figure}

\begin{figure}[t]
    \centering
    \includegraphics[width=1.0\linewidth]{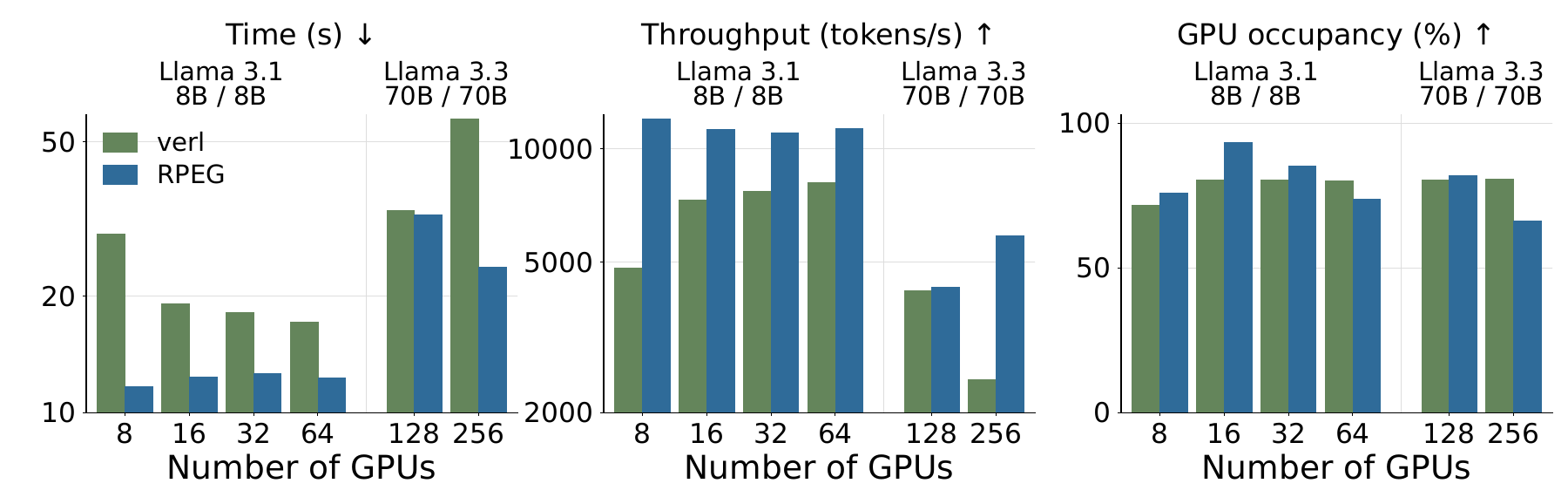}
    \caption{\textbf{RPEG versus \textsc{verl}}. System: GH200s.}
    \label{fig:runtimes-verl}
\end{figure}

We first evaluate RPEG against two existing systems under
synchronous PPO, without bounded asynchrony.
On A100 GPUs, we compare \textbf{RPEG against ReaL}
(Figure~\ref{fig:runtimes}), using models from 7B to 70B.
RPEG is faster in 11 of the 13 jointly executable configurations
and comparable in two (7B/16 GPUs and 32B/32 GPUs).
Throughput improves by $1.4\times$--$2\times$,
with up to 30\% higher effective GPU utilization.
For 32B/16 GPUs, RPEG also finds and executes a
memory-feasible plan where ReaL finds none.
On GH200 GPUs, we compare \textbf{RPEG against \textsc{verl}}
(Figure~\ref{fig:runtimes-verl}), using Llama 3.1 8B
and Llama 3.3 70B with separate actor/critic models.
RPEG achieves $1.024\times$--$2.477\times$ speedup
(median $1.488\times$).

\subsubsection{Joint Pipeline-Structure and Execution Optimization}

We next investigate deeper the benefits of jointly optimizing pipeline structure
$G^\circ$ and execution plan $P$.

\paragraph{Optimization Within A Fixed Training Structure}
We compare feasible, fully colocated baseline configurations
against RPEG-optimized configurations executed using the same
runtime on GH200 GPUs, eliminating cross-framework differences.
For each baseline, we preserve the AC architecture
(shared or separate), synchronous execution, model size,
workload, and GPU budget, while optimizing placement,
DP/PP/TP parallelism, microbatching, and applicable
execution refinements.
Seven Llama 3.1 8B baselines yield 20 measured comparisons
(Table~\ref{pp:tab:before-after}).
All optimized configurations outperform their respective
baselines by $1.250\times$--$1.471\times$
(mean $1.365\times$), reducing graph execution time
by approximately 20--32\%.
This establishes the benefit of optimization within each
structural variant, independently of switching variants.

\begin{table}[h]
\centering
\ssmall
\sf
\renewcommand{\arraystretch}{0.8}
\setlength{\tabcolsep}{3pt}
\caption{\textbf{Benefits of RPEG optimization within fixed
actor--critic architectures}. Model: Llama 3.1 8B. System: GH200s.}
\begin{tabular}{llrrl}
\toprule
\textbf{Nodes} & \textbf{AC structure} &
\makecell[c]{\textbf{Baseline time} \\ \textbf{[seconds/step]}} &
\makecell[c]{\textbf{RPEG time} \\ \textbf{[seconds/step]}} &
\textbf{Speedup} \\
\midrule
2  & Shared   & 11.678 & 8.026--8.158   & 1.432--1.455$\times$ \\
4  & Shared   & 14.311 & 10.726--10.960 & 1.306--1.334$\times$ \\
4  & Separate & 16.752 & 11.392--12.436 & 1.347--1.471$\times$ \\
8  & Shared   & 12.822 & 9.129--9.167   & 1.399--1.404$\times$ \\
8  & Separate & 16.770 & 11.837--12.780 & 1.312--1.417$\times$ \\
16 & Shared   & 13.189 & 10.479--10.552 & 1.250--1.259$\times$ \\
16 & Separate & 16.852 & 12.244--13.031 & 1.293--1.376$\times$ \\
\bottomrule
\end{tabular}
\label{pp:tab:before-after}
\end{table}

\paragraph{Optimization Across Structural Variants}
We summarize the effects of structural optimizations using
Llama 3.1 8B and Llama 3.3 70B on GH200 GPUs.
\textbf{Bounded asynchrony} ($\tau=1$) improves every available
same-architecture comparison against synchronous RPEG,
accelerating separate-model execution by
$1.346\times$--$1.429\times$ at 8B and
$1.217\times$--$1.662\times$ at 70B.
\textbf{AC sharing} generally reduces execution time
by eliminating duplicated model computation and state, by 12--34\% ($1.14\times$--$1.51\times$ speedup).
%
%
%
%
\textbf{Stage fusion} preserves separate actor/critic parameters
while interleaving their execution: a topology-matched 8B test
reduces runtime by 7\%, although expensive resharding
can reverse the benefit.
%

Importantly, for a given fixed PPO configuration, RPEG \textbf{preserves outcome accuracy}: it 
optimizes the scheduling of PPO operations \textit{without changing
the underlying generation, reward computation, advantage estimation,
or loss functions}; bounded asynchrony only changes parameter freshness.
Following the evaluation methodology of AsyncFlow, we compare
RPEG's asynchronous execution ($\tau=1$) against synchronous
\textsc{verl} in terms of mean rollout reward and response length,
using comparable PPO hyperparameters (including update rate,
batch size, learning rate, and maximum sequence lengths),
generation settings, and old-policy log-probability computation.
RPEG achieves a mean reward on the final rollout within
0.47\% relative difference of the \textsc{verl} baseline, providing
empirical evidence of comparable training outcomes under
the evaluated settings.

\input{table-designs}

\subsection{Optimizer Efficiency and Stability}
\label{sec:eval-optimizer}

A richer optimization space is only useful if it can still be searched
efficiently. RPEG remains practical because candidate evaluation is analytical and
heavily cached.

\begin{figure}[h]
    \centering
        \includegraphics[width=0.48\columnwidth]{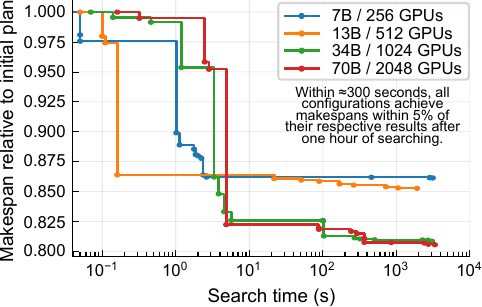}
        \hfill
    \includegraphics[width=0.48\columnwidth]{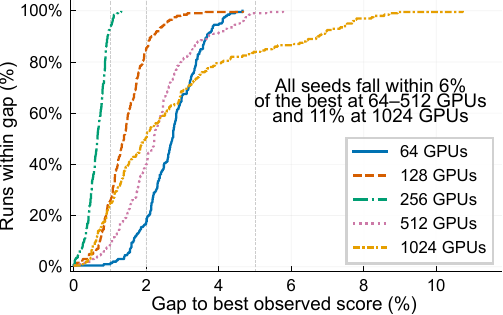}
    \caption{\textbf{(Left: optimizer search effectiveness study)} Best discovered predicted makespan relative
    to the initial plan (lower is better) versus search time,
    across 256--2048 modeled GPUs.
    \textbf{(Right: optimizer stability study)} Cumulative distribution of gaps to the best found outcome
    across 200 independent seeds per configuration (600-second searches).}
    \label{rpegx:fig:eval-optimizer-stability}
\end{figure}

\textbf{Search Cost. }
A convergence study spanning 256--2048 GPUs
(Figure~\ref{rpegx:fig:eval-optimizer-stability}, left) shows that
the optimizer reduces predicted makespan by approximately 14--19\%
relative to the initial plan. Importantly, within just 300 seconds, even at
2048 GPUs, the best discovered plan is already within 5\% of the
makespan achieved after a full hour of searching.
Thus, most optimization benefits are obtained within minutes.
For perspective, amortizing a one-time 300-second (600-second)
search over 1,000 PPO updates at our measured 70B/64-node rate
of 23.7 seconds/update yields an estimated overhead of only
1.3\% (2.5\%) of total runtime.
%

\textbf{Run-to-Run Stability. }
To quantify robustness, we evaluate \(200\) independent runs per
configuration over scales from \(64\) to \(1024\) GPUs.
Figure~\ref{rpegx:fig:eval-optimizer-stability} (right) shows score gaps
relative to the best result found across seeds for each configuration.
The coefficient of variation ranges from 0.27\% to 2.31\%.
All runs at 64--512 GPUs are within 6\% of the best observed score;
at 1024 GPUs they are within 11\%.

%% file: table-designs.tex
\begin{table*}[t]
\centering
\sf
\ssmall
\setlength{\tabcolsep}{2.0pt}
\renewcommand{\arraystretch}{1.08}

\caption{\textbf{Design dimensions of representative RLM-training systems.}
``\faY'' denotes explicit support, ``\faH'' partial or restricted support, and
``\faN'' no first-class support in the cited design. Brief annotations indicate
the mechanism or scope of support.}

\label{tab:taxonomy_rlm_systems}

\begin{tabular}{@{}lllllllll@{}}
\toprule

\textbf{Framework} &
\makecell[c]{\textbf{Cross-iteration}\\\textbf{asynchrony}} &
\makecell[c]{\textbf{Intra-iteration}\\\textbf{overlap}} &
\makecell[c]{\textbf{Sub-invocation}\\\textbf{execution}} &
\makecell[c]{\textbf{Execution}\\\textbf{plan search}} &
\makecell[c]{\textbf{Pipeline}\\\textbf{structure search}} &
\makecell[c]{\textbf{Performance +}\\\textbf{memory model}} &
\makecell[c]{\textbf{Plan}\\\textbf{verification}} &
\makecell[c]{\textbf{Resource-eff.}\\\textbf{objective}} \\

\midrule

\textbf{ReaL~\cite{mei2024real}} &
\faN &
\faH\,(indep.\ ops) &
\faN &
\faY\,(place+3D+mb) &
\faN &
\faY\,(profiled) &
\faN &
\faN \\

\textbf{HybridFlow/\textsc{verl}~\cite{sheng2025hybridflow}} &
\faH\,(\textsc{verl} runtime) &
\faH &
\faN &
\faY\,(place+3D) &
\faN &
\faY\,(simulation) &
\faN &
\faN \\

\textbf{RLHFuse~\cite{zhong2025rlhfuse}} &
\faN &
\faY\,(fusion) &
\faY\,(sample+mb) &
\faH\,(fusion plans) &
\faH\,(fusion) &
\faY\,(simulation) &
\faN &
\faN \\

\textbf{AReaL~\cite{fu2025areal}} &
\faY\,(rollout/train overlap) &
\faN &
\faH\,(samples) &
\faH\,(resource split) &
\faN &
\faN &
\faH\,(version control) &
\faN \\

\textbf{AsyncFlow~\cite{Han2025AsyncFlow}} &
\faY\,(streaming pipeline) &
\faY\,(pipeline) &
\faY &
\faH\,(load balance) &
\faN &
\faN &
\faH\,(staleness checks) &
\faN \\

\textbf{Laminar~\cite{sheng2025laminar}} &
\faY &
\faN &
\faY\,(samples) &
\faH\,(repacking) &
\faN &
\faN &
\faN &
\faN \\

\textbf{StaleFlow~\cite{li2026unleashing}} &
\faY\,(bounded) &
\faH\,(disaggregation) &
\faY\,(samples) &
\faH\,(rollout coord.) &
\faN &
\faN &
\faH\,(staleness checks) &
\faN \\

\textbf{RLinf~\cite{yu2025rlinf}} &
\faH\,(pipelining) &
\faY\,(elastic pipeline) &
\faY\,(microflows) &
\faH\,(place+pipeline) &
\faH\,(flow rewrite) &
\faY\,(profiled) &
\faN &
\faN \\

\textbf{OrchestrRL~\cite{tan2026orchestrl}} &
\faY\,(1-step overlap) &
\faH\,(disaggregation) &
\faH\,(generation) &
\faH\,(generation only) &
\faN &
\faY\,(compute+net sim.) &
\faN &
\faH\,(network cost) \\

\textbf{Libra~\cite{chen2026libra}} &
\faY\,(rollout/train overlap) &
\faH\,(disaggregation) &
\faY\,(requests) &
\faH\,(stage-level) &
\faN &
\faY\,(analytical) &
\faH\,(state sync.) &
\faH\,(resource split) \\

\midrule

\textbf{\nameAS [This work]} &
\faY\,(bounded) &
\faY\,(fusion) &
\faY\,\makecell[l]{(phase+\\microtask)} &
\faY\,(place+3D+mb) &
\faY\,\makecell[l]{(AC+fusion+\\staleness)} &
\faY\,\makecell[l]{(calibrated,\\analytical)} &
\faY\,(hash+versioning) &
\faY\,\makecell[l]{(throughput$\times$\\utilization)} \\

\bottomrule
\end{tabular}
\end{table*}

%% file: rw.tex
\section{Related Work}
\label{sec:related-work}

Recent RLM-training systems can be compared along several complementary architectural features: \emph{cross-iteration asynchrony}, which overlaps different RL iterations; \emph{intra-iteration overlap}, which overlaps or fuses generation, assessment, and training within an iteration; \emph{sub-invocation execution}, which schedules units smaller than a whole model invocation; \emph{execution plan search}, which automatically selects placement, parallelism, or microbatching; \emph{pipeline structure search}, which automates changing the pipeline structure or execution semantics themselves; \emph{performance and memory modeling}, which predicts candidate execution costs; \emph{plan verification}, which checks that the selected logical and physical plan is faithfully realized at runtime; and \emph{resource-efficiency objectives}, which explicitly consider resource use beyond raw runtime or throughput. Existing systems expose different subsets of these capabilities~\cite{mei2024real,sheng2025hybridflow,zhong2025rlhfuse,fu2025areal,Han2025AsyncFlow,sheng2025laminar,li2026unleashing,yu2025rlinf,tan2026orchestrl,chen2026libra}.

Our work extends this landscape along three main axes. First, RPEG exposes a \emph{richer graph configuration} that enables
searching over AC-sharing and bounded-staleness settings together with
per-invocation placement, parallelism, and microbatching, all while
supporting fusion and sub-invocation response-length refinements. Second, RPEG couples this space to a \emph{comprehensive analytical model} that captures heterogeneous computation, communication and resharding, hierarchical hardware, co-resident and versioned state, and explicit throughput--resource-efficiency tradeoffs. Third, RPEG provides an \emph{execution-faithful runtime}: optimized graph structure, placement, model versions, transfers, and resource ownership are reconstructed and verified during distributed execution. Table~\ref{tab:taxonomy_rlm_systems} provides a more fine-grained comparison.

%% file: conc.tex
\section{Conclusion}

RLM training is no longer a single-model scaling problem: it is a heterogeneous, multi-model, asynchronous pipeline optimization problem. RPEG addresses this by making pipeline structure, overlap, placement, and parallelization explicit within one graph-centric representation, and by coupling that representation to calibrated HOC-focused analytical models of time and memory. This enables the optimizer to search jointly over structural and execution decisions, yielding better schedules than prior baselines and translating those gains into measured improvements in throughput, makespan, and utilization on real hardware.

More broadly, RPEG points toward a general systems layer beyond
PPO-style RLM training, jointly optimizing heterogeneous operators,
bounded asynchrony, communication hierarchy, and cost. Such a layer
could support multi-agent and tool-augmented LLM systems, training
with richer verification or search, and inference-time reasoning.
Thus, RPEG is not merely a training optimizer, but a step toward
graph-native orchestration of future large-scale LLM systems.

%% file: acks.tex
\section*{Acknowledgments}
We thank Hussein Harake, Colin McMurtrie, Mark Klein, Angelo Mangili, and the whole CSCS team granting access to the Ault and Alps machines, and for their excellent technical support.
We gratefully acknowledge Polish high-performance computing infrastructure PLGrid (HPC Center: ACK Cyfronet AGH) for providing compute facilities within computational grant no.~PLG/2026/019437; we also thank Łukasz Flis and the whole Cyfronet team for their excellent technical support.
We thank Timo Schneider for help with infrastructure at SPCL.
This project received funding from the European Research Council (Project PSAP, No.~101002047).
This project was supported by the ETH Future Computing Laboratory (EFCL), financed by a donation from Huawei Technologies.
%
%
%
ChatGPT~\cite{openai2026chatgpt} served as an editorial tool for this manuscript, while ideas and content are original work of the authors.

%% file: appendix-pipeline.tex
\clearpage
\section{\nameAS Pipeline: Formulation \& Design Details}
\label{sec:app:pipeline}
\suppressfloats[t]

We now provide extended mathematical details of RPEG.

\subsection{RPEG Pipeline Execution Graph}

\begin{rpegdefinition}[RLM Pipeline Execution Graph]
An RLM Pipeline Execution Graph (RPEG) is a tuple
\[
G^\circ = \left(V^\circ, E^\circ_{\mathrm{data}}, E^\circ_{\mathrm{pub}}, A, M, \tau \right),
\]
where:
\begin{itemize}
    \item $V^\circ$ is a finite set of RPEG vertices (representing model invocations);
    \item $E^\circ_{\mathrm{data}}$ and $E^\circ_{\mathrm{pub}}$ are finite sets of data and publication edges, where each edge $e = (u \to v, a)$ connects a source vertex $u$ to a destination vertex $v$ and carries a single container $a \in A$;
    \item $A$ is the finite set of containers referenced by the RPEG; the boundary subsets $A_{\mathrm{in}}^\circ, A_{\mathrm{out}}^\circ \subseteq A$ (external inputs and outputs) are characterized by well-formedness condition~3; 
    \item $M$ is a set of all models represented in the RPEG pipeline (e.g., actor, critic, reference, reward for PPO-based training);
    \item $M_{\mathrm{train}}\subset M$ is the set of trainable models, where $M$ denotes the full set of models;
    \item $\tau: M_{\mathrm{train}}\to\mathbb N$ assigns a bounded-staleness budget per trainable model.
\end{itemize}    
\end{rpegdefinition}

\begin{figure}[t]
    \centering
    \includegraphics[width=1.0\linewidth]{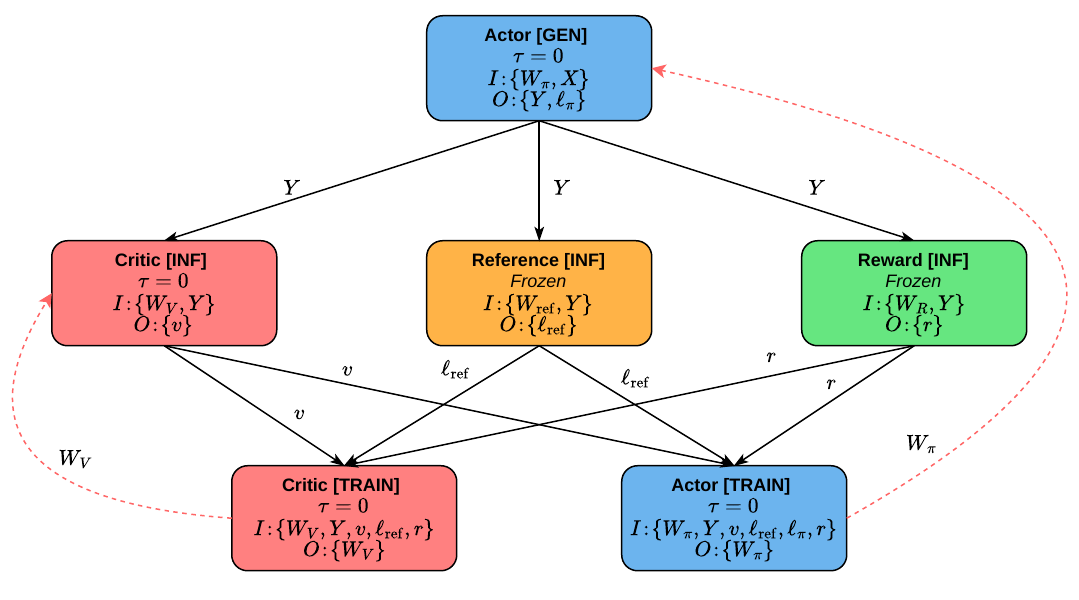}
    \caption{\textbf{RPEG for a PPO-based RLM training pipeline.} 
    The trainable model set is $M_\mathrm{train}=\{\pi_\theta, V_\psi\}$ (actor and critic), both with staleness budget $\tau=0$ (synchronous execution; depicted like this only in this example). 
    External inputs are $A^\circ_\mathrm{in}=\{X, W_\mathrm{ref}, W_R\}$, and no external outputs are exposed ($A^\circ_\mathrm{out}=\emptyset$). 
    Containers: $X$ = prompts, $Y$ = completions, $v$ = per-token value predictions, $r$ = per-completion reward, $\ell_\mathrm{ref}$ = reference log-probs, $\ell_\pi$ = actor log-probs, and $W_m$ = model $m$'s weights.
    Edges encode scheduling dependencies, not all data accesses: boundary inputs $A^\circ_\mathrm{in}$ are accessible to all operators that declare them in their interface, and containers such as $Y$ and $\ell_\pi$ reach training transitively through the assessment stage without requiring direct edges.
    The structure is shown for a fixed workload.}
    \label{ls:fig:rpeg}
\end{figure}

Figure~\ref{ls:fig:rpeg} illustrates the PPO RPEG. A complete formal instantiation -- including all container definitions, operator input and output interfaces, and edge annotations -- is given in Section~\ref{ls:sec:appendix:ppo}.

We now further detail each component of the RPEG.

\vspace{1em}
\subsubsection{\textbf{Vertices and Operator Types}}

Each RPEG vertex $v\in V^\circ$ represents a model invocation. A vertex is characterized by three attributes:
\begin{itemize}
    \item a \emph{role} $r(v)\in  \{\text{generation, inference, training}\}$,
    \item a \emph{model} $\mu(v)\in M$, and
    \item \emph{input/output container interfaces} $\mathcal{I}(v), \mathcal{O}(v)\subseteq A$ declaring which containers the operator may read and write.
\end{itemize}
We write $\kappa(v) = (r(v), \mu(v), \mathcal{I}(v), \mathcal{O}(v))$ for the \emph{operator kind} of~$v$.

The vertex set therefore determines which models participate in the algorithm and how they appear within one iteration. Different policy optimization algorithms correspond to different choices of $V^\circ$ and dependency structure $(E^\circ_\mathrm{data}, E^\circ_\mathrm{pub})$. For example, the base PPO RPEG comprises six vertices -- generation, three inference passes, and two training passes -- while GRPO eliminates the critic and reduces to four. DPO removes online generation entirely, collapsing the graph to three.

\vspace{1em}
\subsubsection{\textbf{Containers}}

A \emph{container} has an identity and size metadata $(\#a,\;s(a))$, where $\#a$ is the element count and $s(a)$ is the element size in bytes. The total byte volume of a container is $\#a \cdot s(a)$.

Containers make communication volume explicit in the IR: the byte volume $\#a \cdot s(a)$ of a container on a connecting edge determines the data transfer cost when source and destination are placed on different devices. The set $A$ collects all containers in the RPEG, covering both within-step data (e.g., completions, log-probabilities, or rewards) and cross-step model weights.

\vspace{1em}
\subsubsection{\textbf{Data and Publication Edges}}

Each RPEG edge carries exactly one container and represents the algorithmic dependence of the destination on the data produced by the source. The RPEG distinguishes two classes of edges:

\begin{itemize}
    \item \textbf{Data edges} $E^\circ_{\mathrm{data}}$ represent within-step dependencies. Generation produces completions consumed by inference; inference produces log-probabilities, rewards, and values consumed by training.
    \item \textbf{Publication edges} $E^\circ_{\mathrm{pub}}$ represent cross-step weight updates. After training updates a model's parameters, a publication edge carries the new weights to the operators that consume them in subsequent iterations. These are the edges whose unrolling semantics are governed by the staleness budget~$\tau$.
\end{itemize}

An important distinction separates \emph{edges} from \emph{operator interfaces}. Each operator declares input and output interfaces ($\mathcal{I}$ and $\mathcal{O}$) that specify which containers it may access, but not every declared input requires a corresponding incoming edge, since edges impose both a data-transfer requirement and a scheduling dependency. Two cases lack edges: boundary inputs $A^\circ_\mathrm{in}$ (e.g., prompts, frozen weights) are accessible to all operators that declare them since they are external to the graph; and containers already resident from a prior stage can be accessed transitively without a direct edge. For example, training operators read $Y$ and $\ell_\pi$ from generation, but the RPEG by default omits direct generation-to-training edges -- unless explicitly determined by the user.

\subsection{Well-Formedness}
\label{ls:sec:rpeg:wellformed}

Not every tuple of vertices, edges, and containers constitutes a valid RPEG. We therefore impose a set of static well-formedness conditions on $G^\circ$ that are fundamental for lowering, scheduling, and costing the RPEG.

\vspace{1em}
\begin{rpegdefinition}[Well-Formedness]
An RPEG $G^\circ$ is well-formed if the following conditions hold:
\begin{enumerate}
    \item \textbf{Interface consistency.}
    For every edge $e=(u \to v, a)\in E^\circ_{\mathrm{data}} \cup E^\circ_{\mathrm{pub}}$ carrying container $a$, the container must satisfy
    \[
    a \in \mathcal{O}(u) \quad \land \quad a \in \mathcal{I}(v).
    \]
    That is, edges may only transfer containers that the source declares as output and the destination declares as input.
    
    \item \textbf{Publication edges carry model parameters.}
    For every publication edge $e = (u \to v, a) \in E^\circ_{\mathrm{pub}}$, the source vertex $u$ must be a training operator and the container $a$ must represent the weight tensor of the trained model.
    
    \item \textbf{Boundary consistency.}
    If a container $a\in A$ is read by some vertex but has no producer, then $a \in A_{\mathrm{in}}^\circ$.
    If a container $a$ is written by some vertex but has no consumer, then $a \in  A_{\mathrm{out}}^\circ$.
    This condition ensures that all data entering or leaving the RPEG is accounted for explicitly.
    
    \item \textbf{Acyclicity of within-step dependencies.}
    The directed graph $(V^\circ, E^\circ_{\mathrm{data}})$ induced by data edges is acyclic.
    Cyclic dependencies may only arise through publication edges, which represent cross-iteration weight updates.
\end{enumerate} 
\end{rpegdefinition}

These conditions are checked statically on the RPEG and ensure that $ G^\circ$ represents a well-defined steady-state computation before unrolling, lowering, scheduling, and costing.

\subsection{Execution Plan and Device Space}
\label{ls:sec:rpeg:execution}

While the RPEG specifies the algorithmic structure of a steady-state RLM training iteration, execution-level decisions determine how this structure is mapped onto a concrete hardware topology. We formalize both the device space and the execution plan in this section.

\vspace{1em}
\subsubsection{\textbf{Device Space}}

\begin{rpegdefinition}[Device Space]
We model the physical hardware as a hierarchical device space
\[
D = [N_r] \times [N_n] \times [N_g]\footnote{Here we interpret $[N_r] := \{0,\ldots,N_r-1\}$ (analogously for $[N_n]$, $[N_g]$), and we use half-open integer intervals $[a,b):=\{a,\ldots,b-1\}$ when specifying contiguous blocks.},
\]
where $N_r$ denotes the number of racks, $N_n$ the number of nodes per rack, and $N_g$ the number of GPUs per node. The total number of devices is therefore $N_r \cdot N_n \cdot N_g$.
\end{rpegdefinition}

\begin{figure}[t] 
    \centering 
    \includegraphics[width=1.0\linewidth]{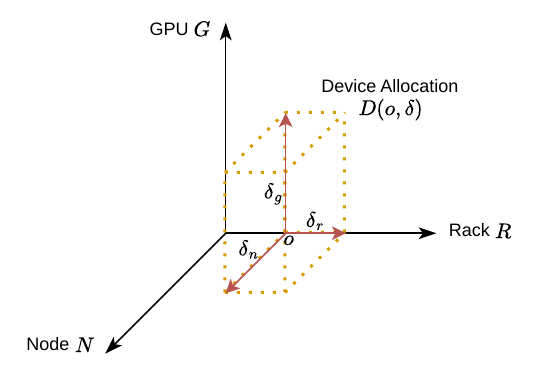} 
    \caption{\textbf{Device space and allocation.} The device space $D = [N_r] \times [N_n] \times [N_g]$ organizes GPUs into a 3D hierarchy. An allocation $D(o, \delta)$ selects a contiguous rectangular region, ensuring that all devices assigned to an operator are physically co-located.} \label{ls:fig:device_space} 
\end{figure}

\vspace{1em}
\begin{rpegdefinition}[Device Allocation]
A \emph{device allocation} is a contiguous hyper-rectangle $D(o,\delta) \subseteq  D$,
defined by an origin $o=(o_r,o_n,o_g)$ and an extent $\delta =(\delta_r,\delta_n,\delta_g)$:
\begin{align*}
    D(o,\delta) = \{ (r,n,g)\in  D \mid
r\in[o_r,o_r+\delta_r),\\
n\in[o_n,o_n+\delta_n),\\
g\in[o_g,o_g+\delta_g)
\}.
\end{align*}
We write $|D(o,\delta)|=\delta_r \cdot \delta_n \cdot \delta_g$ for the number of devices in an allocation. Figure~\ref{ls:fig:device_space} illustrates the device space and a sample allocation.
\end{rpegdefinition}

\vspace{1em}
\begin{rpegexample}[Device Allocation]
Given a cluster with two racks, eight nodes per rack and four GPUs per node, the device space is $D=[2]\times[8]\times[4]$. Allocating the first two nodes in the first rack corresponds to $o=(0,0,0)$ and $\delta=(1,2,4)$.
\end{rpegexample}

\vspace{1em}
\subsubsection{\textbf{Execution Plan}}
The execution plan is a set of functions over $V^\circ$ that assign hardware-dependent attributes -- placement, 3D parallelization strategy, and microbatch count -- to each RPEG vertex. It does not alter the structure of $G^\circ$.

\vspace{1em}
\begin{rpegdefinition}[Execution Plan]
Given an RPEG $G^\circ$, an execution plan is a tuple
\[
P = (\sigma, \ \Pi, \ \mathrm{mb}),
\]
where for every RPEG vertex $v\in V^\circ$:

\begin{itemize}
    \item $\sigma(v)\subseteq D$ is a device allocation;
    \item $\Pi(v)=(P_d,P_p,P_t) \in\mathbb N_{>0}^3$ specifies data, pipeline, and tensor parallel degrees;
    \item $\mathrm{mb}(v)\in\mathbb N_{>0}$ is the number of pipeline microbatches. 
\end{itemize}
\end{rpegdefinition}

The parallelization tuple $\Pi(v)$ determines how operator $v$'s computation is partitioned into shards, while $\sigma(v)$ specifies where these shards are placed in the device space. We require for every vertex $v$ that
\[
|\sigma(v)| = P_d\cdot P_p\cdot P_t,
\]
ensuring that the number of logical shards induced by the parallel degrees matches the number of allocated devices. Furthermore, if $P_p=1$, then $\mathrm{mb}(v)=1$, since microbatching is meaningful only in the presence of pipeline parallelism.

\subsection{PPO RPEG Instantiation}
\label{ls:sec:appendix:ppo}

This section provides the complete formal instantiation of the PPO RPEG.

\vspace{1em}
\subsubsection{\textbf{Models}}

The PPO pipeline uses four models $M = \{\pi_\theta, \pi_\mathrm{ref}, V_\psi, R_\varphi\}$, of which $M_\mathrm{train} = \{\pi_\theta, V_\psi\}$ are trainable. The RPEG has six vertices (Table~\ref{ls:tab:appendix:ppo_vertices}):

\begin{table}[h!]
\centering
\small
\setlength{\tabcolsep}{6pt}
\renewcommand{\arraystretch}{1.15}
\begin{tabularx}{\linewidth}{@{} l l l Y @{}}
    \toprule
    \textbf{Vertex} & \textbf{Role} & \textbf{Model} & \textbf{Description} \\
    \midrule
    $g_\pi$          & generation & $\pi_\theta$      & Autoregressive rollout: produces completions $Y$ and log-probabilities $\ell_\pi$. \\
    $i_\mathrm{ref}$ & inference  & $\pi_\mathrm{ref}$ & Reference forward pass: produces baseline log-probabilities $\ell_\mathrm{ref}$. \\
    $i_R$            & inference  & $R_\varphi$        & Reward scoring: produces per-sequence rewards $r$. \\
    $i_V$            & inference  & $V_\psi$           & Value estimation: produces per-token values $v$. \\
    $t_\pi$          & training   & $\pi_\theta$       & Actor policy-gradient update. \\
    $t_V$            & training   & $V_\psi$           & Critic value-loss update. \\
    \bottomrule
\end{tabularx}
\caption{\textbf{PPO RPEG vertices with operator kinds.} Each vertex is characterized by a role $r(v)$ and a model $\mu(v) \in M$.}
\label{ls:tab:appendix:ppo_vertices}
\end{table}

\subsubsection{\textbf{Containers}}

Table~\ref{ls:tab:appendix:ppo_containers} lists all containers. Let $B$ denote the batch size, $R$ the number of rollouts per prompt, $L_p$ the average prompt length, $L_r$ the average rollout length, $s_b$ the working precision (2\,bytes for BF16), and $\#W_m$ the parameter counts.

\begin{table}[h!]
\centering
\small
\setlength{\tabcolsep}{6pt}
\renewcommand{\arraystretch}{1.15}
\begin{tabularx}{\linewidth}{@{} l l l Y @{}}
    \toprule
    \textbf{Container} & $\boldsymbol{\#a}$ & $\boldsymbol{s(a)}$ & \textbf{Description} \\
    \midrule
    \multicolumn{4}{@{}l}{\textbf{Weight containers (carried by publication edges)}} \\
    \midrule
    $W_\pi$            & $\#W_{\pi_\theta}$      & $s_b$ & Actor parameters. \\
    $W_V$              & $\#W_{V_\psi}$           & $s_b$ & Critic parameters. \\
    $W_\mathrm{ref}$   & $\#W_{\pi_\mathrm{ref}}$ & $s_b$ & Reference parameters (frozen). \\
    $W_R$              & $\#W_{R_\varphi}$        & $s_b$ & Reward parameters (frozen). \\
    \midrule
    \multicolumn{4}{@{}l}{\textbf{Data containers (carried by data edges)}} \\
    \midrule
    $X$                & $BL_p$       & $s_b$ & Prompt tokens. \\
    $Y$                & $BR(L_p{+}L_r)$ & $s_b$ & Completions (prompt + response). \\
    $\ell_\pi$       & $BRL_r$      & $s_b$ & Actor log-probabilities. \\
    $\ell_\mathrm{ref}$ & $BRL_r$   & $s_b$ & Reference log-probabilities. \\
    $r$                & $BR$       & $s_b$ & Rewards (per sequence). \\
    $v$                & $BRL_r$      & $s_b$ & Value estimates (per token). \\
    \bottomrule
\end{tabularx}
\caption{\textbf{PPO RPEG containers.} Boundary inputs: $A_\mathrm{in} = \{X, W_\mathrm{ref}, W_R\}$; no external outputs ($A_\mathrm{out} = \emptyset$).}
\label{ls:tab:appendix:ppo_containers}
\end{table}

\vspace{1em}
\subsubsection{\textbf{Operator Interfaces}}

Each vertex declares which containers it reads and writes (Table~\ref{ls:tab:appendix:ppo_interfaces}). These interfaces determine interface consistency (Section~\ref{ls:sec:rpeg:wellformed}) and are used by the lowering pass to derive communication payloads.

\begin{table}[H]
\centering
\small
\setlength{\tabcolsep}{6pt}
\renewcommand{\arraystretch}{1.15}
\begin{tabularx}{\linewidth}{@{} l l Y @{}}
    \toprule
    \textbf{Vertex} & $\boldsymbol{\mathrm{I}(v)}$ & $\boldsymbol{\mathrm{O}(v)}$ \\
    \midrule
    $g_\pi$          & $\{W_\pi,\; X\}$                                             & $\{Y,\; \ell_\pi\}$ \\
    $i_\mathrm{ref}$ & $\{W_\mathrm{ref},\; Y\}$                                    & $\{\ell_\mathrm{ref}\}$ \\
    $i_R$            & $\{W_R,\; Y\}$                                                & $\{r\}$ \\
    $i_V$            & $\{W_V,\; Y\}$                                                & $\{v\}$ \\
    $t_\pi$          & $\{W_\pi,\; Y,\; \ell_\pi,\; \ell_\mathrm{ref},\; r,\; v\}$ & $\{W_\pi\}$ \\
    $t_V$            & $\{W_V,\; Y,\; \ell_\mathrm{ref},\; r,\; v\}$              & $\{W_V\}$ \\
    \bottomrule
\end{tabularx}
\caption{\textbf{PPO operator input/output interfaces.} Training operators declare containers in their read set that are not delivered by incoming edges (e.g., $t_\pi$ reads $Y$ and $\ell_\pi$ from device memory rather than via a direct edge from $g_\pi$).}
\label{ls:tab:appendix:ppo_interfaces}
\end{table}

\vspace{1em}
\subsubsection{\textbf{Edges}} 

We also formulate PPO edges.

\textbf{Data edges.}
\[
E^\circ_{\mathrm{data}} = \left\{
\begin{array}{l}
g_\pi \xrightarrow{Y} i_{\mathrm{ref}},\quad
g_\pi \xrightarrow{Y} i_R,\quad
g_\pi \xrightarrow{Y} i_V, \\[4pt]
i_{\mathrm{ref}} \xrightarrow{\ell_{\mathrm{ref}}} t_\pi,\quad
i_{\mathrm{ref}} \xrightarrow{\ell_{\mathrm{ref}}} t_V, \\[4pt]
i_R \xrightarrow{r} t_\pi,\quad
i_R \xrightarrow{r} t_V,\quad
i_V \xrightarrow{v} t_\pi,\quad
i_V \xrightarrow{v} t_V
\end{array}
\right\}.
\]
No direct edges connect generation to training.

\textbf{Publication edges.}
\[
E^\circ_{\mathrm{pub}} = \bigl\{\;
t_\pi \xrightarrow{W_\pi} g_\pi,\quad
t_V \xrightarrow{W_V} i_V
\;\bigr\}.
\]

\textbf{Staleness.}
Each publication edge is governed by the staleness budget of its source model: $\tau(\pi_\theta)$ for the actor publication $t_\pi \to g_\pi$, and $\tau(V_\psi)$ for the critic publication $t_V \to i_V$. Setting both to zero recovers synchronous execution; setting $\tau(\pi_\theta) = 1$ allows one step of actor overlap (see Section~\ref{ls:ch:lowering}).

\subsection{Other RLM Training Algorithms}

The RPEG formalism applies to any RLM training algorithm by adjusting the vertex and edge sets. GRPO removes the critic vertices ($i_V$, $t_V$) and derives training signals from relative rewards within groups of sampled responses, reducing the graph to four vertices. DPO removes both the reward model and the online generation loop, collapsing the graph to an offline preference-optimization pipeline. The unrolling, lowering, cost modeling, and optimization procedures from the preceding sections apply without modification to all three variants.


\begin{figure*}[hbtp]
    \centering
    \includegraphics[width=1\textwidth]{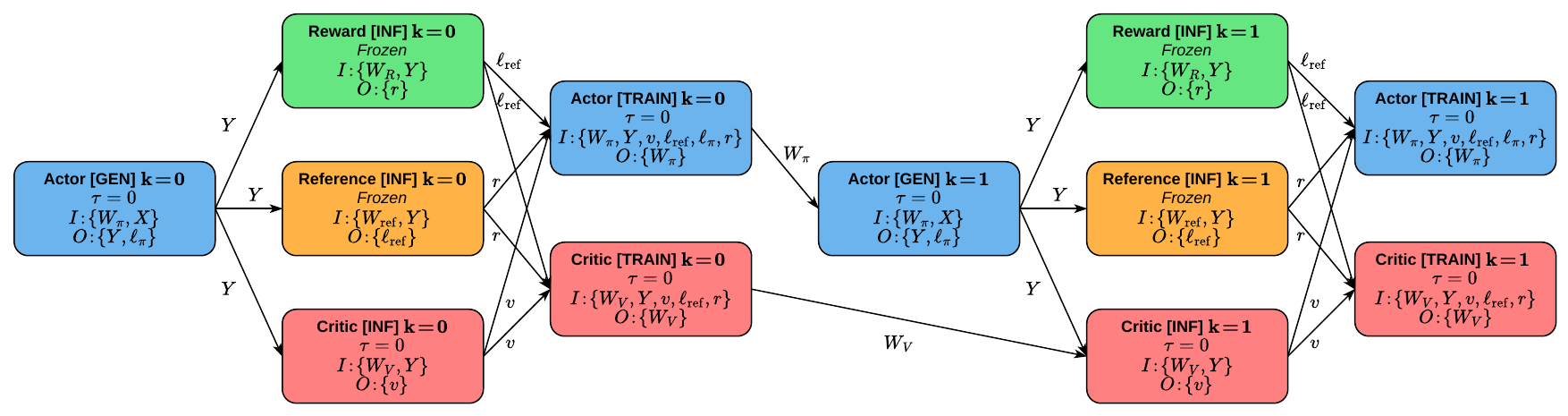}
    \caption{\textbf{Unrolled RPEG for a PPO-based RLM training pipeline.} The cyclic RPEG from \cref{ls:fig:rpeg} is unrolled over $K_\mathrm{eval} = 2$ synchronous steps ($\tau = 0$). Each RPEG vertex is replicated per step, data edges are copied within each step, and publication edges are wired across step boundaries according to the staleness budget $\tau$.}
    \label{ls:fig:unrolled}
\end{figure*}

\begin{figure*}[hbtp]
    \centering
    \includegraphics[width=\textwidth]{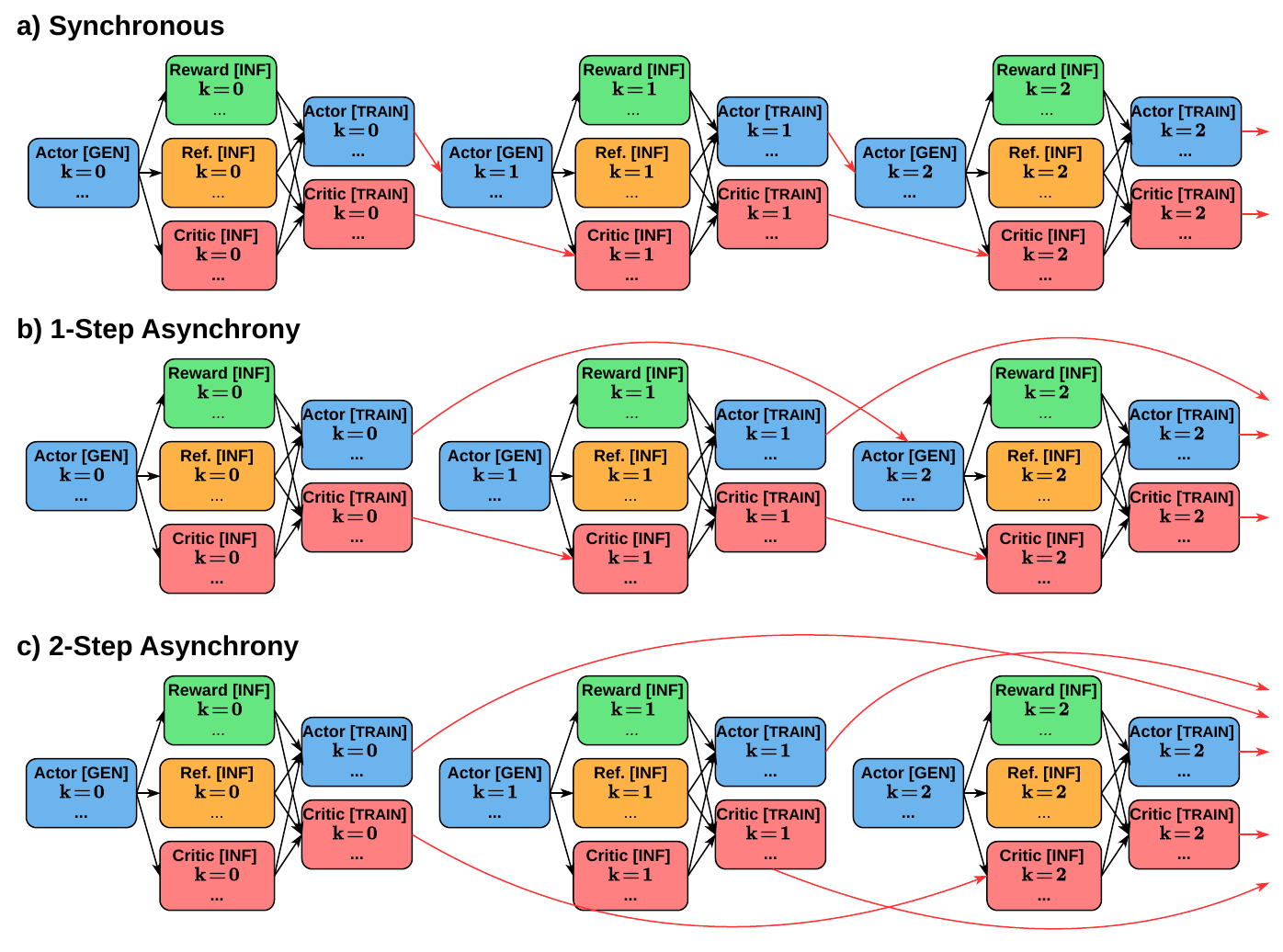}
    \caption{\textbf{Unrolled RPEGs under different staleness budgets} (simplified; operator attributes and edge containers omitted for clarity). (a) Synchronous execution ($\tau = 0$): publication edges connect consecutive steps, enforcing strict ordering. (b) One-step bounded staleness ($\tau(\pi_\theta) = 1$): actor publication edges span two steps, enabling concurrent generation and training. (c) Two-step bounded staleness ($\tau(\pi_\theta) = 2$, $\tau(V_\psi) = 1$): deeper overlap with longer publication arcs. Red arrows denote publication edges carrying weight containers $W_\pi$ and $W_V$; black arrows denote within-step data edges.}
    \label{ls:fig:tau-comparison}
\end{figure*}

\cref{ls:fig:unrolled} shows the synchronous PPO case. \cref{ls:fig:tau-comparison} illustrates how varying $\tau$ changes the structure of the unrolled graph. 

\subsection{Unrolling}
\label{ls:ch:lowering}

Unrolling expands the cyclic RPEG into a finite acyclic graph $\bar{G}$ over $K_{\mathrm{eval}}$ steps, resolving cross-step publication edges according to the staleness budget~$\tau$.
We use PPO as the running example throughout this section, but the same procedures apply to GRPO and DPO graphs.

Before the cyclic RPEG can be unrolled into a finite DAG, we must resolve an ambiguity: the staleness budget $\tau(m)$ bounds how old the consumed weights may be, but does not specify \emph{which} training step provides them. We first formalize this \emph{publisher selection}, then define the unrolled graph that results from it.

\vspace{1em}
\subsubsection{\textbf{Publication and Bounded Staleness}}

When $\tau(m) > 0$ for some model $m \in M_\mathrm{train}$, generation at step $k{+}1$ may overlap with training at step $k$, using slightly stale policy weights for rollout generation.

As noted above, $\tau(m)$ does not determine which specific step provides the weights. To resolve this during unrolling, we introduce a publisher-selection function $s_m(k)$ that maps each step $k$ to the concrete training step whose output it consumes.

\vspace{1em}
\begin{rpegdefinition}[Publisher Selection]
Fix $K_{\mathrm{eval}} > 0$. For each trainable model $m\in M_{\mathrm{train}}$, define
\[
s_m(k)=k-\tau(m)-1,\qquad \tau(m)<k<K_{\mathrm{eval}}.
\]

This choice selects the maximally stale admissible publisher and is deterministic. It satisfies
\[
0 \le s_m(k) < k,
\qquad
(k-1)-s_m(k) \le \tau(m).
\]
\end{rpegdefinition}

The synchronous case is recovered by enforcing $\tau(m)=0$ for all trainable models, yielding $s_m(k)=k-1$.

\vspace{1em}
\subsubsection{\textbf{Unrolled Graph}}

Given the deterministic publisher selection function, we can now formally define the unrolling process. 

\begin{rpegdefinition}[Unrolled Graph]
Given a well-formed RPEG $G^\circ$ and an evaluation horizon $K_{\mathrm{eval}} > 0$, the unrolled graph is
\[
\bar{G} = (\bar{V},\ \bar{E}_{\mathrm{data}},\ \bar{E}_{\mathrm{pub}}),
\]
with the following components:
\begin{itemize}
    \item \textbf{Vertices.} 
    \[
    \bar{V} = V^\circ \times [K_{\mathrm{eval}}],\qquad (v,k)\in\bar{V}.
    \] 
    Each RPEG vertex is replicated once per step index.

    \item \textbf{Within-step edges.} For each data edge $(u\to v, a)\in E^\circ_{\mathrm{data}}$ and each $k\in[K_{\mathrm{eval}}]$,
    \[
    (u,k)\xrightarrow{a}(v,k)\in\bar{E}_{\mathrm{data}},
    \]
    carrying the same container $a$ as the RPEG edge.\footnote{Recall that $[K_\mathrm{eval}] = \{0, \ldots, K_\mathrm{eval}{-}1\}$.}

    \item \textbf{Publication edges.} For each publication edge $(u\to v, a)\in E^\circ_{\mathrm{pub}}$ and each $k\in\{1,\dots,K_{\mathrm{eval}}-1\}$ with $k > \tau(\mu(u))$, where $\mu(u)$ denotes the model invoked by vertex $u$,
    \[
    (u,\;s_{\mu(u)}(k))\xrightarrow{a}(v,k)\in\bar{E}_{\mathrm{pub}},
    \]
    carrying the same weight container $a$ as the RPEG edge. For early steps $k \le \tau(\mu(u))$, no publication edge is created: these steps use the initial parameter version and have no training predecessor. Because no dependency connects them to prior training, the early serving invocations may overlap when the remaining data, ordered-update, and physical-resource constraints permit it; they read the same initial version but process distinct data batches.
\end{itemize}
\end{rpegdefinition}

\vspace{1em}
The key structural property of the unrolled graph is acyclicity, which we verify next.

\vspace{1em}
\begin{rpeglemma}[Acyclicity]
If $( V^\circ, E^\circ_{\mathrm{data}})$ is acyclic, then the unrolled graph $(\bar{ V},\bar{ E}_{\mathrm{data}}\cup\bar{ E}_{\mathrm{pub}})$ is a DAG.
\end{rpeglemma}

\vspace{1em}
\begin{rpegproof}[Proof]
Within-step edges preserve $k$ and are acyclic by the well-formedness assumption on $G^\circ$. Publication edges strictly increase $k$ because $s_m(k) < k$. Together, no directed cycle can exist.
\end{rpegproof}

\vspace{1em}
\subsubsection{\textbf{Ensuring Ordering of Training Updates Under Asynchrony}}

Successive training updates of the same model are
ordered independently of the staleness budget.
For each trainable model $m$ with training operator
$t_m$, unrolling additionally introduces
\[
(t_m,k-1)\xrightarrow{W_m}(t_m,k),
\qquad 1\leq k<K_{\mathrm{eval}}.
\]
These dependencies ensure that consecutive updates
of the same optimizer state remain ordered.

\vspace{1em}
\subsubsection{\textbf{Unrolling Algorithm}}

Algorithm~\ref{ls:alg:unroll} gives the complete unrolling procedure. The runtime is 
\[
O(K_\mathrm{eval} \cdot (|V^\circ| + |E^\circ_\mathrm{data}| + |E^\circ_\mathrm{pub}|)),
\]
as each of the three steps in the pseudocode iterates over one component of the RPEG, replicated $K_\mathrm{eval}$ times.

\begin{algorithm}[t!]
\caption{\textsc{Unroll}: Unroll cyclic RPEG}
\label{ls:alg:unroll}
\KwIn{RPEG $G^\circ = ( V^\circ, E^\circ_{\mathrm{data}}, E^\circ_{\mathrm{pub}},\dots,\tau)$, evaluation horizon $K_{\mathrm{eval}}$}
\KwOut{Unrolled DAG $\bar{ G}=(\bar{ V},\bar{ E}_{\mathrm{data}},\bar{ E}_{\mathrm{pub}})$}
\BlankLine

$\bar{ V} \gets \emptyset$, $\bar{ E}_{\mathrm{data}} \gets \emptyset$, $\bar{ E}_{\mathrm{pub}} \gets \emptyset$\;

\tcc{Step 1: replicate vertices}
\For{$k \gets 0$ \KwTo $K_{\mathrm{eval}}-1$}{
    \ForEach{$v \in  V^\circ$}{
        add $(v,k)$ to $\bar{ V}$\;
    }
}

\tcc{Step 2: replicate within-step data edges}
\For{$k \gets 0$ \KwTo $K_{\mathrm{eval}}-1$}{
    \ForEach{$(u \to v) \in  E^\circ_{\mathrm{data}}$}{
        add $(u,k) \to (v,k)$ to $\bar{ E}_{\mathrm{data}}$\;
    }
}

\tcc{Step 3: instantiate publication edges with bounded staleness}
\For{$k \gets 1$ \KwTo $K_{\mathrm{eval}}-1$}{
    \ForEach{$(u \to v) \in  E^\circ_{\mathrm{pub}}$}{
        $m \gets \mu(u)$\;

        \If{$k \le \tau(m)$}{
            continue \tcc*{step $k$ uses initial weights of model $m$}
        }

        $k' \gets \max\{0,\ k - (\tau(m) + 1)\}$\;

        add $(u,k') \to (v,k)$ to $\bar{ E}_{\mathrm{pub}}$\;
    }
}

\Return{$\bar{ G}$}\;
\end{algorithm}

Each unrolled edge inherits the container from its RPEG edge. The algorithm omits this for brevity.

\begin{figure*}[t]
    \centering
    \includegraphics[width=1\textwidth]{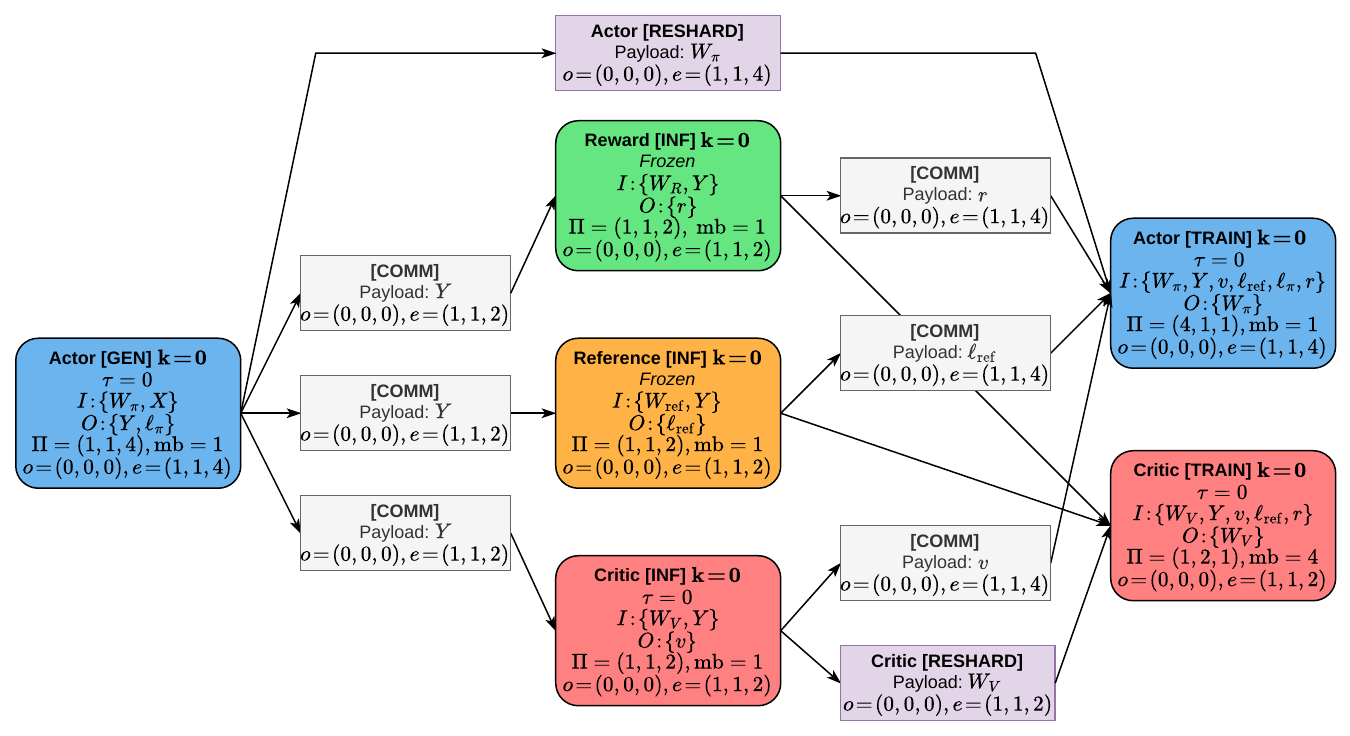}
    \caption{\textbf{Executable graph for one step of a PPO-based RLM training pipeline.} Derived from the RPEG in \cref{ls:fig:rpeg} via unrolling and lowering. \textsc{Comm} and \textsc{Reshard} nodes are inserted wherever the execution plan requires data movement between different device allocations or parallel layouts.}
    \label{ls:fig:executable}
\end{figure*}

\subsection{Lowering} \label{ls:sec:lowering}

Lowering incorporates the execution plan $P$ to produce the executable graph $\hat{G}$, inserting explicit communication and resharding nodes wherever placement or strategy differences require data movement.
Thus, lowering is the stage at which the unrolled graph $\bar{G}$ and the execution plan $P$ are combined into a single hardware-aware DAG.
The resulting executable graph $\hat{G}$ serves as input to the cost model and scheduler defined further. Both transformations are defined generically over any RPEG. We use PPO as the running example throughout this section, but the same procedures apply to GRPO and DPO graphs.

Concretely, lowering takes the pair $(\bar{G}, P)$ and constructs the executable graph $\hat{G}$ by traversing the nodes and edges of $\bar{G}$ and inserting \textsc{Comm} and \textsc{Reshard} nodes wherever the execution plan requires data movement or weight redistribution.

\vspace{1em}
\begin{rpegdefinition}[Executable Graph]
The executable graph is a DAG
\[
\hat{G} = (\hat{V}, \ \hat{E}),
\]
whose nodes are of three kinds:
\begin{itemize}
    \item \textsc{Op} nodes correspond to unrolled RPEG vertices, each attributed with a device allocation, parallel strategy, and microbatch count from the execution plan.
    \item \textsc{Comm} nodes represent data transfer between \textsc{Op} nodes placed on different device allocations (on both data and publication edges).
    \item \textsc{Reshard} nodes represent weight redistribution when successive invocations of the same model use different parallel layouts on the same devices.
\end{itemize}
\end{rpegdefinition}

\vspace{1em}
\cref{ls:fig:executable} shows the PPO-based executable graph for one step. Before describing the lowering procedure itself, we specify the attributes that distinguish the three node kinds.

\vspace{1em}
\textbf{Allocation.}
Each \textsc{Op} node occupies exactly one device allocation (as specified by the execution plan). \textsc{Comm} and \textsc{Reshard} nodes are attributed with both the source and destination allocations.

\vspace{1em}
\textbf{Payload.}
Every node in $\hat{G}$ carries a payload in bytes:
\begin{itemize}
    \item each \textsc{Comm} node carries a payload equal to the byte volume of the container on the corresponding data edge,
    \item each \textsc{Reshard} node carries a payload equal to the full parameter size of the associated model,
    \item \textsc{Op} nodes carry zero payload.
\end{itemize}

\begin{figure*}[hbtp]
    \centering
    \includegraphics[width=\textwidth]{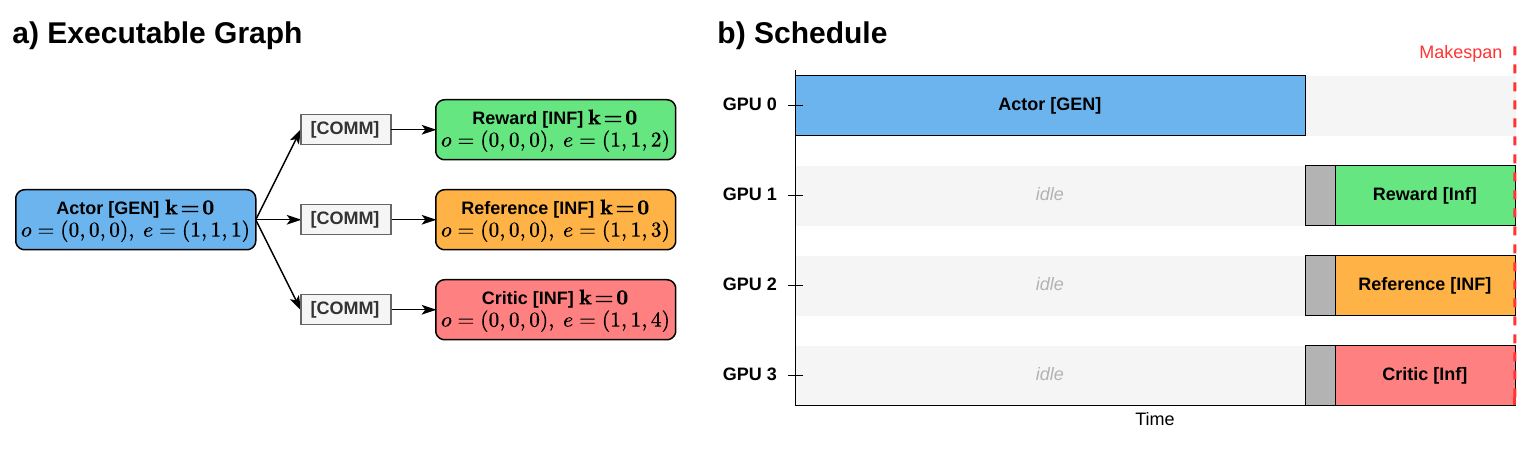}
    \caption{\textbf{Without sub-node splitting.} (a) Executable graph: generation on GPU~0 feeds three inference operators on GPUs~1--3 via \textsc{Comm} nodes. (b) Schedule: inference GPUs are idle for the entire duration of generation, since no partial results are available until the full batch completes. For clarity, only device allocations (origin $o$, extent $\delta$) are shown on each node; parallel strategies, microbatch counts, and read/write interfaces are omitted.}
    \label{ls:fig:batched}
\end{figure*}

\begin{figure*}[hbtp]
    \centering
    \includegraphics[width=\textwidth]{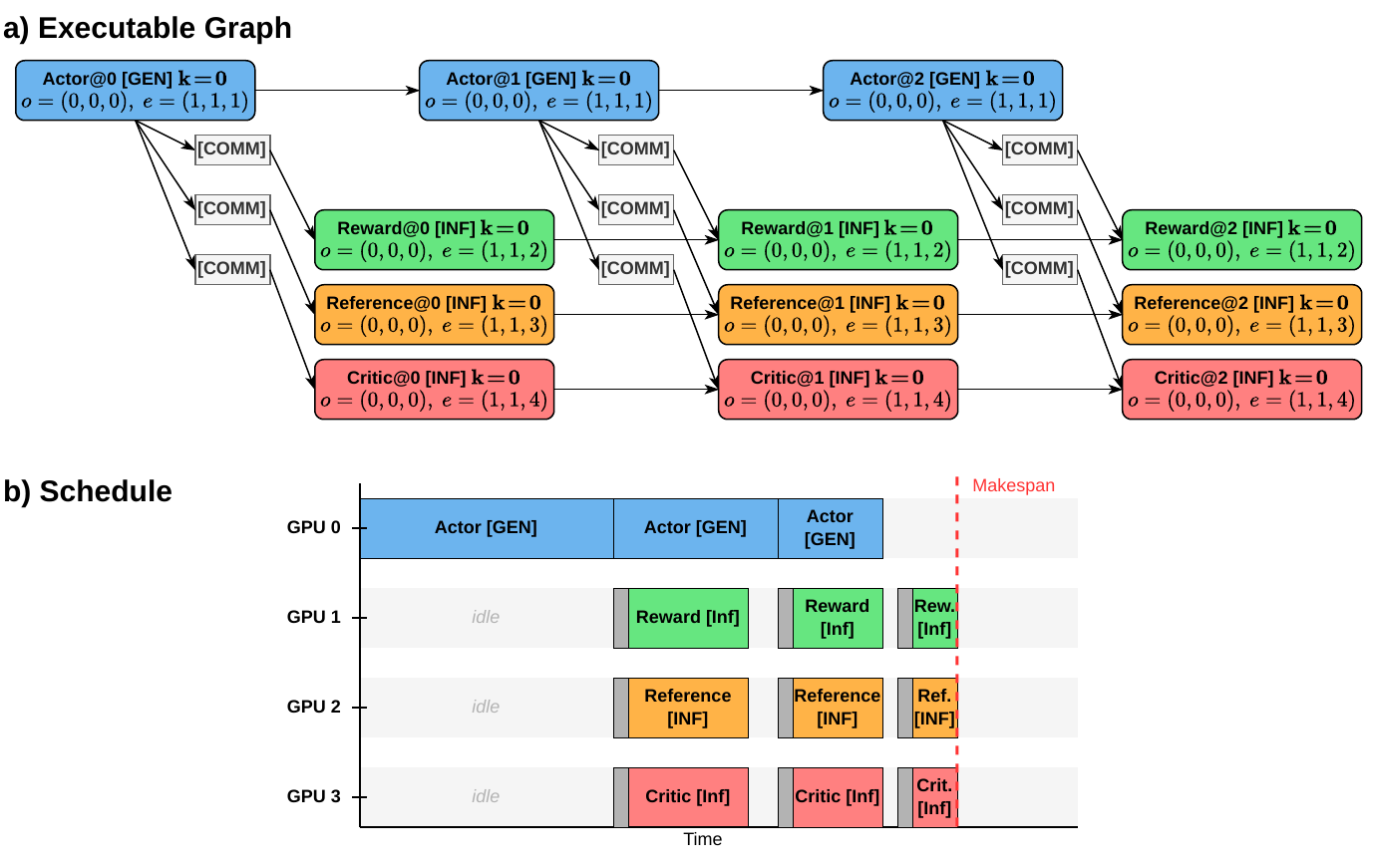}
    \caption{\textbf{With quantile-based sub-node splitting ($Q = 3$).} (a) Executable graph: generation and each inference operator are split into three sub-nodes chained sequentially. Each generation sub-node feeds the corresponding inference sub-nodes via \textsc{Comm} nodes. (b) Schedule: inference sub-nodes from early-finishing sequences overlap with later generation phases, substantially reducing idle time and makespan. For clarity, only device allocations (origin $o$, extent $\delta$) are shown on each node; parallel strategies, microbatch counts, and read/write interfaces are omitted.}
    \label{ls:fig:microbatched}
\end{figure*}

We now describe how the executable graph is constructed by \emph{lowering} the unrolled graph with the execution plan. 

\vspace{1em}
\subsubsection{\textbf{Stage 1: Operator Instantiation}}

For every unrolled vertex $(v,k)\in\bar{V}$, we create an \textsc{Op} node $x_{v,k} \in \hat{V}$ with device allocation $\sigma(v)$, parallel strategy $\Pi(v)$, and microbatch count $\mathrm{mb}(v)$.

When generation quantile splitting is enabled (see below), generation and inference vertices may be replaced by multiple sub-nodes during this stage.

\vspace{1em}
\textbf{Response-length--aware sub-node splitting.}
Autoregressive generation produces responses of highly variable length. In reasoning workloads, response-length distributions are heavily skewed: many sequences finish quickly while a small number of long-tail rollouts dominate wall-clock time~\cite{zhong2025rlhfuse,sheng2025laminar, besta2026performance}. In a monolithic generation operator, the entire batch must finish before any downstream inference can begin -- the longest response delays all subsequent work. When generation and inference are placed on \emph{disjoint} devices, the lowering pass can exploit this skewness by splitting generation into phases based on the response-length distribution, allowing inference to start processing early-finishing sequences while longer ones continue decoding. This is related to the inter-stage fusion mechanism proposed by RLHFuse~\cite{zhong2025rlhfuse}, but expressed here as a structural graph transformation within the lowering pass rather than a runtime scheduling decision.

\vspace{1em}
\begin{rpegdefinition}[Response-Length Quantiles]
\label{ls:def:quantiles}
The response-length distribution is specified by $Q$ quantile breakpoints
\[
(f_1, t_1),\; (f_2, t_2),\; \ldots,\; (f_Q, t_Q),
\]
where each $f_q \in (0, 1]$ is a cumulative fraction of sequences and $t_q$ is the number of generated tokens at which that fraction completes. We require $f_Q = 1$ (all sequences finish by $t_Q = L_r$) and $f_1 < f_2 < \cdots < f_Q$.
\end{rpegdefinition}

\vspace{1em}
\begin{rpegexample}[Response-Length Quantiles]
With $Q = 4$ and breakpoints
\[
\bigl((0.25, 512),\; (0.50, 1024),\; (0.85, 1536),\; (1.0, 2048)\bigr),
\]
25\% of sequences finish within 512 tokens, 50\% within 1024, 85\% within 1536, and all by 2048.
\end{rpegexample}

\vspace{1em}
\textbf{Obtaining the response-length distribution.}
The quantile breakpoints $(f_q, t_q)$ are obtained empirically by profiling generation lengths on a representative sample of prompts. Before optimization, a small subset of the training prompts (e.g., a few hundred) is processed through the current model checkpoint, and the empirical response-length distribution is recorded. The quantile breakpoints are then fitted directly from this sample. The number of quantiles $Q$ controls the granularity of the sub-node splitting: $Q = 1$ recovers the monolithic generation operator, while larger $Q$ enables finer-grained overlap at the cost of additional nodes in the executable graph. In practice, $Q \in \{3, 4, 5\}$ captures the relevant structure of the distribution without introducing excessive graph complexity.

\vspace{1em}
\textbf{Splitting mechanism.}
The lowering pass splits a generation vertex into sub-nodes only when the execution plan places generation and at least one downstream inference operator on \emph{disjoint} devices -- otherwise no overlap is possible and the operator remains an indivisible \textsc{Op} node. Concretely, for each generation vertex $(v, k)$ with a disjoint downstream inference vertex:

\begin{enumerate}
    \item The single \textsc{Op} node $x_{v,k}$ is replaced by $Q$ \emph{generation sub-nodes} $x_{v,k}^{g_1}, \ldots, x_{v,k}^{g_Q}$, chained sequentially on the same devices. Sub-node $x_{v,k}^{g_q}$ represents the decode phase during which the fraction $f_q - f_{q-1}$ of sequences complete (with $f_0 = 0$). The first sub-node additionally includes the prefill phase.

    \item Each downstream inference vertex $w$ on disjoint devices is similarly split into $Q$ \emph{inference sub-nodes} $x_{w,k}^{i_1}, \ldots, x_{w,k}^{i_Q}$, also chained sequentially on their own devices.

    \item Generation sub-node $q$ is connected to inference sub-node $q$ via a \textsc{Comm} node with payload $\#a \cdot s(a) \cdot (f_q - f_{q-1})$ bytes -- the container volume scaled by the fraction of sequences completing in that phase.
\end{enumerate}

All sub-nodes inherit the device allocation, parallel strategy, and microbatch count from the original RPEG vertex.

\cref{ls:fig:batched,ls:fig:microbatched} illustrate the effect of sub-node splitting on both the executable graph and the resulting schedule for a 4-GPU example.

Crucially, this wiring makes gen-inf overlap (inter-stage fusion) \emph{possible} but does not enforce it. The optimizer controls whether overlap occurs by choosing the device allocations: if it places generation and inference on disjoint devices, the scheduler can execute inference sub-nodes concurrently with later generation sub-nodes. If the optimizer places them on the same devices, the scheduler serializes them and no overlap occurs -- in this case, no splitting is performed, since sub-nodes on shared devices would add graph complexity without enabling any concurrency.

\vspace{1em}
\subsubsection{\textbf{Stage 2: Communication Node Insertion}}

\noindent When sub-node splitting has been applied, the edges between generation and inference sub-nodes created in Stage~1 are processed by the following stages in the same way as edges between unsplit \textsc{Op} nodes -- \textsc{Comm} and \textsc{Reshard} insertion does not distinguish between them.

We traverse the data edges of $\bar{G}$. For each edge $(u,k)\xrightarrow{a}(v,k)\in\bar{E}_{\mathrm{data}}$ with container $a$:

\begin{itemize}
    \item \textbf{Same allocation} ($\sigma(u) = \sigma(v)$): no transfer is needed. We add a precedence edge $x_{u,k} \to x_{v,k}$.

    \item \textbf{Different allocation} ($\sigma(u) \neq \sigma(v)$): data must move between device allocations. We insert a \textsc{Comm} node $c$ with payload $\#a \cdot s(a)$ bytes and add edges $x_{u,k} \to c \to x_{v,k}$.
\end{itemize}

After this stage, all cross-allocation data movement is represented explicitly.

\vspace{1em}
\subsubsection{\textbf{Stage 3: Intra-Iteration Weight Redistribution}}

When two operators of the same trainable model appear within one iteration, the lowering pass must ensure that the second operator receives weights in a compatible layout. To identify such pairs, we compute a topological ordering $\mathrm{Topo}(G^\circ)$ of $(V^\circ, E^\circ_{\mathrm{data}})$, collect all vertices with $\mu(v) = m$ for each trainable model~$m$ (frozen models have fixed weights and are never resharded), and define $\mathrm{Pairs}(m)$ as the consecutive pairs in this ordering.

Not every pair requires resharding. Depending on the staleness budget and device placement, a pair falls into one of two categories -- \emph{shadow pairs} and \emph{non-shadow pairs} -- which the lowering pass handles differently.

\vspace{1em}
\textbf{Shadow pairs.}
In synchronous RLM training, generation and training of the same model execute sequentially and may share a single weight copy that is resharded between them as needed~\cite{mei2024real}. When $\tau(m) > 0$ and the optimizer places the two operators on disjoint device allocations, a more efficient pattern becomes possible: each device allocation maintains its own persistent weight copy, avoiding intra-step resharding between those copies while retaining cross-step publication and enabling concurrent execution. We call this a \emph{shadow pair}.

\vspace{1em}
\begin{rpegdefinition}[Shadow Pair]
\label{ls:def:shadow}
A consecutive pair $(u, v) \in \mathrm{Pairs}(m)$ of a trainable model $m \in M_\mathrm{train}$ is a \emph{shadow pair} under execution plan $P$ if:
\begin{enumerate}
    \item $\tau(m) > 0$ \emph{(bounded asynchrony)}, and
    \item $\sigma(u) \cap \sigma(v) = \emptyset$ \emph{(device disjointness)}.
\end{enumerate}
\end{rpegdefinition}

These conditions identify the shadow-pair optimization. Synchronous persistent serving workers can still retain a copy to avoid reconstruction; therefore a single-copy policy is not universally faster. Overlapping allocations restrict concurrency, but physical copy lifetime is determined by actual workers and required versions, not solely by the shadow-pair predicate.

When a shadow pair is active, two weight instances coexist in the device space:
\begin{itemize}
    \item \textbf{Training instance} on $\sigma(t_m)$: holds up-to-date weights, optimizer state (FP32 master weights, first and second moments), and gradients. This is the source of parameter updates.
    \item \textbf{Shadow instance} on the consumer's devices (e.g., $\sigma(g_m)$ for generation or $\sigma(i_m)$ for inference): holds a read-only BF16 weight snapshot. This copy is periodically refreshed by a \textsc{Comm} node that transfers updated weights from the training instance (Stage~4).
\end{itemize}

Because no intra-step reshard or precedence edge connects the two sides, they are free to execute concurrently. The idea of maintaining stale parameter replicas for concurrent execution appears in recent asynchronous RLM training systems~\cite{fu2025areal}; our framework exposes it as a structural property of the execution plan that the optimizer can discover automatically.

For shadow pairs, the lowering pass skips the intra-step reshard entirely: no \textsc{Reshard} node and no precedence edge are inserted between $x_{u,k}$ and $x_{v,k}$ within any step~$k$. 

\vspace{1em}
\textbf{Non-shadow pairs.}
For each non-shadow pair $(u, v) \in \mathrm{Pairs}(m)$ and each $k \in [K_\mathrm{eval}]$:

\begin{itemize}
    \item \textbf{Same allocation and strategy} ($\sigma(u) = \sigma(v)$, $\Pi(u) = \Pi(v)$): the weight layout is already compatible; no action.

    \item \textbf{Same allocation, different strategy} ($\sigma(u) = \sigma(v)$, $\Pi(u) \neq \Pi(v)$): the weights reside on the same devices but in an incompatible parallel layout. We create a \textsc{Reshard} node $r \in \hat{V}$ with payload $\#W_m \cdot s_b$ bytes (the full parameter count of model~$m$) and add edges $x_{u,k} \to r \to x_{v,k}$ to $\hat{E}$.

    \item \textbf{Different allocation} ($\sigma(u) \neq \sigma(v)$): the two operators run on different (possibly overlapping) device allocations -- for example, inference on 8~GPUs and training on 16~GPUs with partial overlap. Since this is not a shadow pair, $\tau(m) = 0$ or the allocations are not fully disjoint. Each side holds its own weight copy: training devices have weights from the optimizer state, while inference devices load them transiently via a prior publication transfer and free them after the operator completes. No redistribution is needed, but a bare precedence edge $x_{u,k} \to x_{v,k}$ enforces that the first operator completes before the second begins.
\end{itemize}

\vspace{1em}
\subsubsection{\textbf{Stage 4: Cross-Iteration Publication Edges}}

Stages~1--3 handle within-step structure. The final stage wires the cross-step publication edges that carry updated weights from training at an earlier step to consumers at a later step. For each publication edge $(u,k') \xrightarrow{a} (v,k) \in \bar{E}_{\mathrm{pub}}$ with $k' < k$, the lowering pass selects one of three actions based on the relative placement and strategy of source and destination:

\begin{itemize}
    \item \textbf{Same allocation and strategy} ($\sigma(u) = \sigma(v)$, $\Pi(u) = \Pi(v)$): a bare precedence edge $x_{u,k'} \to x_{v,k}$ suffices.

    \item \textbf{Same allocation, different strategy} ($\sigma(u) = \sigma(v)$, $\Pi(u) \neq \Pi(v)$): we insert a \textsc{Reshard} node as in Stage~3.

    \item \textbf{Different allocation} ($\sigma(u) \neq \sigma(v)$): the updated weights must be sent from the training devices to the consumer's devices. We insert a \textsc{Comm} node $c$ with edges $x_{u,k'} \to c \to x_{v,k}$; the source retains its copy. When the source and destination allocations partially overlap, devices in the intersection can produce their weight shard locally, reducing the transfer to $\#W_m \cdot s_b \cdot (n_\mathrm{max} - n_\mathrm{overlap}) / n_\mathrm{max}$, where $n_\mathrm{max} = \max(|\sigma(u)|, |\sigma(v)|)$ and $n_\mathrm{overlap} = |\sigma(u) \cap \sigma(v)|$.
\end{itemize}

\vspace{1em}
\subsubsection{\textbf{Lowering Algorithm}}

Algorithm~\ref{ls:alg:lower} gives the complete lowering procedure. The runtime is
\[
O\left(K_\mathrm{eval} \cdot (|V^\circ| + |E^\circ_\mathrm{data}| + |E^\circ_\mathrm{pub}|) \right),
\]
since the number of intra-step model pairs is bounded by $|V^\circ|$.

\begin{algorithm}[!t]
\footnotesize
\caption{\textsc{Lower}: lower unrolled DAG to executable graph}
\label{ls:alg:lower}
\KwIn{Unrolled DAG $\bar{G}$, execution plan $P$, RPEG $G^\circ$, quantiles $(f_q, t_q)_{q=1}^Q$ (optional)}
\KwOut{Executable DAG $\hat{G}=(\hat{V},\hat{E})$}
\BlankLine
$\hat{V} \gets \emptyset$, $\hat{E} \gets \emptyset$\;

\BlankLine
\tcc{Stage 1: \textsc{Op} nodes (optional quantile splitting)}
\ForEach{$(v,k) \in \bar{V}$}{
    \eIf{$r(v) = \mathrm{generation}$ \textbf{and} quantiles given \textbf{and} $\exists$ disjoint downstream inf}{
        replace $x_{v,k}$ by sub-nodes $x_{v,k}^{g_1}, \ldots, x_{v,k}^{g_Q}$ and
        chain $x_{v,k}^{g_1} \to \cdots \to x_{v,k}^{g_Q}$\;
        split each disjoint downstream inf into $x_{w,k}^{i_1}, \ldots, x_{w,k}^{i_Q}$\;
        connect $x_{v,k}^{g_q} \to \textsc{Comm} \to x_{w,k}^{i_q}$ with scaled payload\;
    }{
        create \textsc{Op} node $x_{v,k}$ with $\sigma(v), \Pi(v), \mathrm{mb}(v)$\;
    }
}

\BlankLine
\tcc{Stage 2: \textsc{Comm} insertion on data edges}
\tcc{For refined operators, resolve the corresponding input/output interfaces; do not duplicate phase edges already created in Stage 1.}
\ForEach{$(u,k) \xrightarrow{a} (v,k) \in \bar{E}_{\mathrm{data}}$}{
    \eIf{$\sigma(u) = \sigma(v)$}{
        add $x_{u,k} \to x_{v,k}$ to $\hat{E}$\;
    }{
        create \textsc{Comm} node $c$ with payload $\#a \cdot s(a)$\;
        add $x_{u,k} \to c \to x_{v,k}$ to $\hat{E}$\;
    }
}

\BlankLine
\tcc{Stage 3: intra-step weight redistribution}
Compute $\mathrm{Pairs}(m)$ from $\mathrm{Topo}(G^\circ)$ for each trainable $m$\;

\ForEach{$(u,v,m) \in \mathrm{Pairs}$ \textbf{and} $k \in [K_{\mathrm{eval}}]$}{
    \If{$\tau(m) > 0$ \textbf{and} $\sigma(u) \cap \sigma(v) = \emptyset$}{
        skip \tcc*{shadow pair}
    }
    \If{$\sigma(u) = \sigma(v)$ \textbf{and} $\Pi(u) = \Pi(v)$}{skip\;}
    \eIf{$\sigma(u) = \sigma(v)$}{
        create \textsc{Reshard} $r$ with payload $\#W_m \cdot s_b$\;
        add $x_{u,k} \to r \to x_{v,k}$\;
    }{
        add $x_{u,k} \to x_{v,k}$ \tcc*{precedence only}
    }
}

\BlankLine
\tcc{Stage 4: cross-step publication edges}
\ForEach{$(u,k') \xrightarrow{a} (v,k) \in \bar{E}_{\mathrm{pub}}$}{
    \If{the exact required version already has an equivalent installation dependency}{reuse that dependency\;}
    \If{$\sigma(u) = \sigma(v)$ \textbf{and} $\Pi(u) = \Pi(v)$}{
        add $x_{u,k'} \to x_{v,k}$\;
    }
    \ElseIf{$\sigma(u) = \sigma(v)$}{
        create \textsc{Reshard} $r$ with payload $\#W_m \cdot s_b$\;
        add $x_{u,k'} \to r \to x_{v,k}$\;
    }
    \Else{
        create \textsc{Comm} $c$ with payload $\#W_m \cdot s_b \cdot (n_\mathrm{max} - n_\mathrm{overlap}) / n_\mathrm{max}$\;
        add $x_{u,k'} \to c \to x_{v,k}$\;
    }
}

\Return{$\hat{G}$}\;
\end{algorithm}


\subsection{Scheduling}
\label{ls:ch:scheduling}

Given the executable graph $\hat{G}$ and the node-level time map $\mathcal{T}$ (delivered by our performance model), this section details the derivation of the execution time of the entire graph and derives a throughput-based score that serves as the optimization objective.

The key challenge is that the graph's makespan cannot be obtained by simply summing node execution times: nodes may execute concurrently when their precedence constraints are satisfied and their device allocations do not conflict, but must serialize when they compete for the same devices. Computing an optimal schedule under these constraints is NP-hard; we therefore use a greedy list-scheduling heuristic.

\vspace{1em}
\begin{rpegdefinition}[Schedule]
A schedule $\left(s(\cdot), f(\cdot)\right)$ for $\hat{G} = (\hat{V}, \hat{E})$ assigns each node $v \in \hat{V}$ a start time $s(v) \in \mathbb{R}_{\ge 0}$ and finish time $f(v) = s(v) + \mathcal{T}(v)$ such that:
\begin{enumerate}
  \item \textbf{Precedence.} 
  \[
  \forall (u \to v) \in \hat{E}: f(u) \le s(v).
  \]
  That is, a node may only start after all its predecessors have finished.
  \item \textbf{Device reservation.}
  \begin{align*}
    \forall u \ne v \in \hat{V}: \sigma(u) \cap \sigma(v) \neq \emptyset \\ \implies [s(u), f(u)) \cap [s(v), f(v)) = \emptyset.
  \end{align*}
  That is, nodes that reserve overlapping devices may not execute concurrently. Here $\sigma(v)$ denotes the reserved device allocation: the full allocation for \textsc{Op} nodes, and the destination allocation for \textsc{Comm}/\textsc{Reshard} nodes (source devices are free after send initiation).
\end{enumerate}
\end{rpegdefinition}

\vspace{1em}
\begin{rpegdefinition}[Makespan]
The makespan of a schedule  $\left(s(\cdot), f(\cdot)\right)$ is the latest finish time across all nodes:
\[
\mathrm{MS}(\hat{G}) = \max_{v \in \hat{V}} f(v).
\]
\end{rpegdefinition}

\vspace{1em}
\subsubsection{\textbf{Scheduling Heuristic}}
\label{ls:sec:cost:scheduling}

The scheduler constructs a feasible schedule by processing nodes greedily, assigning each node the earliest start time that satisfies both constraints from the schedule definition above.
When multiple nodes are ready simultaneously, the scheduler must decide which to schedule first. We prioritize nodes that lie on or near the critical path, since delaying them is most likely to increase the makespan. This is captured by the downstream rank:

\vspace{1em}
\begin{rpegdefinition}[Downstream Rank]
\[
\mathrm{rank}(v) = \mathcal{T}(v) + \max_{(v \to w) \in \hat{E}} \mathrm{rank}(w),
\]
with $\max \emptyset = 0$. The rank approximates the remaining work on the longest path from $v$ to any sink.
\end{rpegdefinition}

\vspace{1em}
\begin{rpegdefinition}[Earliest Feasible Start Time]
For a ready node $v$ (all predecessors scheduled), the earliest feasible start time is
\begin{equation}
\mathrm{EST}(v) = \max\Bigl\{
\max_{(u \to v) \in \hat{E}} f(u),\;
\max_{d \in \sigma(v)} \mathrm{avail}(d)
\Bigr\},
\label{ls:eq:graphcost:est}
\end{equation}
where $\mathrm{avail}(d)$ tracks when device $d$ becomes free. The first term enforces precedence; the second ensures all blocking devices are available.
\end{rpegdefinition}

\vspace{1em}
\textbf{Selection and tie-breaking.}
At each step, the scheduler selects the ready node with the smallest $\mathrm{EST}$. Ties are broken by:
\begin{enumerate}
    \item \textsc{Comm}/\textsc{Reshard} nodes before \textsc{Op} nodes (to avoid artificial serialization of transfers),
    \item larger downstream rank $\mathrm{rank}(v)$.
\end{enumerate}
After scheduling $v$ at $s(v) = \mathrm{EST}(v)$, the availability of all devices in $\sigma(v)$ is updated to $f(v)$. Two additional tie-breakers (sub-node rank inheritance and RPEG-order consistency) are described after the algorithm.

\vspace{1em}
\subsubsection{\textbf{Scheduling Algorithm}}

Algorithm~\ref{ls:alg:schedule} returns $\mathrm{MS}(\hat{G})$. Since $|\hat{V}| = O(K_\mathrm{eval} \cdot |V^\circ|)$ and $|\hat{E}| = O(K_\mathrm{eval} \cdot |E^\circ|)$, the runtime is 
\[
O(K_\mathrm{eval} \cdot |V^\circ| \log(K_\mathrm{eval} \cdot |V^\circ|) + K_\mathrm{eval} \cdot |E^\circ|).
\] 
The implementation uses a min-heap with lazy deletion, which accounts for the $\log$ factor.

\begin{algorithm}[t]
\caption{\textsc{GreedySchedule}$(\hat{G}, \mathcal{T})$}
\label{ls:alg:schedule}
\KwIn{Executable DAG $\hat{G}=(\hat{V},\hat{E})$, time map $\mathcal{T}$, device space $D$.}
\KwOut{Schedule $(s, f)$ and makespan.}
\BlankLine
Compute $\mathrm{rank}(v)$ for all $v$ in reverse topological order\;

Initialize $\mathrm{avail}(d) \gets 0$ for all $d \in D$\;

$\mathcal R \gets \{v \in \hat{V} : v \text{ has no predecessors}\}$\;

\While{$\mathcal R \neq \emptyset$}{
  Select $v^\star \in \arg\min_{v \in \mathcal R} \mathrm{EST}(v)$\;

  \tcc{Tie-break: \textsc{Comm}/\textsc{Reshard} before \textsc{Op}, then larger rank}

  $s(v^\star) \gets \mathrm{EST}(v^\star)$,\quad
  $f(v^\star) \gets s(v^\star) + \mathcal{T}(v^\star)$\;

  \ForEach{$d \in \sigma(v^\star)$}{$\mathrm{avail}(d) \gets f(v^\star)$\;}

  Update $\mathcal R$: remove $v^\star$, add newly-ready successors\;
}
\Return{$\max_{v \in \hat{V}} f(v)$}\;
\end{algorithm}

\vspace{1em}
\textbf{Sub-node rank inheritance.}
When generation quantile splitting (\cref{ls:sec:lowering}) produces a chain of sub-nodes $x_{v,k}^{g_1} \to \cdots \to x_{v,k}^{g_Q}$ on the same devices, rank naturally decreases along the chain (later sub-nodes have shorter remaining paths). Without correction, the scheduler may interleave unrelated operators between generation phases, which disrupts KV-cache residency (the cache built during earlier phases would be evicted before later phases use it). To prevent this, all sub-nodes in a chain inherit the rank of the chain head (the first sub-node).

\vspace{1em}
\textbf{RPEG-order tie-breaking.}
Beyond rank, the scheduler uses the RPEG vertex index as an additional tie-breaker: operators from the same RPEG position across different unrolled steps are scheduled in the same relative order. This produces steady-state regularity in the schedule, which is important for asynchronous execution where overlapping iterations should exhibit consistent device-usage patterns.

\subsection{Optimization Objective}\label{ls:sec:graphcost:objectives}

The executable graph $\hat{G}$ spans $K_\mathrm{eval}$ unrolled iterations. Under bounded staleness, work from different iterations may overlap in time, so a per-iteration latency metric is not applicable: it cannot capture how consecutive iterations overlap. We instead use global throughput over the full evaluation horizon as the optimization objective.

\vspace{1em}
\subsubsection{\textbf{Throughput Objective}}

Since $W_\mathrm{step}$ is constant for a fixed workload, maximizing throughput is equivalent to minimizing the makespan. The optimizer uses $\mathrm{Score}_\mathrm{penalized}(C)$ (Equation~\ref{ls:eq:cost:penalty}) as its search objective; when the utilization objective is enabled, the penalty is applied to $\mathrm{Score}_\mathrm{eff}(C)$ (Equation~\ref{ls:eq:cost:eff}) instead. The score captures the end-to-end effect of all optimization decisions: the RPEG structure $G^\circ$, the staleness budget $\tau$, the execution plan $P = (\sigma, \Pi, \mathrm{mb})$, and structural choices such as stage fusion and placement strategies.

\vspace{1em}
\begin{rpegdefinition}[Throughput Score]
\label{ls:def:score}
The throughput score of a configuration $C = (G^\circ, P)$ is
\begin{equation}
\mathrm{Score}(C) = \frac{K_\mathrm{eval} \cdot W_\mathrm{step}}{\mathrm{MS}(\hat{G})},
\label{ls:eq:cost:score}
\end{equation}
where $W_\mathrm{step} = BR(L_p + L_r)$ is the number of tokens processed per iteration and $\mathrm{MS}(\hat{G})$ is the makespan from Algorithm~\ref{ls:alg:schedule}.
\end{rpegdefinition}

\vspace{1em}
\textbf{Memory penalty.}
Not all configurations are memory-feasible. Rather than rejecting infeasible configurations outright (which would create hard boundaries in the search landscape), the optimizer applies a soft exponential penalty to the raw throughput score:

\begin{equation}
\mathrm{Score}_\mathrm{penalized}(C) = \mathrm{Score}(C) \cdot e^{-\beta_\mathrm{oom} \cdot \max(0,\, r_\mathrm{peak} - 1)},
\label{ls:eq:cost:penalty}
\end{equation}
\normalsize

where $r_\mathrm{peak} = \max_d\, \mathrm{peak}(d) / M_\mathrm{eff}$ is the worst-case peak-to-capacity ratio across all devices and $\beta_\mathrm{oom} = 40$ controls the penalty steepness. When $r_\mathrm{peak} \le 1$ (feasible), the penalty is $1$ and the score is unchanged. For moderately infeasible configurations ($1$--$3\%$ overflow), the penalty is small enough that the optimizer can temporarily accept high-throughput configurations that slightly exceed memory, then refine them toward feasibility through subsequent moves. At 10\% overflow, the factor is $e^{-4}\approx0.018$; it decreases further for larger violations. The optimizer never returns an infeasible configuration as its final result; the penalty only guides intermediate search decisions.

\vspace{1em}
\subsubsection{\textbf{Cost-Efficiency (Throughput$\times$Utilization) Objective}}

Plans that partition devices across operators may achieve high throughput but leave a fraction of devices idle between operators. For example, two plans on the same 32-GPU cluster may achieve similar throughput, but one keeps all devices busy while the other leaves 8 devices idle whenever generation is not active. To steer the optimizer toward plans that are both fast and resource-efficient, we define an optional efficiency-aware score. Let $t_\mathrm{busy}(d)$ denote the total time device $d$ spends executing vertices of $\hat{G}$, and define the \emph{mean device utilization} as
\[
\mathrm{Util}(C) = \frac{1}{|D|} \sum_{d \in D} \frac{t_\mathrm{busy}(d)}{\mathrm{MS}(\hat{G})}.
\]

\vspace{1em}
\begin{rpegdefinition}[Throughput$\times$Utilization Score] 
\label{ls:def:eff-score}
Given the mean device utilization of configuration $C$, we define the efficiency score (throughput$\times$utilization) as

\small
\begin{equation}
\mathrm{Score}_\mathrm{eff}(C) = \mathrm{Score}(C) \cdot
\begin{cases}
\mathrm{Util}(C), & \text{if } \mathrm{Score}(C) \ge (1 - \zeta) \cdot \mathrm{Score}^\star, \\[4pt]
\lambda_\mathrm{floor}, & \text{otherwise},
\end{cases}
\label{ls:eq:cost:eff}
\end{equation}
\normalsize
where $\mathrm{Score}^\star$ is the best raw throughput found so far during the search, $\zeta$ is a user-specified maximum throughput loss, and $\lambda_\mathrm{floor} = 10^{-12}$ penalizes plans that fall below the throughput floor.
\end{rpegdefinition}

\subsection{Search and Optimization}
\label{ls:ch:search}

The preceding sections defined a deterministic evaluation pipeline that compiles any configuration $C = (G^\circ, P)$ into a quantitative score. This section defines how the optimizer searches over the configuration space to maximize that score.

The search operates on two levels. An \emph{outer loop} enumerates structural variants of the RPEG (staleness budgets, actor-critic sharing), while an \emph{inner loop} optimizes the execution plan $P = (\sigma, \Pi, \mathrm{mb})$ for each fixed RPEG using simulated annealing with multi-vertex moves and elite-seeded warm restarts. The procedure is an anytime heuristic: the best feasible configuration found so far can be returned at any point, making it practical for users who simply set a time or evaluation budget.

\vspace{1em}
\subsubsection{\textbf{Optimization Problem}}

Throughout this section, we fix the workload $\mathcal W$, model specifications $\mathcal S$, hardware $\mathcal H$, and evaluation horizon $K_\mathrm{eval}$. The optimization variable is the configuration $C = (G^\circ, P)$.

\vspace{1em}
\begin{rpegdefinition}[Optimization Objective]
\label{ls:def:opt-objective}
Given the scoring function from Definition~\ref{ls:def:score} (or Definition~\ref{ls:def:eff-score}), the optimizer seeks
\begin{equation}
C^\star = \underset{C \in \Omega}{\mathrm{argmax}}\; \mathrm{Score}(C),
\qquad
\Omega = \{C : \textsc{Valid}(C)\}.
\label{ls:eq:opt:objective}
\end{equation}
\end{rpegdefinition}

\vspace{1em}
A configuration $C = (G^\circ, P)$ is \emph{valid} if:
\begin{enumerate}[label=\textbf{(V\arabic*)}]
    \item \textbf{Well-formedness.} $G^\circ$ is well-formed (\cref{ls:sec:rpeg:wellformed}).
    \item \textbf{Plan feasibility.} For every $v \in V^\circ$: $(\sigma(v), \Pi(v), \mathrm{mb}(v)) \in X_\mathrm{cand}(v)$.
    \item \textbf{Staleness feasibility.} $\tau \in Y_\mathrm{cand}(G^\circ)$ (\cref{ls:sec:search:intermodel-space}).
    \item \textbf{Memory feasibility.} The peak memory on every device fits within GPU capacity. The final returned configuration must satisfy this condition. During search, however, the hard constraint is relaxed via the soft penalty from Equation~\ref{ls:eq:cost:penalty}, which allows the optimizer to temporarily explore near the memory boundary and refine toward feasibility.
\end{enumerate}

\vspace{1em}
\subsubsection{\textbf{Search Space}}

The configuration space is a union of structure-specific plan spaces: sharing or fusion can change the invocation set and therefore the available plan assignments. It has two conceptual components: the \emph{structural space} over RPEGs $G^\circ$ and the \emph{plan space} over execution plans $P = (\sigma, \Pi, \mathrm{mb})$.

\vspace{1em}
\paragraph{\textbf{Plan-Level Search Space}} 
\label{sec:app:search:plan-space}

Using the nested 3D parallelization model, the plan for each vertex $v$ must satisfy several constraints to ensure that the logical parallel structure maps cleanly onto the physical device topology. Let $B$ be the batch size, $L_m$ the layer count of the model invoked at $v$, and $B_\mathrm{local} = BR / P_d(v)$ the per-replica sequence count.

\vspace{1em}
\textbf{Parallelization degrees.}

\begin{enumerate}[label=\textbf{(P\arabic*)}]
    \item \textbf{Tensor parallelism.} $P_t(v) \mid N_g:$ TP is restricted to intra-node placement, so the group size must divide the GPU count per node.
    \item \textbf{Pipeline parallelism.} $P_p(v) \mid L_m$ and $P_p(v)$ is $1$ or even: each stage receives an equal share of layers. Odd depths $>1$ cause asymmetric packing and are excluded.
    \item \textbf{Data parallelism.} $P_d(v) \mid B$ and $P_d(v) \le N_r \cdot N_n \cdot N_g$ (the total GPU count). Every replica processes the same number of samples. When the pipeline fits on a single node, intra-node DP is permitted.
\end{enumerate}

\vspace{1em}
\textbf{Microbatch count.}

\begin{enumerate}[label=\textbf{(P\arabic*)},resume]
    \item \textbf{No pipeline, no microbatching.} If $P_p = 1$, then $\mathrm{mb} = 1$.
    \item \textbf{Microbatch alignment.} If $P_p > 1$:
    \[
    \mathrm{mb}(v) \mid B_\mathrm{local},
    \qquad
    \mathrm{mb}(v) \ge P_p(v),
    \qquad
    \frac{B_\mathrm{local}}{\mathrm{mb}(v)} \ge 8.
    \]
    The first ensures equal-sized microbatches. The second provides enough microbatches to fill the pipeline. The third maintains a minimum per-device microbatch size for SM utilization.
\end{enumerate}

\vspace{1em}
\textbf{Device packing.}
Given $(P_d, P_p, P_t)$, each node can host $s = \min(P_p, \lfloor N_g / P_t \rfloor)$ pipeline stages (limited by either the pipeline depth or the available GPUs after TP). A single data-parallel replica spans $n_\mathrm{rep} = P_p / s$ nodes (divisibility guaranteed by P8 below), so the operator's total allocation requires $n_\mathrm{need} = P_d \cdot n_\mathrm{rep}$ nodes.

\vspace{1em}
\begin{enumerate}[label=\textbf{(P\arabic*)},resume]
    \item \textbf{Feasible layout.} Either $n_\mathrm{need} \le N_n$ (all required nodes fit within a single rack) or $N_n \mid n_\mathrm{need}$ with $n_\mathrm{need} / N_n \le N_r$ (the allocation spans one or more complete racks).
    \item \textbf{Aligned origin.} The allocation origin is a multiple of the extent in each dimension, so allocations tile the device space without overlap.
    \item \textbf{Exact packing.} $s \mid P_p$, so each allocated GPU hosts exactly one logical shard of the model: $|\sigma(v)| = P_d \cdot P_p \cdot P_t$.
\end{enumerate}

Combining all of the above, the per-vertex plan candidate set collects every admissible combination of placement, parallelization strategy, and microbatch count:

\vspace{1em}
\begin{rpegdefinition}[Plan Candidate Set]
\label{ls:def:plan-cand}
\[
X_\mathrm{cand}(v) = \{(\sigma, \Pi, \mathrm{mb}) \mid \text{P1--P8 hold}\}.
\]
Together, these constraints guarantee a unique deterministic mapping of logical shards to physical GPUs.
\end{rpegdefinition}

\vspace{1em}
\paragraph{\textbf{Structure-Level Search Space}}
\label{ls:sec:search:intermodel-space}

Plan-level modifications optimize intra-model parallelism and placement for a fixed RPEG. Inter-model modifications, in contrast, change the RPEG $G^\circ$ itself. The optimizer re-runs the plan-level search for each structural variant.

The base RPEG (e.g., PPO) is specified chosen by the user and fixed for a given optimization run. Within that graph, the optimizer searches over two structural dimensions: (i) whether actor and critic parameters are shared, and (ii) the staleness budgets $\tau$ that control asynchronous execution. A third structural transformation, response-length quantile splitting, is not represented as an explicit structural move but applied automatically during lowering whenever the optimizer places generation and inference operators on disjoint device allocations.

We represent inter-model choices by an \emph{inter-model configuration}
\[
y = (\mathit{ac},\tau),
\qquad
\mathit{ac} \in \{0,1\},
\qquad
\tau:M_{\mathrm{train}}\to\mathbb N,
\]
where $\mathit{ac} = 1$ indicates that actor--critic parameter sharing is enabled and $\tau$ is the staleness budget.

\textbf{Structural and inter-model constraints.}
We enforce the following constraints on $y$:

\begin{enumerate}[label=\textbf{(S\arabic*)}]
    \item \textbf{Applicability of AC-sharing.}
    The actor--critic shared variant ($\mathit{ac} = 1$) may only be selected if the base RPEG contains both an actor and a critic model. 

    \item \textbf{Staleness candidate set.}
    Since $\tau: M_\mathrm{train} \to \mathbb N$ is a function, the search enumerates its image componentwise: for each trainable model $m \in M_\mathrm{train}$,
    \[
    \tau(m) \in T_{\mathrm{cand}}(m),
    \]
    where $T_{\mathrm{cand}}(m)$ is a finite set of allowable budgets (typically $\{0,1\}$).
\end{enumerate}

\vspace{1em}
\begin{rpegdefinition}[Inter-Model Candidate Set]
The inter-model candidate set collects all admissible combinations of sharing flag and staleness budget:
\[
Y_{\mathrm{cand}}
=
\Bigl\{
(\mathit{ac},\tau)
\ \Bigm|\
\text{constraints (S1)--(S2) hold}
\Bigr\}.
\]
Selecting $(\mathit{ac},\tau) \in Y_{\mathrm{cand}}$ determines the structural RPEG variant and the staleness budget, which controls how publication edges are instantiated during unrolling (\cref{ls:ch:lowering}).    
\end{rpegdefinition}

\paragraph{\textbf{Evaluation Pipeline}}
\label{ls:sec:search:evaluate}

Having defined what the optimizer searches over, we now describe how it scores each candidate configuration. Since local search frequently proposes configurations that have already been evaluated (e.g., a candidate that was explored in an earlier restart), results are memoized in a hash-based cache. The cache key is a canonical hash $\textsc{Hash}(C)$ derived from the RPEG structure (vertices, edges, containers, staleness budget $\tau$) and the execution plan $P=(\sigma, \Pi, \mathrm{mb})$. The evaluation budget $E$ counts only cache \emph{misses} (full pipeline executions); cache hits are free.

\begin{algorithm}[t]
\caption{\textsc{EvaluateCached}: configuration scoring}
\label{ls:alg:search:evaluate}
\KwIn{Configuration $C=(G^\circ,P)$, evaluation horizon $K_\mathrm{eval}$, cache.}
\KwOut{Penalized score $\mathrm{Score}_\mathrm{pen}(C)$.}
\BlankLine
$h\leftarrow \textsc{Hash}(C)$\;

\If{$h\in\mathrm{Cache}$}{
    \Return{$\mathrm{Cache}[h]$}\;
}

$\bar{G}\leftarrow \textsc{Unroll}(G^\circ,\, K_{\mathrm{eval}})$ \tcp*{Alg.~\ref{ls:alg:unroll}}

$\hat{G}\leftarrow \textsc{Lower}(\bar{G},\,P)$ \tcp*{Alg.~\ref{ls:alg:lower}}

$(s,f)\leftarrow \textsc{GreedySchedule}(\hat{G})$ \tcp*{Alg.~\ref{ls:alg:schedule}}

$\mathrm{Score}(C) \leftarrow K_\mathrm{eval} \cdot W_{\mathrm{step}} \;/\; \mathrm{MS}(\hat{G})$\;

Apply memory penalty (Equation~\ref{ls:eq:cost:penalty}) to obtain $\mathrm{Score}_\mathrm{pen}(C)$\;

$\mathrm{Cache}[h]\leftarrow \mathrm{Score}_\mathrm{pen}(C)$\;

\Return{$\mathrm{Score}_\mathrm{pen}(C)$}\;
\end{algorithm}

The algorithm first computes the raw score via the standard pipeline (unroll, lower, schedule), then applies the soft memory penalty from Equation~\ref{ls:eq:cost:penalty}.

\textbf{Runtime.}
On a cache miss, \textsc{EvaluateCached} executes the full pipeline. The dominant step is scheduling (\cref{ls:sec:cost:scheduling}), so the total cost per evaluation is
\[
O\bigl(K_\mathrm{eval} \cdot |V^\circ| \log(K_\mathrm{eval} \cdot |V^\circ|) + K_\mathrm{eval} \cdot |E^\circ|\bigr).
\]
Cache hits are $O(1)$ (hash lookup).

\vspace{1em}
\paragraph{\textbf{Optimization Strategy}}
\label{ls:sec:search:strategy}

The inter-model candidate set $Y_\mathrm{cand}$ is small and can be enumerated exhaustively. The combinatorial difficulty lies in the plan space: each vertex $v$ admits $|X_\mathrm{cand}(v)|$ valid choices, and the joint plan space grows as $\prod_{v \in V^\circ} |X_\mathrm{cand}(v)|$, which is exponential in the number of RPEG vertices. Similar to ReaL~\cite{mei2024real}, we associate each configuration with a cost defined by its (penalized) score and formulate the search as sampling from an energy-based distribution over configurations. We augment this approach with multi-vertex proposals, best-of-$K_x$ candidate selection, and elite-seeded warm restarts. We therefore adopt a two-level strategy:

\begin{enumerate}
    \item \textbf{Outer loop (structural enumeration).}
    Enumerate all $y \in Y_\mathrm{cand}$. For each structural variant, run an independent plan-level search and retain the best feasible configuration found.

    \item \textbf{Inner loop (plan search).}
    For a fixed RPEG $G^\circ$, optimize the execution plan $P$ via multi-start simulated annealing (SA). Each iteration selects one or two RPEG vertices, proposes alternative plan assignments, and accepts or rejects the best proposal according to the Metropolis criterion. Multiple SA chains (restarts) are run per structural variant, seeded from an elite set of previously discovered plans.
\end{enumerate}

The overall procedure is \emph{anytime}: it maintains the best feasible configuration across all structural variants and can return it whenever a wall-clock time limit or evaluation budget $E_\mathrm{max}$ is reached.

\vspace{1em}
\paragraph{\textbf{Score-Normalized Acceptance.}}
\label{ls:sec:search:formulation}

We use a stochastic acceptance rule to balance exploration
and exploitation. Given the current configuration $C$
with penalized score $S>0$ and a proposed configuration
$C'$ with score $S'>0$, define the relative score change
\begin{equation}
\Delta(C,C')=\frac{S'-S}{S}.
\end{equation}
The proposal is accepted with probability
\begin{equation}
\label{ls:eq:sa:accept}
p_{\mathrm{acc}}(C\to C')
=
\min\left\{1,
\exp\left(\frac{\Delta(C,C')}{t}\right)
\right\}.
\end{equation}
The temperature $t>0$ controls exploration: improving
proposals are always accepted, while worsening proposals
may also be accepted. Normalization by $S$ makes the
acceptance rule largely independent of the absolute score
scale. This is a stochastic search heuristic rather than
exact sampling from a prescribed target distribution.

Since $\mathrm{Score}(C)$ is a product of a positive throughput and a positive penalty factor, $S > 0$ always holds. The normalization by $|S|$ makes the acceptance threshold scale-invariant: the same temperature $t$ produces comparable acceptance rates regardless of the absolute score magnitude. This differs from the standard Metropolis ratio $\exp((S'-S)/t)$, which would require re-tuning $t$ for every workload.

Improving proposals are always accepted. Crucially, worsening proposals are accepted with non-zero probability, which allows the chain to escape local optima by temporarily traversing lower-scoring regions of the landscape. This is the central advantage over greedy search: a purely greedy optimizer can become permanently trapped in a suboptimal basin, whereas the stochastic acceptance mechanism provides a principled way to trade short-term score regression for long-term exploration. As detailed in \cref{ls:sec:search:temperature}, $t$ is held constant throughout each chain rather than decreased as in classical SA. The exploration--exploitation tradeoff is instead controlled through the multi-start mechanism described in \cref{ls:sec:search:restarts}.

\vspace{1em}
\paragraph{\textbf{Move Proposals}}
\label{ls:sec:search:moves}

Evaluating a full configuration through the pipeline is expensive relative to enumerating valid strategy tuples. To avoid wasting evaluations on invalid or obviously poor candidates, the optimizer pre-computes a \emph{candidate directory} for each RPEG vertex before the SA search begins. The directory contains all valid strategy tuples satisfying constraints P1--P8, filtered by per-vertex memory feasibility (strategies that exceed GPU capacity even in isolation are excluded; the soft penalty from Equation~\ref{ls:eq:cost:penalty} handles joint infeasibility across co-resident operators during search) and sorted by estimated execution time. Candidates are sampled uniformly from the full directory.

Each SA iteration proposes up to $K_x=6$ neighbor configurations by modifying the plan of one or two RPEG vertices while keeping $G^\circ$ fixed. The proposal mechanism uses two move kernels.

\vspace{1em}
\textbf{Single-vertex kernel.}
A target vertex is selected with a bias toward scheduling bottlenecks: with probability $0.6$, the vertex is sampled proportionally to its execution time on the critical path (the longest path through the current schedule). Otherwise it is chosen uniformly at random. For the selected vertex, $K_x$ candidate strategies are sampled uniformly from the directory and each is paired with a randomly chosen aligned device origin. 

\vspace{1em}
\textbf{Donor-receiver pair kernel.}
Single-vertex moves cannot redistribute devices between operators in a single step: shrinking one vertex's allocation without simultaneously expanding another leaves devices idle. To address this, a \emph{donor-receiver} kernel proposes coordinated two-vertex moves. A donor is selected with probability proportional to its device count, and a receiver is chosen among vertices whose allocation overlaps with the donor's. The donor is shrunk to a smaller contiguous region, and the receiver is re-sampled to potentially expand into the freed devices. For example, if generation occupies two full racks while inference is constrained to a single node, a pair move can transfer one rack from generation to inference in one step. No sequence of single-vertex moves can reach this configuration without passing through intermediate states where devices are left unassigned. The pair kernel is selected with probability $p_\mathrm{pair} = 0.28$.

\vspace{1em}
\textbf{Best-of-$K_x$ selection.}
All $K_x$ proposals are evaluated via \textsc{EvaluateCached} (\cref{ls:sec:search:evaluate}), with cache misses dispatched to a thread pool for parallel evaluation. Only the highest-scoring proposal $C_\mathrm{best}$ is subjected to the acceptance step (Equation~\ref{ls:eq:sa:accept}). This best-of-$K_x$ strategy amortizes evaluation cost across multiple local directions before committing to a single chain update.

\vspace{1em}
\paragraph{\textbf{Temperature and Chain Termination}}
\label{ls:sec:search:temperature}

Unlike classical simulated annealing, which decreases $t$ over the course of a single run, our optimizer holds $t$ \emph{constant} within each chain. The first chain runs at $t = t_\mathrm{init}$; warm-restarted chains run at $t = t_\mathrm{warm} > t_\mathrm{init}$, which allows broader exploration around the elite seed. The higher restart temperature is safe because the chain starts from a known-good plan rather than an arbitrary initial point. The exploration--exploitation tradeoff is therefore managed across chains (via the two temperature levels and the multi-start mechanism in \cref{ls:sec:search:restarts}) rather than within a single chain. The implementation also supports geometric cooling ($t \leftarrow \max\{t_\mathrm{min}, \gamma \cdot t\}$) as an alternative; our ablation finds that it degrades throughput by $1$--$5\%$ compared to constant temperature.  

Chains are not run for a fixed number of iterations. Instead, the optimizer monitors consecutive rejections: if $w_\mathrm{stag}=10$ proposals are rejected in a row, the chain is terminated and a new restart begins. This ensures that productive chains continue while stuck chains are recycled quickly.

\vspace{1em}
\paragraph{\textbf{Restarts and Elite Seeding}}
\label{ls:sec:search:restarts}

Random restarts are a standard technique to escape local optima in SA, but they waste budget rediscovering plans that earlier chains already found. We instead use \emph{warm restarts}: the first chain ($r = 0$) is seeded from a \emph{concurrent baseline heuristic} that groups operators into concurrent groups -- sets of operators with no data dependencies that can run in parallel -- and partitions the device space so that each operator within a group receives a disjoint allocation, selecting the partition that minimizes the estimated step time. All subsequent chains begin from the best feasible plan discovered so far (the elite seed) and run at $t_\mathrm{warm} > t_\mathrm{init}$, allowing broader exploration around the known-good starting point. The optimizer maintains an elite set of the top-$K_\mathrm{elite} = 3$ feasible plans per vertex set, shared across staleness variants that share the same vertices (since a good plan under one $\tau$ is a structurally sound seed for another).

\vspace{1em}
\paragraph{\textbf{Outer Loop and Budget}}
\label{ls:sec:search:outer}

The outer loop iterates over all inter-model candidates $y = (\mathit{ac}, \tau) \in Y_\mathrm{cand}$. Termination is governed by either a wall-clock time limit or an evaluation budget $E_\mathrm{max}$ (cache misses), whichever is reached first. In time-limited mode, the remaining wall-clock time is split evenly among the remaining structural variants; in budget-limited mode, each variant receives a fixed budget $E_\mathrm{struct}$. The candidate directory (\cref{ls:sec:search:moves}) is rebuilt only when the vertex set changes (e.g., when transitioning from the base RPEG to the AC-shared variant); staleness variants that share the same vertices reuse the same directory.

The best feasible configuration $C^\star$ is maintained across all structural variants. In practice, a 10-minute time budget suffices for workloads up to 1024 GPUs, producing $75$K--$220$K evaluations depending on cluster and model size. After the SA search completes, a deterministic \emph{polish phase} (\cref{ls:sec:search:polish}) performs greedy coordinate descent to extract any remaining local improvements. Algorithm~\ref{ls:alg:search:optimize} gives the complete procedure.

\begin{algorithm}[t]
\caption{\textsc{Optimize}: multi-start SA with structural enumeration}
\label{ls:alg:search:optimize}
\KwIn{Baseline $C_0$, candidates $Y_{\mathrm{cand}}$, time limit or budgets
$E_\mathrm{max}$/$E_\mathrm{struct}$, temperature $t$, stagnation window $w_\mathrm{stag}$, proposal count $K_x$.}
\KwOut{Best feasible configuration $C^\star$.}
\BlankLine

\ForEach{$(\mathit{ac}, \tau) \in Y_{\mathrm{cand}}$}{
  \For{each restart $r$ until budget exhausted}{
    Seed $C$ from baseline ($r{=}0$) or elite plan ($r{\ge}1$)\;

    \Repeat{$w_\mathrm{stag}$ consecutive rejections or budget exhausted}{
      Propose $K_x$ neighbors (single-vertex or pair kernel) \tcp*{\cref{ls:sec:search:moves}}

      Evaluate proposals via \textsc{EvaluateCached} \tcp*{Alg.~\ref{ls:alg:search:evaluate}}

      $C_\mathrm{best} \gets \mathrm{argmax}$ among proposals\;

      Accept $C_\mathrm{best}$ with prob.\ $p_\mathrm{acc}(C \to C_\mathrm{best})$ \tcp*{Eq.~\ref{ls:eq:sa:accept}}

      Update elite set and global best $C^\star$\;
    }
  }
}
$C^\star \gets \textsc{Polish}(C^\star)$ \tcc*{\cref{ls:sec:search:polish}}
\Return{$C^\star$}\;
\end{algorithm}

\vspace{1em}
\paragraph{\textbf{Polish Phase}}
\label{ls:sec:search:polish}

The SA search is stochastic: even when the chain has converged near a good basin, it may not have evaluated every neighbor of the final plan. The \emph{polish phase} addresses this by performing deterministic greedy coordinate descent~\cite{wright2015coordinate} on the best plan returned by \textsc{Optimize}. It visits each RPEG vertex in a fixed order and, for each vertex, evaluates all feasible strategies in the candidate directory. If any strategy strictly improves the score, the best one is applied before moving to the next vertex. This process repeats in passes over all vertices until a full pass yields no improvement.

Two properties make the polish phase efficient. First, isolated local operator costs can be reused across identical strategy/extent choices on homogeneous hardware. Second, many candidates will already reside in the evaluation cache from the preceding SA search, so the polish phase primarily performs hash lookups rather than full pipeline evaluations. Cache misses are dispatched to the thread pool in parallel, as in the SA inner loop.

The polish phase is guaranteed to terminate (the score increases strictly at each accepted move, and the candidate set is finite) and is deterministic (no random choices). It cannot discover plans that require coordinated multi-vertex changes, but it reliably extracts any single-vertex improvements that the stochastic search may have missed.

%% file: appendix-eval.tex
\clearpage
\section{Evaluation: Supplementary Details}
\label{rpegx:appendix-evaluation}
\label{app:eval}

This appendix specifies different evaluation configurations.

\textbf{Platforms.}
GH200 execution and hardware calibration use Clariden at CSCS,
with four NVIDIA GH200 superchips per node (each with a 72-core Grace
CPU and 120\,GB HBM3), intra-node NVLink connectivity
($\approx$160\,GB/s per GPU pair), and four HPE Slingshot-11
interfaces per node. The runtime uses persistent PyTorch/Megatron
workers launched through Slurm, NCCL over OFI/libfabric/CXI for
GPU collectives and weight transfers, and Gloo for semantic data.
The A100 experiments use Athena at AGH Cyfronet, with dual 64-core
AMD EPYC 7742X CPUs, 1\,TB RAM, and eight NVIDIA A100 40-GB GPUs
per node, interconnected via NVLink/NVSwitch and four 200-Gb/s HDR
InfiniBand interfaces.
The A100 comparison with ReaL uses a ReaL-derived runtime
(Python 3.10, CUDA 12.1; one worker per GPU) and matched
synchronous PPO workloads, without response-length splitting
or bounded asynchrony.
Analytical optimizer searches run on Piora CPU nodes at CSCS,
each with dual 64-core AMD EPYC 7773X processors and 1\,TB RAM,
using six CPU cores per concurrent search task.

\textbf{Workload hyperparameters.}
We use one PPO epoch, actor/critic learning rates $10^{-6}$, zero weight
decay, gradient clipping at 1.0, PPO clip 0.2, $\gamma=1$,
GAE $\lambda=0.95$, and KL coefficient 0.01. 

\textbf{Analytical model validation.}
The 123 configurations comprise 34 generation, 51 inference, and
38 training cases. These cases span 3B, 7B, and 13B models, 1--128 GPUs,
batch sizes of 32--2048, and 25 DP/PP/TP strategies. Measurements use BF16, FlashAttention-2,
$L_p=512$, $L_r=128$, and one rollout per prompt; training uses
activation checkpointing and ZeRO-2 for $P_d>1$. The memory
comparison accounts for checkpointing and ZeRO-2 profiling conventions;
$P_d=1$ traces recorded before Adam-state materialization are excluded.

\textbf{Predicted-schedule configurations.}
Schedules unroll $K_{\mathrm{eval}}=8$ steps with one-step bounded
staleness and three response-length phases: completion fractions
$(0.50,0.85,1.0)$ at lengths $(L_r/4,L_r/2,L_r)$.
The throughput--utilization objective uses $\delta=0.10$.
Three scaling sweeps cover actor-only, critic/reward, and joint
model growth. The first two use 8/16/32/64 GPUs: the scaled models
grow as 7/13/34/70B, while unscaled models remain at 7B. The joint
sweep uses 16/32/64/128 GPUs, with actor and critic sizes
7/13/34/70B and reward sizes 7/7/13/13B. Each configuration is
evaluated at $(B,L_r)\in\{(512,128),(256,384),(128,896)\}$,
yielding 36 analytical cases. The strong-scaling study fixes
13B models and the workload specified in the main evaluation.

\textbf{Optimizer stability.}
Table~\ref{ls:tab:stability-setup} specifies five modeled configurations
used in the 200-seed stability study. Each CPU-side analytical search
runs for 600 seconds; its throughput is the number of evaluated plans,
not the number of physical training runs.
A separate convergence study evaluates 7B/256,
13B/512, 34B/1024, and 70B/2048 model/GPU configurations
over 3,600 seconds, tracking the best predicted makespan
relative to the initial plan.

\begin{table}[h]
\centering
\scriptsize
\setlength{\tabcolsep}{2pt}
\renewcommand{\arraystretch}{1.12}
\caption{Optimizer-stability workloads. A/C/R denotes actor/critic/reward
model size (billions of parameters); topology denotes racks $\times$
nodes/rack $\times$ GPUs/node.}
\label{ls:tab:stability-setup}
\begin{tabular}{@{}rlcrcr@{}}
\toprule
\textbf{GPUs} & \textbf{Topology} & \textbf{A/C/R (B)} &
\textbf{$B$} & \textbf{$L_p/L_r$} & \textbf{Evals/600 s} \\
\midrule
64   & $2{\times}8{\times}4$   & 13/13/7  & 1024 & 128/384 & $\sim$220K \\
128  & $2{\times}16{\times}4$  & 34/13/13 & 1024 & 128/384 & $\sim$200K \\
256  & $4{\times}16{\times}4$  & 70/34/13 & 1024 & 128/384 & $\sim$160K \\
512  & $8{\times}16{\times}4$  & 70/70/34 & 2048 & 128/384 & $\sim$105K \\
1024 & $16{\times}16{\times}4$ & 175/70/34 & 2048 & 128/384 & $\sim$75K \\
\bottomrule
\end{tabular}
\end{table}

\textbf{Ablations.}
Each experiment disables one mechanism, with five seeds per
scenario. The 50\,GB HBM limit in \textsc{Mem-tight} is imposed
analytically to evaluate memory-constrained planning. Table~\ref{ls:tab:ablation-setup} identifies the four
workloads; results are reported in Table~\ref{tab:ablation}.

\begin{table}[h]
\centering
\scriptsize
\setlength{\tabcolsep}{2pt}
\renewcommand{\arraystretch}{1.12}
\caption{Analytical ablation workloads. A/C/R denotes actor/critic/reward
model size (billions of parameters); topology denotes racks $\times$
nodes/rack $\times$ GPUs/node.}
\label{ls:tab:ablation-setup}
\resizebox{\columnwidth}{!}{%
\begin{tabular}{@{}lrclcl@{}}
\toprule
\textbf{Scenario} & \textbf{GPUs} & \textbf{Topology} &
\textbf{HBM/GPU} & \textbf{A/C/R (B)} & \textbf{$B;L_p/L_r$} \\
\midrule
\textsc{Mem-tight} & 16  & $1{\times}4{\times}4$   & 50\,GB  & 13/13/13 & 128;128/896 \\
\textsc{Wide}      & 128 & $1{\times}32{\times}4$  & 102\,GB & 7/7/7    & 1024;128/384 \\
\textsc{Hetero}    & 32  & $1{\times}8{\times}4$   & 102\,GB & 34/7/7   & 512;128/128 \\
\textsc{Deep}      & 64  & $2{\times}8{\times}4$   & 102\,GB & 7/70/70  & 512;128/128 \\
\bottomrule
\end{tabular}%
}
\end{table}

A1 disables response-length splitting; A2 disables bounded
asynchrony; A3 fixes microbatch counts to $P_p$ for generation/inference
and $2P_p$ for training; A4 removes coordinated donor--receiver moves;
A5 disables warm restarts; A6 uses uniform rather than critical-path
invocation selection; A7 removes final polishing; A8 replaces soft
memory penalties with hard rejection; and A9 uses geometric cooling
instead of constant-temperature chains.

%% file: appendix-model.tex
\clearpage
\section{Analytical Performance and Memory Model: Additional Details}
\label{rpegx:appendix-model}

This appendix complements Section~\ref{sec:model} with the assumptions,
calibration procedure, and implementation choices behind the analytical
performance model. We follow the same order as the main paper: model
configuration and communication, intra-model execution time, inter-model
communication, and device-level memory feasibility. The role-specific work--depth--memory foundations are given in
\cite{besta2026performance}. The RPEG-specific execution-time,
communication, and physical device-memory equations are derived in
Appendix~\ref{app:ad-derivations}; we do not duplicate them here.
The analytical model {assigns node times} \(\mathcal T(x)\) {and
checks memory feasibility} \(\mathrm{MemFeasible}(C)\) {to rank
execution plans, not to prescribe physical start times.}

\subsection{Model Configuration and Calibration}
\label{ls:ch:cost-modeling}

\subsubsection{Inputs and Modeling Assumptions}
\label{ls:sec:cost:hw}

The analytical model receives three groups of inputs. The workload
\(\mathcal W=(B,R,L_p,L_r)\) specifies input prompts per step, rollouts per prompt,
average prompt length, and average rollout length. Each invoked model
\(m\in\mathcal M\) supplies its transformer architecture, model role,
and working scalar precision \(s_b\). The hardware description supplies
the hierarchical device topology, GPU memory capacity, and calibrated
compute and communication parameters. The candidate execution plan \(P=(\sigma,\Pi,mb)\) assigns
each invocation \(v\) its GPU allocation \(\sigma(v)\), DP/PP/TP degrees
\(\Pi(v)=(P_d,P_p,P_t)\), and microbatch count \(mb(v)\).

We use mixed-precision BF16 computation, FP32 Adam master weights and moment
arrays, activation checkpointing, and ZeRO-2 sharding for optimizer state
and reduced gradients. The estimator targets GPU-resident training rather
than CPU offloading. These choices determine both the HBM accounting and
the communication primitives costed within a training invocation.
Parameter formulas retain the dominant attention projections, FFN weights,
embeddings, and task heads; small bias and normalization parameters are
omitted in the baseline.

The nominal hardware topology is given as
\(D=[N_r]\times[N_n]\times[N_g]\).
Device placements are mapped to physical nodes and racks, rather than
assuming a flat, uniform interconnect. The time model distinguishes three
communication tiers: intra-node, inter-node within a rack, and inter-rack.
When a transfer spans more than one tier, it uses the slowest applicable
tier as a conservative proxy. For tensor parallelism, the legal plan
space constrains each TP group to one node.

These assumptions are specific to the evaluated modeling implementation;
they are not architectural requirements of RPEG. In particular,
architectures with grouped-query attention, gated FFNs, vocabulary-sharded
heads, or nonstandard pipeline partitions require architecture- and
layout-specific dimensions in the cited work terms and our physical
residency estimator (Appendix~\ref{app:wshard}).

\subsubsection{Automated Hardware Calibration}
\label{ls:sec:cost:calibration}

Rather than substituting advertised GPU FLOP/s or network line rates,
RPEG measures effective parameters with an automated suite on the target
system. Its calibration campaign, typically on the order of minutes,
comprises the following measurements: NCCL AllReduce message-size sweeps;
NCCL send/receive sweeps; actual transformer-forward timing; contiguous
HBM transfers; GEMM-shape sweeps; and non-GEMM layer timings at a reference
batch size. The resulting configuration is stored in a hardware-parameter
file and supplied to the optimizer together with the workload, model
specifications, and available devices.

The measured quantities have distinct meanings. Transformer-forward
benchmarks provide an effective compute ceiling \(F_{\mathrm{peak}}\),
while memory-copy tests provide an effective \(\mathrm{BW}_{\mathrm{mem}}\).
Shape sweeps fit the saturation constants \(\gamma_M\) and \(\gamma_K\);
non-GEMM measurements fit \(c_{\mathrm{ng}}\) at the reference batch size
\(B_{\mathrm{cal}}\). Communication measurements fit separate
\((\alpha,\beta)\) pairs for collective and P2P transfers at each
hardware tier; \(\alpha\) is latency and \(\beta\) is inverse effective
bandwidth, not bandwidth itself.

The AllReduce benchmark calibrates a collective-family proxy used by the
ring-style expressions for AllReduce, ReduceScatter, and AllGather.
NCCL may select different algorithms as message sizes,
rank counts, and network topology change, so a fitted \((\alpha,\beta)\)
pair should be understood as an empirical approximation over the measured
range. Similarly, P2P and collective parameters are not interchangeable:
a high aggregate collective bandwidth does not imply equivalent point-to-
point transfer performance.

Note that the model
can evaluate new plans cheaply, but its parameters need not transfer
accurately across different kernel implementations, model architectures,
message-size regimes, or cold versus warmed process states.

\subsubsection{Hierarchical Communication Model}
\label{ls:sec:cost:comm-model}

Communication appears in two places: {the intra-operator term}
\(T_{\mathrm{comm}}(v)\) {for} TP, PP, and DP transfers embedded in
\textsc{Op} nodes, and {the explicit transfer costs}
\(T_{\mathrm{xfer}}\) {and} \(T_{\mathrm{reshard}}\) {for}
\textsc{Comm}/\textsc{Reshard} nodes inserted by lowering. Both use the \(\alpha\)--\(\beta\) model and the primitive
estimates in Table~\ref{tab:cost-model-collectives} of the main paper.
For those estimates, \(M\) is the per-GPU input buffer for AllReduce and
ReduceScatter, the final gathered buffer for AllGather, the aggregate
per-GPU exchange size for AlltoAll, and the transferred P2P payload.
Collective costs vanish when the participant count is one.

The table uses ring-based reduction/gather expressions and a pairwise-
exchange upper-bound estimate for AlltoAll. The all-to-all formula counts
\(M\beta\) as an aggregate bandwidth term, including any locally retained
share; a perfectly balanced pairwise exchange instead sends
\((p-1)M/p\) remote bytes when \(M\) includes the self-share. This and
tier selection make the model a cost proxy rather than a packet-exact
simulation of the network. Per-message control overhead, dynamic
communicator setup, and transport-specific tensor packing are outside
this simple expression.

\subsection{Intra-Model Runtime Details}
\label{ls:sec:cost:node}

\subsubsection{Operator Work and Effective Throughput}
\label{ls:sec:cost:op}
\label{ls:sec:cost:gemm}

For every \textsc{Op} node \(v\), the {five-term operator-time model}
\(T_{\mathrm{op}}(v)\) in Section~\ref{sec:model-intra-model}
separates GEMM compute \(T_{\mathrm{gemm}}(v)\), pipeline bubbles
\(T_{\mathrm{bubble}}(v)\), intra-operator communication
\(T_{\mathrm{comm}}(v)\), non-GEMM overhead
\(T_{\mathrm{non\mbox{-}gemm}}(v)\), and Adam update traffic
\(T_{\mathrm{opt}}(v)\). The role-specific FLOP expressions for generation, teacher-forced
inference, and training are adopted from
\cite{besta2026performance}.
They distinguish policy, reference, value, reward, and shared-model heads.

{The sum defining} \(T_{\mathrm{op}}(v)\) {is an additive prediction.}
Actual computation, collectives, and pipeline transfers can partially
overlap, so {it} is not an exact critical-path reconstruction of the
kernels executed inside a model invocation.

The {roofline-limited GEMM throughput} \(F_{\mathrm{eff}}(v)\)
{depends on the GEMM-shape efficiency} \(\eta(v)\) {and
arithmetic intensity} \(I(v)\){, not just nominal work} \(W(v)\). In particular, increasing \(P_d\) reduces the local batch,
while increasing \(P_t\) narrows local projection matrices. Either can
lower achieved per-GPU GEMM throughput enough to offset the benefit of
additional devices.

{Arithmetic intensity} \(I(v)=W_{\mathrm{dev}}(v)/(\Omega_W+\Omega_A)\)
{uses estimated HBM bytes read/written per microbatch}
\(\Omega_W+\Omega_A\){, whereas} \(\mathrm{peak}(d)\)
{measures simultaneous GPU-memory occupancy: a resident tensor
can be accessed repeatedly. For training,} \(I(v)\) {uses
forward-equivalent FLOPs and byte traffic for forward, recomputation,
and backward.}
{Generation uses separate effective throughputs}
\(F_{\mathrm{eff}}^{\mathrm{prefill}}(v)\) {and}
\(F_{\mathrm{eff}}^{\mathrm{decode}}(v)\) {in}
\(T_{\mathrm{gemm}}^{\mathrm{gen}}(v)\) {because the two phases
have different memory-access patterns.} In particular, decode repeatedly
loads or accesses parameters and historical KV state while producing
relatively little new arithmetic work. {The evaluated estimator
uses representative lengths} \(L_p,L_r\) {rather than tracking
time-varying active-sequence counts or per-rank KV-head layouts during
decoding.}

\subsubsection{Pipeline Bubbles and Intra-Operator Transfers}
\label{ls:sec:cost:bubble}
\label{ls:sec:cost:comm}

{The 1F1B bubble multiplier}
\((mb(v)+P_p(v)-1)/mb(v)\) {applied to}
\(T_{\mathrm{gemm}}(v)\) models fill and drain under balanced
execution. It decreases as the microbatch count grows but ignores
unbalanced stage durations and communication contention. The search
therefore optimizes {microbatch count} \(mb(v)\) {jointly
with the parallel degrees} \(\Pi(v)=(P_d,P_p,P_t)\)
{and placement} \(\sigma(v)\){, rather than fixing it
solely from} \(P_p(v)\). A larger number of microbatches
can reduce bubbles while increasing latency-sensitive collective calls.

Tensor-parallel AllReduce {contributes to} \(T_{\mathrm{tp}}(v)\)
{by aggregating partial layer results; TP groups are restricted
to one node because the repeated per-layer reductions favor low-latency,
high-bandwidth local links.} Pipeline P2P operations
move activations between consecutive stages {and contribute to}
\(T_{\mathrm{pp}}(v)\) according to the boundaries crossed by the
selected placement. Data-parallel ReduceScatter aggregates and partitions
gradients for ZeRO-2, while the associated AllGather restores the updated
parameter view across replicas; {both contribute to}
\(T_{\mathrm{dp}}(v)\). The latter is a parameter-publication cost and need not be
adjacent to gradient reduction in the actual training schedule.

Table~\ref{ls:tab:comm-summary} records the primitive counts used by
the simplified estimator. {TP assumes two AllReduces per transformer
layer and forward pass (attention and FFN), or three forward-equivalent
passes for training. PP counts approximate critical-path fill/drain steps:}
\(mb+P_p-2=(mb-1)+(P_p-1)\) {for inference and}
\(mb+2P_p-3=(mb-1)+2(P_p-1)\) {for 1F1B training.
ZeRO-2 contributes one gradient ReduceScatter and one updated-weight
AllGather per update (Appendix~\ref{app:intra-op-comm}).}
Generation TP is split into prompt prefill and decode; the decode count
assumes that the full local DP batch is processed together. A decoder that independently pipelines distinct
microbatches needs the corresponding alternate count, as explained in
Appendix~\ref{app:intra-op-comm}.

\begin{table*}[t]
\centering
\scriptsize
\setlength{\tabcolsep}{4pt}
\renewcommand{\arraystretch}{1.13}
\caption{Simplified intra-operator communication costs per GPU.
Here \(\ell=L_p+L_r\) for teacher-forced inference/training; prefill has
\(\ell=L_p\). PP counts approximate critical-path steps.
The generation decode row assumes one decode step over the full local
DP batch. All payloads are per participating GPU.}
\label{ls:tab:comm-summary}
\begin{tabularx}{\textwidth}{@{}l l l Y Y@{}}
\toprule
\textbf{Role} & \textbf{Parallelism} & \textbf{Primitive} &
\textbf{Count} & \textbf{Message bytes} \\
\midrule
Generation prefill & TP & AllReduce & \(2(L_m/P_p)\,mb\) &
\(BR L_p d_m s_b/(P_d mb)\) \\
Generation decode & TP & AllReduce & \(2(L_m/P_p)(L_r-1)\) &
\(BR d_m s_b/P_d\) \\
Inference & TP & AllReduce & \(2(L_m/P_p)\,mb\) &
\(BR\ell d_m s_b/(P_d mb)\) \\
Training & TP & AllReduce & \(6(L_m/P_p)\,mb\) &
\(BR\ell d_m s_b/(P_d mb)\) \\
Inference & PP & P2P & \(mb+P_p-2\) &
\(BR\ell d_m s_b/(P_dP_tmb)\) \\
Training & PP & P2P & \(mb+2P_p-3\) &
\(BR\ell d_m s_b/(P_dP_tmb)\) \\
Training & DP & ReduceScatter + AllGather & One each &
\(N_m s_b/(P_pP_t)\) \\
\bottomrule
\end{tabularx}
\end{table*}

The pipeline-transfer byte count \(M_{\mathrm{pp}}(\ell)\) presumes
TP-sharded stage-boundary activations; frameworks that send replicated hidden states need a
different transfer volume. All PP costs vanish for \(P_p=1\), and
TP/DP collectives vanish for a single participating TP/DP rank.

\subsubsection{Non-GEMM Kernels and Optimizer Updates}
\label{ls:sec:cost:nongemm}
\label{ls:sec:cost:opt}

The GEMM work formulas {define} \(W(v)\) {using dominant
matrix multiplications and exclude} normalization, residual operations,
nonlinearities, positional operations, and other non-GEMM kernels{,
whose residual latency is represented by}
\(T_{\mathrm{non\mbox{-}gemm}}(v)\) {rather than added to the
GEMM FLOP count.}
At \(P_t=1\), {the full-transformer-calibrated compute ceiling}
\(F_{\mathrm{peak}}\) {already absorbs part of this overhead.}
The separate calibrated non-GEMM term
\(T_{\mathrm{non\mbox{-}gemm}}(v)\) therefore adds only a residual
correction as tensor parallelism \(P_t\) increases.
Its batch scaling and four-pass-equivalent training multiplier are
heuristics; launch latency or fused-kernel behavior can depend on
microbatch count and sequence length in ways not captured by one
constant. The baseline sets \(T_{\mathrm{non\mbox{-}gemm}}^{\mathrm{gen}}(v)=0\),
which is an approximation rather than an assertion that prefill
and decode contain no such operations.

{The bandwidth-bound Adam estimate is}
\(T_{\mathrm{opt}}(v)=18N_m/(P_dP_pP_t\,\mathrm{BW}_{\mathrm{mem}})\).
{Here} \(N_m/(P_dP_pP_t)\) {approximates the
optimizer-owned parameters per GPU, and 18 bytes/parameter is an
effective coefficient, not the physical read/write total: an uncached
update requires 24 bytes for FP32 master/moment reads and writes, 2 for
the BF16 gradient read, and 2 for the BF16 weight write (28 bytes per
parameter). Replicated boundary embeddings or heads can make the
worst-stage count larger (Appendix~\ref{app:optimizer-step}).}

\subsection{Inter-Model Communication and Resharding Details}
\label{ls:sec:cost:comm-reshard}

A \textsc{Comm} node {has predicted time}
\(T_{\mathrm{xfer}}(u,v,a)\) {for container} \(a\)
{between allocations} \(\sigma(u),\sigma(v)\).
The estimator divides {its payload} \(|a|s(a)\)
{by the assumed number of concurrent P2P pairs}
\(\min(|\sigma(u)|,|\sigma(v)|)\) {and applies the
communication tier's} \((\alpha_{\mathrm{p2p}},\beta_{\mathrm{p2p}})\).
This assumes balanced transfers and ideal overlap among those pairs.

A \textsc{Reshard} node redistributes parameters when the same model
is invoked with incompatible parallel layouts on an identical GPU
allocation. The analytical proxy treats the transformation as a
symmetric all-to-all \(T_{\mathrm{A2A}}(N_m s_b/|\sigma(u)|,|\sigma(u)|)\)
over that allocation. {Actual} \(T_{\mathrm{reshard}}(u,v,m)\)
{depends on the layout and backend: multiple collectives,
nonuniform slices, or host-mediated copies can be much slower than
this idealized NCCL all-to-all, even at equal payload.}

These costs are especially important for bounded asynchrony
\(\tau(m)>0\) and training/serving handoffs. The runtime may hold two parameter versions,
reserve both source and destination worker allocations, and require
redistribution before an invocation can begin. Their exact physical
constraints must be checked during lowering and execution; the
analytical model supplies a plan-ranking approximation, not a guarantee
that a transfer follows its idealized route. The formula-level details
appear in Appendix~\ref{app:comm-reshard}.

\subsection{Memory Model Details}
\label{ls:sec:cost:memory}

\subsubsection{Parameter Shards and Persistent State}
\label{ls:sec:cost:parameters}

The memory model distinguishes {the logical model parameter count}
\(N_m\) {from the worst-stage per-GPU parameter count}
\(W_{\mathrm{shard}}(v)\). Transformer layers are partitioned across
pipeline stages and tensor-parallel ranks, whereas boundary embeddings
and task heads may remain replicated in the simplified layout. The
worst-stage parameter shard \(W_{\mathrm{shard}}(v)\) is therefore not
obtained by dividing the model's total parameter count uniformly by
all GPUs. Tied input/output embeddings may share storage when both
are on one stage; separate boundary stages can retain physically
separate copies of one logically tied matrix. The full stage-level
count is derived in Appendix~\ref{app:wshard}.

For a training device, the persistent analytical footprint
\(S(m,d)\) comprises ZeRO-2-sharded FP32 master weights and Adam moments. BF16 working weights
are counted when materialized, along with any live gradient buffers.
A frozen-model or disjoint asynchronous serving device keeps the
appropriate BF16 parameter shard persistently. A synchronous serving
allocation that only receives parameters transiently does not, under
the simplified lifetime policy, retain a separate full persistent
copy between invocations.

Distinct models colocated on one GPU contribute their persistent
footprints \(S(m,d)\) additively. For a shared actor--critic backbone, shared
parameters must be counted once, with separate policy/value heads
added according to their actual residence. For partial overlap or
multiple versions of one model, residence and copy lifetimes must be
determined by exact allocations and version dependencies, not by the
number of logical invocation vertices alone.

\subsubsection{Transient Activations, Gradients, and KV Cache}
\label{ls:sec:cost:transient}

Training activation memory \(M_{\mathrm{act}}^{T}(v)\) depends on
{up to} \(\min(mb(v),P_p(v))\) {live pipeline
microbatches}, sequence length \(L_p+L_r\), activation-checkpoint
interval \(c\), and intermediate tensor widths. The square-root checkpoint interval is a
practical heuristic rather than the exact optimum for arbitrary
activation-buffer coefficients. The baseline activation expression
conservatively reserves a recomputation buffer for each live warm-up
microbatch; a tighter expression is also derived in
Appendix~\ref{app:transient-memory}, but was not substituted into the
reported baseline calculations.

Teacher-forced inference retains only local forward intermediates
\(M_{\mathrm{act}}^I(v)\), whereas autoregressive generation
additionally retains a KV cache \(M_{\mathrm{kv}}(v)\) whose peak
scales with the number of active sequences, their admitted context
lengths, and the physical KV-head width on each GPU. Average
lengths alone need not bound worst-case cache residence. In grouped-query attention,
KV storage cannot generally be estimated by dividing the full hidden
width by tensor-parallel degree: ranks may own different numbers of
KV heads, including replicated groups. The role-specific head also
matters for transient buffers: a critic or reward scalar head does
not generally require a vocabulary-wide output allocation.

The transient footprint \(A(v,d)\) additionally contains
materialized BF16 weights, live gradients and buckets,
transfer/resharding workspaces, and kernel scratch buffers when they
are not already classified as persistent. Double counting must be avoided when BF16 serving
weights are already resident. The simple activation/KV expressions
are only components of the complete transient bound, not its entirety.

\subsubsection{Co-Residency and Device Feasibility}
\label{ls:sec:coresidency}

As described in Section~\ref{sec:coresidency}, {the analytical
device peak} \(\mathrm{peak}(d)\) adds all co-resident persistent
model state \(S(m,d)\) and the largest transient allocation
\(A(v)\) assigned to GPU \(d\). The maximum is appropriate when
operations assigned to one GPU execute exclusively and their
transient buffers have the assumed lifetime. When several worker
processes persist on one GPU, allocator reservations, retained model
versions, and transfer buffers can overlap beyond that bound.

The device-feasibility predicate \(\mathrm{MemFeasible}(C)\)
checks every GPU against
\(M_{\mathrm{eff}}=(1-\varepsilon_{\mathrm{rt}})M_{\mathrm{GPU}}\),
with a calibrated safety fraction for fragmentation and other runtime
overheads. The nominal 10\% reserve used in the analytical setup is a
modeling parameter, not a universal physical guarantee; actual
allocated and reserved HBM and the effective per-run capacity govern
runtime admission. Search may assign a soft penalty to memory-violating
proposals, but only feasible final plans are selected for execution.

The five placement cases in Figure~\ref{fig:memory-model} of the main
paper illustrate the distinction: synchronous disjoint serving uses
transient weights; asynchronous disjoint serving maintains shadow
copies; partial overlaps mix the two; and fully colocated execution
may require layout resharding.
The equations for \(S(m,d)\), \(A(v,d)\), per-device peaks, and the
memory-feasibility predicate are given in
Appendix~\ref{app:device-peak}.

%% file: appendix-derivations.tex
\clearpage
\section{Analytical Derivations}
\label{app:ad-derivations}

This appendix gives the RPEG-specific performance-model derivations omitted from Section~\ref{sec:model}.
{We reuse the work--depth--memory foundations of \cite{besta2026performance} instead of repeating that paper's per-role FLOP and logical-storage derivations. Our focus is the calibrated execution-time estimator and its physical per-GPU memory accounting.}
We denote node runtime by \(\mathcal{T}\). For an operator node \(v\), let

\[
\Pi(v)=(P_d,P_p,P_t), \qquad P=P_dP_pP_t,
\]
and \(m=\mu(v)\) be the invoked model. The model architecture is
\[
(L_m,d_m,d_{\mathrm{ff},m},V_m,d_{h,m}),
\]
where \(L_m\) is the number of transformer layers, \(d_m\) is the hidden dimension, \(d_{\mathrm{ff},m}\) is the FFN dimension, \(V_m\) is the vocabulary size, and \(d_{h,m}\) is the attention head dimension. The runtime tensor precision uses \(s_b\) bytes per scalar.
We use \(B\) prompts per step, \(R\) rollouts per prompt,
and representative average prompt and rollout lengths \(L_p\) and \(L_r\).

\subsection{Operator-Node Runtime}
\label{app:operator-runtime}

For an operator node \(v\in\hat V\), the runtime model decomposes execution time into five terms:
\[
\begin{aligned}
\mathcal{T}(v)&=T_{\mathrm{gemm}}(v)+T_{\mathrm{bubble}}(v)
+T_{\mathrm{comm}}(v)\\
&\quad+T_{\mathrm{non\mbox{-}gemm}}(v)+T_{\mathrm{opt}}(v).
\end{aligned}
\]

The terms capture GEMM compute, pipeline bubbles, intra-operator communication, non-GEMM overheads, and optimizer traffic, respectively. The decomposition is necessary because a single peak-throughput estimate
\[
\mathcal{T}(v)\approx \frac{W(v)}{P_dP_pP_tF_{\mathrm{peak}}}
\]
does not capture small-batch GEMM under-utilization, non-GEMM operations that do not shrink with tensor parallelism, or memory-bandwidth-bound optimizer traffic.

\paragraph{GEMM time.}
Under 3D parallelism, the total GEMM work \(W(v)\) is distributed over
the \(P\) devices. Assuming balanced work across the 3D mesh, the raw GEMM time is
\[
T_{\mathrm{gemm}}(v)
=
\frac{W(v)}{P\,F_{\mathrm{eff}}(v)},
\]
where \(F_{\mathrm{eff}}(v)\) is the achieved per-device throughput. We model it by a roofline-style bound corrected for calibrated GEMM utilization:
\[
F_{\mathrm{eff}}(v)
=
\min\!\left(F_{\mathrm{peak}}\eta(v),\, I(v)BW_{\mathrm{mem}}\right).
\]
Here, \(F_{\mathrm{peak}}\) is workload-effective peak throughput, \(BW_{\mathrm{mem}}\) is HBM bandwidth, \(I(v)\) is arithmetic intensity, and \(\eta(v)\) is GEMM utilization.

\paragraph{Arithmetic intensity.}
For inference and training operators processing sequences of length
\[
\ell = L_p+L_r,
\]
the per-device arithmetic intensity is
\[
I(v)=\frac{W_{\mathrm{dev}}(v)}{\Omega_W+\Omega_A},
\]
where \(W_{\mathrm{dev}}(v)\) is per-device work for one microbatch.
For inference, \(W_{\mathrm{dev}}=W_I(v)/(P\,mb)\); for training, \(W_{\mathrm{dev}}=W_T(v)/(4P\,mb)\), representing one forward-equivalent pass.
The HBM-traffic terms \(\Omega_W\) and \(\Omega_A\) count weight and activation bytes, respectively, per device and microbatch:
\begin{align*}
\Omega_W
&=
\frac{L_m}{P_pP_t}
\left(4d_m^2+2d_md_{\mathrm{ff},m}\right)s_b,\\
\Omega_A
&=
2\,b_{\mathrm{loc}}\,\ell\,d_m\frac{L_m}{P_p}s_b,
\end{align*}
with local microbatch size $b_{\mathrm{loc}}=\frac{BR}{P_d\,mb}$.

The factor four above converts training FLOPs to a single forward-equivalent pass. \(\Omega_W\) assumes one weight read, and \(\Omega_A\) one activation read and write per local layer. For generation, prefill uses \(\ell=L_p\) and decode uses \(\ell=1\) per token.

\paragraph{GEMM utilization and generation.}
The calibrated GEMM utilization is
\[
\eta(v)
=
\frac{b_{\mathrm{loc}}\ell}{b_{\mathrm{loc}}\ell+\gamma_M}
\cdot
\frac{
\min(d_m/P_t,d_{\mathrm{ff},m}/P_t)
}{
\min(d_m/P_t,d_{\mathrm{ff},m}/P_t)+\gamma_K
}.
\]
The first factor captures under-utilization from small row dimension \(b_{\mathrm{loc}}\ell\); the second captures under-utilization from tensor-parallel narrowing of the hidden or FFN dimension. The constants \(\gamma_M,\gamma_K\) are obtained from calibration.

For generation, we evaluate the two phases separately:
\[
T_{\mathrm{gemm}}^{\mathrm{gen}}(v)
=
\frac{W_{\mathrm{prefill}}(v)}{P\,F_{\mathrm{eff}}^{\mathrm{prefill}}(v)}
+
\frac{W_{\mathrm{decode}}(v)}{P\,F_{\mathrm{eff}}^{\mathrm{decode}}(v)}.
\]

\subsection{Role-Specific Work Terms}
\label{app:work-terms}

The work--depth--memory analysis of generation, inference,
and training (including the actor, reference, value, reward, and
shared-model heads) is already derived in
\cite{besta2026performance}; we do not reproduce it here. We use the
corresponding leading-order role-specific FLOP terms
\(W_G(m)\), \(W_I(m)\), and \(W_T(m)\) to define \(W(v)\) for each
executable invocation, expressed in the workload notation
\((B,R,L_p,L_r)\). The following sections instead derive the
RPEG-specific translation of this work into predicted GPU time,
communication costs, and physical memory feasibility.

\subsection{Pipeline Bubble}
\label{app:pipeline-bubble}

When \(P_p>1\), the \(P_p\) pipeline stages process \(mb\) microbatches. Under a 1F1B schedule, the fill/drain bubble contributes \(P_p-1\) idle slots out of \(mb+P_p-1\) total slots. Equivalently, the combined GEMM and bubble time is
\[
T_{\mathrm{gemm}}(v)+T_{\mathrm{bubble}}(v)
=
T_{\mathrm{gemm}}(v)
\left(
\frac{mb+P_p-1}{mb}
\right).
\]
When \(P_p=1\), the factor is one and \(T_{\mathrm{bubble}}(v)=0\).

\subsection{Intra-Operator Communication}
\label{app:intra-op-comm}

The intra-operator communication term is
\[
T_{\mathrm{comm}}(v)
=
T_{\mathrm{tp}}(v)+T_{\mathrm{pp}}(v)+T_{\mathrm{dp}}(v).
\]
The three terms correspond to tensor-parallel AllReduce, pipeline-parallel point-to-point activation transfer, and data-parallel gradient synchronization. For generation and inference, \(T_{\mathrm{dp}}(v)=0\).

\paragraph{Tensor-parallel AllReduce}
Each transformer layer requires two tensor-parallel AllReduces per forward pass: one after the attention block and one after the FFN block. For inference over length \(\ell=L_p+L_r\),
\[
T_{\mathrm{tp}}^{I}(v)
=
2\frac{L_m}{P_p}\,mb\,
T_{\mathrm{AllReduce}}
\left(
\frac{BR\,\ell\,d_m\,s_b}{P_d\,mb},
P_t
\right).
\]
For training, the simplified forward/recompute/backward count uses three forward-equivalent phases, giving \(6L_m/P_p\) collectives per microbatch:
\[
T_{\mathrm{tp}}^{T}(v)
=
6\frac{L_m}{P_p}\,mb\,
T_{\mathrm{AllReduce}}
\left(
\frac{BR(L_p+L_r)d_m s_b}{P_d\,mb},
P_t
\right).
\]
For generation, prefill and decode are evaluated separately:
\begin{align*}
T_{\mathrm{tp}}^{G}(v)
&=
2\frac{L_m}{P_p}\cdot\\
&\biggl[
mb\,
T_{\mathrm{AllReduce}}
\left(
\frac{BR\,L_p\,d_m\,s_b}{P_d\,mb},
P_t
\right)
+\\
&(L_r-1)
T_{\mathrm{AllReduce}}
\left(
\frac{BR\,d_m\,s_b}{P_d},
P_t
\right)
\biggr].
\end{align*}
The above decode term represents one
batched decode step over all local DP sequences, while prefill is
microbatched. If decode is also executed as \(mb\) distinct microbatches,
consistency instead requires
\[
T_{\mathrm{tp,decode}}^G(v)
=2\frac{L_m}{P_p}\,mb(L_r-1)\,
T_{\mathrm{AllReduce}}\!\left(
\frac{BR d_m s_b}{P_d mb},P_t\right).
\]
The appropriate expression depends on the implementation's
decode batching; the two must not be combined.

\paragraph{Pipeline-parallel P2P}
Pipeline stages exchange TP-sharded activations. Let
\[
M_{\mathrm{pp}}(\ell)
=
\frac{BR\,\ell\,d_m\,s_b}{P_dP_t\,mb}
\]
be the per-microbatch activation message size at sequence length \(\ell\). The division by \(P_t\) assumes that the communicated
hidden tensor is physically TP-sharded. Replicated-output pipeline
layouts require the corresponding actual message size. A pipeline with \(P_p\) stages has \(P_p-1\) stage boundaries. Let \(n_{\mathrm{intra}}\) and \(n_{\mathrm{inter}}\) be the number of intra-node and inter-node boundaries, respectively, with
\[
n_{\mathrm{intra}}+n_{\mathrm{inter}}=P_p-1.
\]
Under nested placement, if a node contains \(N_g\) GPUs and each TP group has size \(P_t\), then approximately \(N_g/P_t\) pipeline stages fit per node, so
\begin{align*}
n_{\mathrm{inter}}
&=
\max\left(
0,\left\lceil \frac{P_p}{N_g/P_t}\right\rceil-1
\right),\\
n_{\mathrm{intra}}&=P_p-1-n_{\mathrm{inter}}.
\end{align*}
For inference,
\begin{align*}
T_{\mathrm{pp}}^{I}(v)
&=
(mb+P_p-2) \cdot \\
&\biggl[
\frac{n_{\mathrm{intra}}}{P_p-1}
\left(\alpha^{\mathrm{p2p}}_{\mathrm{intra}}+M_{\mathrm{pp}}(L_p+L_r)\beta^{\mathrm{p2p}}_{\mathrm{intra}}\right)
+\\
&\frac{n_{\mathrm{inter}}}{P_p-1}
\left(\alpha^{\mathrm{p2p}}_{\mathrm{tier}}+M_{\mathrm{pp}}(L_p+L_r)\beta^{\mathrm{p2p}}_{\mathrm{tier}}\right)
\biggr].
\end{align*}
For training under 1F1B, the number of pipeline communication steps is \(mb+2P_p-3\):
\begin{align*}
T_{\mathrm{pp}}^{T}(v)
&=
(mb+2P_p-3) \cdot \\
&\biggl[
\frac{n_{\mathrm{intra}}}{P_p-1}
\left(\alpha^{\mathrm{p2p}}_{\mathrm{intra}}+M_{\mathrm{pp}}(L_p+L_r)\beta^{\mathrm{p2p}}_{\mathrm{intra}}\right)
+\\
&\frac{n_{\mathrm{inter}}}{P_p-1}
\left(\alpha^{\mathrm{p2p}}_{\mathrm{tier}}+M_{\mathrm{pp}}(L_p+L_r)\beta^{\mathrm{p2p}}_{\mathrm{tier}}\right)
\biggr].
\end{align*}
When \(P_p=1\), we set \(T_{\mathrm{pp}}(v)=0\).

\paragraph{Data-parallel synchronization}
For training operators, the simplified ZeRO-2 communication estimate comprises a ReduceScatter and an AllGather. More precisely, ReduceScatter aggregates/partitions the
gradients; AllGather communicates the subsequently updated parameter
partitions so that data-parallel replicas agree. These collectives can
occur at different phases and need not form an adjacent back-to-back pair. Let \(N_m\) be the total number of scalar parameters of model \(m\). The BF16 shard size per DP group is
\[
M_{\mathrm{dp}}
=
\frac{N_m s_b}{P_pP_t}.
\]
Thus,
\[
T_{\mathrm{dp}}(v)
=
T_{\mathrm{ReduceScatter}}(M_{\mathrm{dp}},P_d)
+
T_{\mathrm{AllGather}}(M_{\mathrm{dp}},P_d).
\]
The collective tier is selected from the full data-parallel allocation.
The common payload \(N_m s_b/(P_pP_t)\) treats parameters
as uniformly PP/TP-sharded. Replicated boundary embeddings or heads
and mixed-precision gradient buckets require the appropriate stage-
specific byte count in a layout-exact implementation.

\subsection{Non-GEMM Overhead}
\label{app:non-gemm}

The work terms \(W(v)\) count matrix multiplication FLOPs. Transformer layers also execute normalization, residual additions, positional embeddings, FlashAttention score computation, and activation functions. These operations have low arithmetic intensity and are therefore assumed to be memory-bandwidth-bound.

At \(P_t=1\), their cost is assumed to be absorbed into the calibrated \(F_{\mathrm{peak}}\), because calibration uses full transformer operations. At \(P_t>1\), GEMM work shrinks with tensor parallelism, whereas some non-GEMM operations, especially normalization and residual additions, do not shrink proportionally. We model the residual TP-invariant overhead with a calibrated per-layer constant \(c_{\mathrm{ng}}\):
\[
T_{\mathrm{non\mbox{-}gemm}}(v)
=
\frac{L_m}{P_p}\,
c_{\mathrm{ng}}
\left(1-\frac{1}{P_t}\right)
\frac{b_{\mathrm{loc}}}{B_{\mathrm{cal}}},
\]
where \(B_{\mathrm{cal}}\) is the per-device batch size used when calibrating \(c_{\mathrm{ng}}\). The expression assumes that the fitted \(c_{\mathrm{ng}}\)
has the same aggregation scope as this operator-level correction.
For training, we multiply this term by \(4\), corresponding to forward, recompute, and backward phases. For generation, we set
\[
T_{\mathrm{non\mbox{-}gemm}}(v)=0,
\]
because decode processes one token per step and its runtime is assumed to be dominated by weight and KV-cache traffic.

\subsection{Optimizer Step}
\label{app:optimizer-step}

For training, we approximate the bandwidth-bound Adam update using an effective traffic coefficient of 18 bytes per parameter:
\[
T_{\mathrm{opt}}(v)
=
\frac{N_m}{P_dP_pP_t}
\cdot
\frac{18}{BW_{\mathrm{mem}}},
\]
where \(N_m/(P_dP_pP_t)\) is the average logical number of parameters per device. For generation and inference operators,
\[
T_{\mathrm{opt}}(v)=0.
\]

\subsection{Explicit Communication and Reshard Nodes}
\label{app:comm-reshard}

The executable DAG also contains explicit \textsc{Comm} and \textsc{Reshard} nodes.

\paragraph{Comm nodes}
A \textsc{Comm} node is inserted when the source and destination operators ($u, v$) are placed on different allocations. If container \(a\) has element count \(|a|\) and element size \(s(a)\), the byte volume is \(|a|s(a)\). Since concurrent send/recv pairs are limited by the side with fewer devices, the transfer time is
\[
T_{\mathrm{xfer}}(u,v,a)
=
\alpha_{\mathrm{p2p}}
+
\frac{|a|s(a)}{\min(|\sigma(u)|,|\sigma(v)|)}
\beta_{\mathrm{p2p}},
\]
where \((\alpha_{\mathrm{p2p}},\beta_{\mathrm{p2p}})\) are selected at the communication tier spanned by \(\sigma(u)\) and \(\sigma(v)\). This is a balanced, fully parallel transfer approximation. It
omits per-pair setup, serialization, packing/repartitioning, and asymmetric
source/destination routes; the physical controller may also reserve
both endpoint allocations while the analytical scheduler is more
optimistic.

\paragraph{Reshard nodes}
A \textsc{Reshard} node is inserted when two consecutive invocations of the same model use the same allocation but incompatible parallel layouts. All devices in the allocation participate symmetrically in the layout transformation. We model this as an all-to-all:
\[
T_{\mathrm{reshard}}(u,v,m)
=
T_{\mathrm{A2A}}
\left(
\frac{N_m s_b}{|\sigma(u)|},
|\sigma(u)|
\right),
\]
where \(N_m s_b\) is the full parameter size of model \(m\) in Bytes, and \(T_{\mathrm{A2A}}(M,p)\) is the all-to-all cost for per-GPU message size \(M\) over \(p\) GPUs.

\subsection{Parameter Accounting and \(W_{\mathrm{shard}}\)}
\label{app:wshard}

We now derive \(W_{\mathrm{shard}}(v)\), the per-GPU parameter footprint of operator \(v\) on its worst pipeline stage. The memory model counts dominant weight matrices: attention projections, FFN projections, embeddings, and task heads. Biases and normalization parameters are omitted.

The logical parameter counts of the policy, reference,
critic, reward, and shared actor--critic are derived in
\cite{besta2026performance}. Here we require only the dominant
per-layer count to determine physical residency on a particular
pipeline stage and GPU:
\[
W_m^{\mathrm{layer}}=4d_m^2+2d_md_{\mathrm{ff},m}.
\]
This leading-order expression assumes standard attention and a two-projection FFN.
The remaining calculation is placement-specific: pipeline stages
hold different layers and boundary embeddings/heads, which is not
captured by a logical per-model parameter total.

Under pipeline parallelism, transformer layers are split across \(P_p\) stages, so each stage receives \(L_m/P_p\) layers (assuming \(P_p\mid L_m\)). However, embeddings and task heads are pinned to boundary stages rather than spread uniformly. Let
\[
E_m=V_md_m
\]
be the embedding matrix size. Define two indicators:
\begin{align*}
\chi_m^{\mathrm{LM}}=
\begin{cases}
1, & m\in\{\pi_\theta,\pi_{\mathrm{ref}},\pi_{\mathrm{AC}}\},\\
0, & \text{otherwise},
\end{cases}\\
\chi_m^{\mathrm{scal}}=
\begin{cases}
1, & m\in\{V_\psi,R_\phi,\pi_{\mathrm{AC}}\},\\
0, & \text{otherwise}.
\end{cases}
\end{align*}
Then the non-layer resident component on pipeline stage \(p\in\{1,\ldots,P_p\}\) is
\begin{align*}
W^{\mathrm{nonsplit}}_m(p)
&=
\mathbf{1}[p=1]\,E_m \\
&+
\mathbf{1}[p=P_p]
\left(
\chi_m^{\mathrm{LM}}E_m+\chi_m^{\mathrm{scal}}d_m
\right).
\end{align*}

The transformer-layer component on stage \(p\) is
\[
W^{\mathrm{split}}_m(p)
=
\frac{L_m}{P_p}W^{\mathrm{layer}}_m.
\]
Tensor parallelism shards transformer-layer weights across \(P_t\) ranks, but embeddings and heads are replicated on each tensor-parallel rank. Hence the per-GPU resident parameter footprint on stage \(p\) is
\[
W^{\mathrm{gpu}}_m(p)
=
\frac{W^{\mathrm{split}}_m(p)}{P_t}
+
W^{\mathrm{nonsplit}}_m(p).
\]
The worst pipeline stage determines the per-GPU parameter footprint:
\[
W_{\mathrm{shard}}(v)
=
\max_{p\in\{1,\ldots,P_p\}}
W^{\mathrm{gpu}}_m(p).
\]
Equivalently,
\begin{align*}
W_{\mathrm{shard}}(v)
&=
\max_{p}
\biggl[
\frac{1}{P_t}
\frac{L_m}{P_p}
(4d_m^2+2d_md_{\mathrm{ff},m})\\
&\text{+}\;\mathbf{1}[p=1]V_md_m\\
&\text{+}\;\mathbf{1}[p=P_p]
\left(
\chi_m^{\mathrm{LM}}V_md_m+\chi_m^{\mathrm{scal}}d_m
\right)
\biggr].
\end{align*}
If tied input/output embeddings are physically shared on the same pipeline stage, the duplicate embedding term can be stored once. The formula above is the conservative resident-footprint model used for boundary-stage memory pressure.

From \(W_{\mathrm{shard}}(v)\), the model-state components follow:
\[
M_w(v)=W_{\mathrm{shard}}(v)s_b
\]
for one BF16 weight copy, $M_{\mathrm{opt}}(v)=\frac{12W_{\mathrm{shard}}(v)}{P_d(v)}$
for ZeRO-2-sharded FP32 Adam state, and $M_{\nabla}(v)=\frac{W_{\mathrm{shard}}(v)s_b}{P_d(v)}$
for ZeRO-2-sharded BF16 gradients.

\subsection{Transient Memory}
\label{app:transient-memory}

Transient memory is allocated when an operator starts and released when it finishes. We denote the transient peak of operator \(v\) by \(A(v)\). Let
\[
b_{\mathrm{loc}}=\frac{BR}{P_d\,mb}
\]
be the per-device microbatch size.

\paragraph{Training activations}
Training retains activations for the backward pass. With activation checkpointing, only every \(c\)-th layer output is retained, and the missing activations are recomputed during backward. We use
\[
c=\max\!\left\{1,\left\lfloor \sqrt{\frac{L_m}{P_p}}\right\rfloor\right\}.
\]

The checkpoint memory stores
\[
\left\lceil \frac{L_m/P_p}{c}\right\rceil
\]
hidden states of width \(d_m\). The recomputation buffer must hold the largest single-layer intermediate. We approximate the per-token peak by
\[
M_{\mathrm{buf}}^{\mathrm{layer}}
=
\max\left(
2d_m+\frac{6d_m}{P_t},
\,
2d_m+\frac{2d_{\mathrm{ff},m}}{P_t}
\right).
\]
Under a 1F1B schedule, the warmup retains activations for \(\min(mb,P_p)\) microbatches. Thus
\begin{align*}
M_{\mathrm{act}}^{T}(v)
&=
s_b\,
\min(mb,P_p)\,
b_{\mathrm{loc}}\,
(L_p+L_r) \cdot \\
&\left[
\left\lceil \frac{L_m/P_p}{c}\right\rceil d_m
+
c\,M_{\mathrm{buf}}^{\mathrm{layer}}
\right].
\end{align*}
The displayed expression conservatively reserves both the
checkpoint and recomputation-buffer contributions for up to
\(\min(mb,P_p)\) live microbatches. 

\paragraph{Inference activations}
For inference and generation, only one layer's intermediates are live at a time. The peak is the maximum of the attention intermediate, FFN intermediate, and LM-head intermediate: The baseline uses the LM-head width as a conservative
upper bound. For critic/reward-only scalar heads, a role-specific
estimator can replace \(d_m+V_m\) with \(d_m+1\).
\footnotesize\begin{align*}
M_{\mathrm{act}}^{I}(v)
=
s_b\,b_{\mathrm{loc}}\,\ell\, \cdot \max\left(
2d_m+\frac{6d_m}{P_t},
\,
2d_m+\frac{2d_{\mathrm{ff},m}}{P_t},
\,
d_m+V_m
\right),
\end{align*}
\normalsize
where \(\ell=L_p+L_r\) for teacher-forced inference over the full prompt--response sequence and \(\ell=L_p\) for generation prefill.

\paragraph{KV cache for generation}
Autoregressive generation maintains a key-value cache. At the last decode step, the cache covers a representative \(L_p+L_r\) tokens. Each layer stores one key and one value vector per token, sharded across pipeline and tensor parallelism:
\[
M_{\mathrm{kv}}(v)
=
2s_b
\cdot
\frac{B}{P_d}
\cdot R
\cdot (L_p+L_r)
\cdot
\frac{L_m}{P_p}
\cdot
\frac{d_m}{P_t}.
\]
The factor \(2s_b\) accounts for both key and value caches.  

This expression assumes ordinary multi-head attention and
an evenly sharded KV dimension. With grouped-query attention, the
physical per-TP-rank KV width \(d_{\mathrm{kv,local}}\)
replaces \(d_m/P_t\), giving
\[
M_{\mathrm{kv}}^{\mathrm{GQA}}(v)
=2s_b\frac{BR}{P_d}(L_p+L_r)\frac{L_m}{P_p}
\,d_{\mathrm{kv,local}}.
\]
Replicated KV groups must be counted as replicated rather
than divided again by \(P_t\).

Training additionally includes transient BF16 weights and sharded gradients.
Overall, we have:
\[
A(v)=
\begin{cases}
\begin{aligned}M_{\mathrm{act}}^{T}(v)\\+M_w(v)+M_{\nabla}(v)\end{aligned}, & r(v)=\mathrm{training},\\[2mm]
M_{\mathrm{act}}^{I}(v), & r(v)=\mathrm{inference},\\[2mm]
M_{\mathrm{act}}^{I}(v)+M_{\mathrm{kv}}(v), & r(v)=\mathrm{generation}.
\end{cases}
\]

\subsection{Co-Residency-Aware Device Peaks}
\label{app:device-peak}

The formulas above describe memory for an isolated operator. RLM training pipelines colocate multiple models and roles on the same physical devices, so memory feasibility must account for persistent state that remains resident across the schedule and transient state that appears only during operator execution.

For a model \(m\) on device \(d\), define \(S(m,d)\) as the persistent footprint, excluding transient activations and KV cache: For a device holding neither training, shadow, nor frozen
state for model \(m\), set \(S(m,d)=0\). A training device
stores the modeled sharded master/moment arrays; any BF16 copy not
persistently retained must appear in the transient term.
\[
S(m,d)=
\begin{cases}
\dfrac{12\,W_{\mathrm{shard}}(m,d)}{P_d}, &
\begin{aligned}
&\text{if \(d\) serves \(m\) as a}\\
&\text{training device},
\end{aligned}\\
W_{\mathrm{shard}}(m,d)s_b, &
\begin{aligned}
&\text{if \(d\) serves \(m\) as a shadow}\\
&\text{or frozen device}.
\end{aligned}
\end{cases}
\]
Here, \(W_{\mathrm{shard}}(m,d)\) denotes the per-GPU resident parameter shard of model \(m\) on device \(d\), derived from the corresponding operator-level \(W_{\mathrm{shard}}(v)\).

A training device stores persistent FP32 Adam state; BF16 weights are materialized from the FP32 master weights on demand and released afterward. A shadow device stores a persistent BF16 copy for asynchronous generation or inference. A frozen device stores persistent BF16 weights for a non-trainable model such as the reference or reward model.

The per-device peak memory is persistent co-resident state plus the worst transient operator mapped to that device:
\[
\mathrm{peak}(d)
=
\sum_{m\ \mathrm{placed\ on}\ d} S(m,d)
+
\max_{v\ \mathrm{placed\ on}\ d} A(v).
\]

A configuration is memory-feasible iff
\[
\forall d\in D:\quad \mathrm{peak}(d)\le M_{\mathrm{eff}},
\]
where
\[
M_{\mathrm{eff}}=(1-\varepsilon_{\mathrm{rt}})M_{\mathrm{GPU}}
\]
reserves a safety margin for runtime overhead and fragmentation.